\documentclass[preprint,12pt]{elsarticle}
\usepackage{amssymb}

\usepackage{array}
\usepackage{makecell}
\usepackage{amsmath}
\usepackage{rotating}
\usepackage{multirow}
\usepackage{booktabs}
\usepackage{tablefootnote}
\usepackage{array}
\usepackage{longtable} 
\newcolumntype{C}[1]{>{\centering\arraybackslash}p{#1}}
\newcolumntype{L}[1]{>{\raggedright\arraybackslash}p{#1}}
\newcolumntype{R}[1]{>{\raggedleft\arraybackslash}p{#1}}
\usepackage{colortbl}
\usepackage[table, svgnames, dvipsnames]{xcolor}
\usepackage{todonotes}

\usepackage{caption}
\usepackage{url}
\usepackage{hyperref}

\usepackage{lscape} 
\usepackage{pdflscape}

\usepackage[
  draft,
  markup=margin,
  authormarkup=none
]{changes}
\usepackage{xcolor}

\usepackage{csquotes}

\definechangesauthor[name={}, color=blue]{r1}
\definechangesauthor[name={}, color=blue]{r2}
  {\begin{added}[comment={#2}]}%
  {\end{added}}

\journal{AI-Enhanced Business Process Management}

\begin{document}

\begin{frontmatter}



\title{Assessing the Business Process Modeling Competences of Large Language Models}


\author[1,2]{Chantale Lauer}
\author[1,3]{Peter Pfeiffer}
\author[1,2]{Alexander Rombach}
\author[1,2]{Nijat Mehdiyev} %% Author name

\affiliation[1]{organization={German Research Center for Artificial Intelligence (DFKI)},%Department and Organization
            addressline={Campus D3 2}, 
            city={Saarbrücken},
            postcode={66123}, 
            %state={Saarland},
            country={Germany}}
\affiliation[2]{organization={Saarland University},%Department and Organization
            addressline={Campus D3 2}, 
            city={Saarbrücken},
            postcode={66123}, 
            %state={Saarland},
            country={Germany}}
\affiliation[3]{organization={5Plus GmbH},%Department and Organization
            addressline={Goethestraße 1}, 
            city={Würzburg},
            postcode={97072}, 
            %state={Bavaria},
            country={Germany}}
%% Abstract
\begin{abstract}
The creation of Business Process Model and Notation (BPMN) models is a complex and time-consuming task requiring both domain knowledge and proficiency in modeling conventions. Recent advances in large language models (LLMs) have significantly expanded the possibilities for generating BPMN models directly from natural language, building upon earlier text-to-process methods with enhanced capabilities in handling complex descriptions.
However, there is a lack of systematic evaluations of LLM-generated process models. Current efforts either use LLM-as-a-judge approaches or do not consider established dimensions of model quality. 
To this end, we introduce BEF4LLM, a novel LLM evaluation framework comprising four perspectives: syntactic quality, pragmatic quality, semantic quality, and validity. 
Using BEF4LLM, we conduct a comprehensive analysis of open-source LLMs and benchmark their performance against human modeling experts. Results indicate that LLMs excel in syntactic and pragmatic quality, while humans outperform LLMs in semantic aspects; however, the differences in scores are relatively modest, highlighting LLMs' competitive potential despite challenges in validity and semantic quality. 
The insights highlight current strengths and limitations of using LLMs for BPMN modeling and guide future model development and fine-tuning. Addressing these areas is essential for advancing the practical deployment of LLMs in business process modeling.
\end{abstract}

\begin{highlights}
\item We introduce the BEF4LLM framework for evaluating LLMs in text-to-BPMN tasks
\item A large-scale benchmark of open-source LLMs shows significant differences
\item Larger LLMs do not necessarily generate higher-quality BPMN models
\item Some LLMs can create BPMN models similar to modeling experts
\item Valid BPMN-XML generation remains a major challenge for current LLMs
\end{highlights}

%% Keywords
\begin{keyword}
Conversational Process Modeling \sep Business Process Management \sep Large Language Models \sep BPMN


\end{keyword}

\end{frontmatter}



\section{Introduction}

The modeling of business processes using the \emph{Business Process Model and Notation} (BPMN) is fundamental to organizational analysis, communication, and automation \cite{Dumas2018, Mendling_Metrics_2008}.  
BPMN allows practitioners to capture complex procedural knowledge in a clear, standardized form \cite{Dumas2018, Dijkman_2007Formal_SA}.  
Creating high-quality BPMN diagrams is, however, cognitively demanding and time-consuming; it typically requires rare experts who master both the application domain and BPMN model's syntax and semantics \cite{Mendling_Metrics_2008, Dumas2018}.  
This reliance on scarce expertise remains a persistent challenge in business process management (BPM) \cite{Compagnucci2024}.

Recent advances in artificial intelligence, especially large language models (LLMs), now make it possible to generate structured artifacts, including process models, directly from natural language text \cite{grohs_large_2023,kourani_process_2024,Kourani_Benchmark_2024,Lauer_Generating_2025}.  
Their capabilities in interpreting and generating text underpin LLM-assisted process modeling and have led to conversational BPMN modeling \cite{klievtsova_conversational_2023}, where process models are iteratively co-constructed through dialogue between humans and an LLM-powered assistant \cite{Lauer_Generating_2025}. This allows faster process model construction and enables non-experts to obtain accurate process models.

While initial experiments have shown promising results in generating process models from textual descriptions, e.g., BPMN models \cite{grohs_large_2023,kourani_process_2024}, an extensive evaluation of LLMs' capabilities for BPMN modeling remains open.
Throughout the years, several process modeling guidelines \cite{Becker_GOM_2000,Mendling_7PMG_2010}, quality assessment frameworks \cite{Krogstie_sequal_BPM_2016,Reijers_Bsuiness_2010}, and a large amount of metrics, e.g., size, density, sequentiality, or cyclicity to assess the quality of process models in various aspects have been developed \cite{Dijkman_2007Formal_SA,Mendling_Metrics_2008,Schoknecht_SimilaritySOTA_2017}. These provide a valuable source for constructing an assessment framework for LLM-driven BPMN modeling, which can be used to get an objective picture of the capabilities of LLMs in this task. Thereby, the strengths and weaknesses of LLMs, also in comparison to human modelers, can be identified. This also supports the further development of LLMs and their application to BPM tasks that demand a processual understanding.

Although these tools exist, they have not been applied in a systematic manner to LLM-based BPMN modeling yet. This work aims to fill this gap by making the following main contributions:

\begin{enumerate}
    \item We propose the BPMN Evaluation Framework for LLMs (BEF4LLM) framework, building on the SIQ framework~\cite{Reijers_Bsuiness_2010}, which comprises 39 metrics for assessing BPMN models across four quality dimensions: syntactic quality, pragmatic quality, semantic quality, and validity. The framework is specifically designed to enable automated, large-scale evaluation of LLMs in BPMN modeling.

    \item Using the BEF4LLM framework, a large-scale benchmark of open-source LLMs is conducted, including 17 different-sized LLMs of various families on 105 curated text-BPMN model pairs. This marks the first extensive benchmark of open-source LLMs in generating BPMN models based on objective process quality metrics, obtaining rich insights into the current LLM landscape with regard to this task. 

    \item We perform a detailed analysis of the experiment results, comparing LLMs across quality dimensions and parameter counts using statistical tests. Further, we compare the performance of LLMs with human experts on a smaller subset of the data. 
\end{enumerate}

The findings offer a standardized procedure for evaluating LLMs in BPMN modeling as well as concrete guidance on LLM selection and their further development, e.g., fine-tuning requirements and open research challenges. 

The rest of the paper is structured as follows. 
In the next section, we introduce and define key concepts, including business process modeling, particularly BPMN and its specification, process model quality, and large language models (LLMs). \autoref{chap:related_work} elaborates on related work, including quality frameworks and assessment for process models, LLM applications in BPM, and LLM benchmarks. \autoref{chap:framework} introduces the BEF4LLM framework, including its quality measurement components and metric calculation. The experimental setting is introduced in \autoref{chap:exp_settings}, followed by the presentation of the results in \autoref{chap:results}. \autoref{chap:discussion} discusses the results, highlighting strengths and weaknesses of LLMs in BPMN model generation as well as limitations of our work. Finally, the paper is concluded in \autoref{chap:conclusion}.

\section{Preliminaries}

In this section, we present and define the basic knowledge required to understand this paper, including business process modeling with BPMN, process model quality, and LLMs.

\subsection{Business Process Modeling Notation (BPMN)}

Business process modeling refers to the systematic abstraction and representation of organizational activities as structured process models. Such process models capture those aspects of business processes that are pertinent to analysis, communication, or automation objectives, deliberately omitting extraneous detail. 

\paragraph{Foundations and Modeling Elements}
Among available modeling approaches, BPMN is widely adopted in the information systems (IS) discipline due to its expressive power and its standardized graphical notation, which is intended to be readily understandable to both business and technical stakeholders~\cite{Mendling_Metrics_2008, Compagnucci2024}. BPMN specifies four principal categories of modeling constructs~\cite{Dijkman_2007Formal_SA, Saber_formalization_2014}: (i) \textbf{flow objects} (events, activities, gateways), (ii) \textbf{connecting objects} (sequence flows, message flows, associations), (iii) \textbf{swimlanes} (pools and lanes), and (iv) \textbf{artifacts} (data objects, groups, annotations). In the following, we formalize only the BPMN elements required for our framework and experiments. A complete BPMN formalization is beyond the scope of this paper. Consequently, we do not cover artifacts, associations, subprocesses, or groups, and we omit an exhaustive formalization of all task and event subtypes.

\begin{figure}[htbp]
    \centering
    \includegraphics[width=0.95\linewidth]{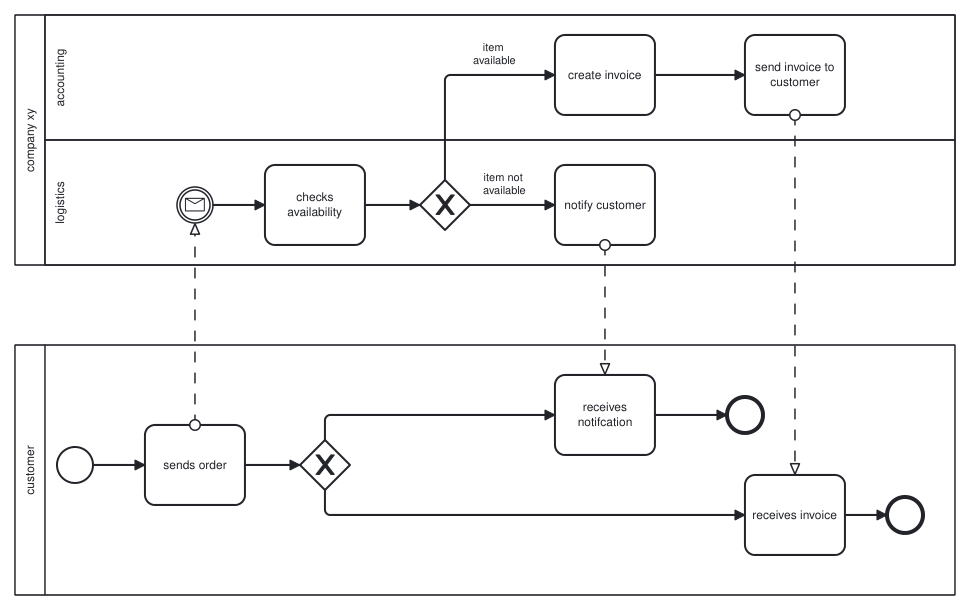}
    \caption{Illustrative example of a BPMN model.}
    \label{fig:example_bpmn}
\end{figure}

\paragraph{Structural Formalization and Notation}
Figure~\ref{fig:example_bpmn} presents a BPMN model featuring two processes — one for a customer and another for an organizational entity (``company xy''). Each process is encapsulated in a pool $\mathcal{PO}$, which may be subdivided into lanes $\mathcal{L}$ to represent internal subunits such as ``logistics'' or ``accounting''. Collectively, pools and lanes are termed \emph{swimlanes}. 

Formally, a BPMN process (set $\mathcal{P}$) is modeled as a directed graph composed of flow objects ($\mathcal{FO}$) and flow connections ($\mathcal{F}$). Flow objects include: \textbf{events} ($\mathcal{E}$), partitioned into start ($\mathcal{E}^{\mathrm{S}}$), intermediate ($\mathcal{E}^{\mathrm{I}}$), and end ($\mathcal{E}^{\mathrm{E}}$) events — with message events denoted as $\mathcal{E}^{\mathrm{S}_{\!M}}$, $\mathcal{E}^{\mathrm{I}_{\!M}}$, and $\mathcal{E}^{\mathrm{E}_{\!M}}$;
\textbf{activities} ($\mathcal{A}$), comprising atomic tasks ($\mathcal{T}$) and subprocesses; and \textbf{gateways} ($\mathcal{G}$), classified as parallel ($\mathcal{G}_\mathrm{AND}$), exclusive ($\mathcal{G}_\mathrm{XOR}$), inclusive ($\mathcal{G}_\mathrm{OR}$), or event-based ($\mathcal{G}_\mathrm{EVENT}$), and further as split ($\mathcal{G}^{\mathrm{S}}$) or join ($\mathcal{G}^{\mathrm{J}}$) gateways. Corresponding split and join gateways are formally related via a mapping function $\text{Matchgate}: \mathcal{G}^{\mathrm{S}} \rightarrow \mathcal{G}^{\mathrm{J}}$. Exception handling is modeled by the partial function $\text{Excp}: \mathcal{E}^{\mathrm{I}} \rightharpoonup \mathcal{A}$, associating \textbf{interrupting} intermediate events with the activities they abort. For any flow object $x \in \mathcal{FO}$, $\mathit{label}(x)$ retrieves its textual label (or returns $\varnothing$ if unlabeled). Tasks (rectangles), events (circles), and gateways (diamonds) are visually distinguished, with tasks typically labeled (e.g., ``send order'').

\paragraph{Process Connectivity and Control Flow}
Flow connections are described by the set $\mathcal{F} \subseteq (\mathcal{FO} \cup \mathcal{PO}) \times (\mathcal{FO} \cup \mathcal{PO})$, partitioned into \textbf{sequence flows} ($\mathcal{F}^{\mathrm{S}}$, solid arrows) specifying intra-process execution order, and \textbf{message flows} ($\mathcal{F}^{\mathrm{M}}$, dashed arrows) representing inter-pool communication. 
Message flows depict communication between pools or between a flow object and a pool (e.g., a customer sending an order to a company).
For any $x \in \mathcal{FO}$, the set of incoming sequence flows is $in(x) = \{y \in \mathcal{FO} \mid (y, x) \in \mathcal{F}^{\mathrm{S}}\}$, and the set of outgoing sequence flows is $out(x) = \{y \in \mathcal{FO} \mid (x, y) \in \mathcal{F}^{\mathrm{S}}\}$. A \emph{path} is defined as a non-empty sequence of flow objects connected by sequence flows, from a start to an end event. 
%Message flows depict communication between pools or between a flow object and a pool (e.g., a customer sending an order to a company).

\subsection{Process Model Quality} \label{subsec:PMQ}
With process model quality, we refer to the extent to which a process model meets certain standards and criteria. Thereby, frameworks define the scope and terms of process model quality, while metrics provide concrete measurements for assessing it. Those metrics enable the measurement of certain aspects of the process model complexity, such as size, density, partitionability, connector interplay, cyclicity, and concurrency, edge style, crossing edges, edge angles, symmetry in blocks, as well as the consistency flow \cite{Sorg_complexity_2025, Mendling_Metrics_2008}. These evaluate the process specifications encoded in the process model as well as the process model layout and representation. Those metrics now allow us to analyze and judge the process model and its quality aspects. 

Several frameworks defining process model quality, like SEQUAL \cite{Krogstie_2006_quality_framework, Krogstie_sequal_BPM_2016} or SIQ \cite{Reijers_Bsuiness_2010}, exist. We build upon the SIQ framework \cite{Reijers_Bsuiness_2010}, which defines process model quality in three dimensions: syntactic, pragmatic, and semantic quality. The abbreviation SIQ also hints at its characteristics, as "it is \textit{S}imple enough to be practically applicable, yet \textit{I}ntegrates the most relevant insights from the BPM field, while it deals with \textit{Q}uality" \cite[~p.171]{Reijers_Bsuiness_2010}. Thus, the intention behind building the SIQ framework was to establish a simple process quality definition that can easily be used in practice without neglecting important aspects of BPM field. 

\begin{figure}
    \centering
    \includegraphics[width=0.7\linewidth]{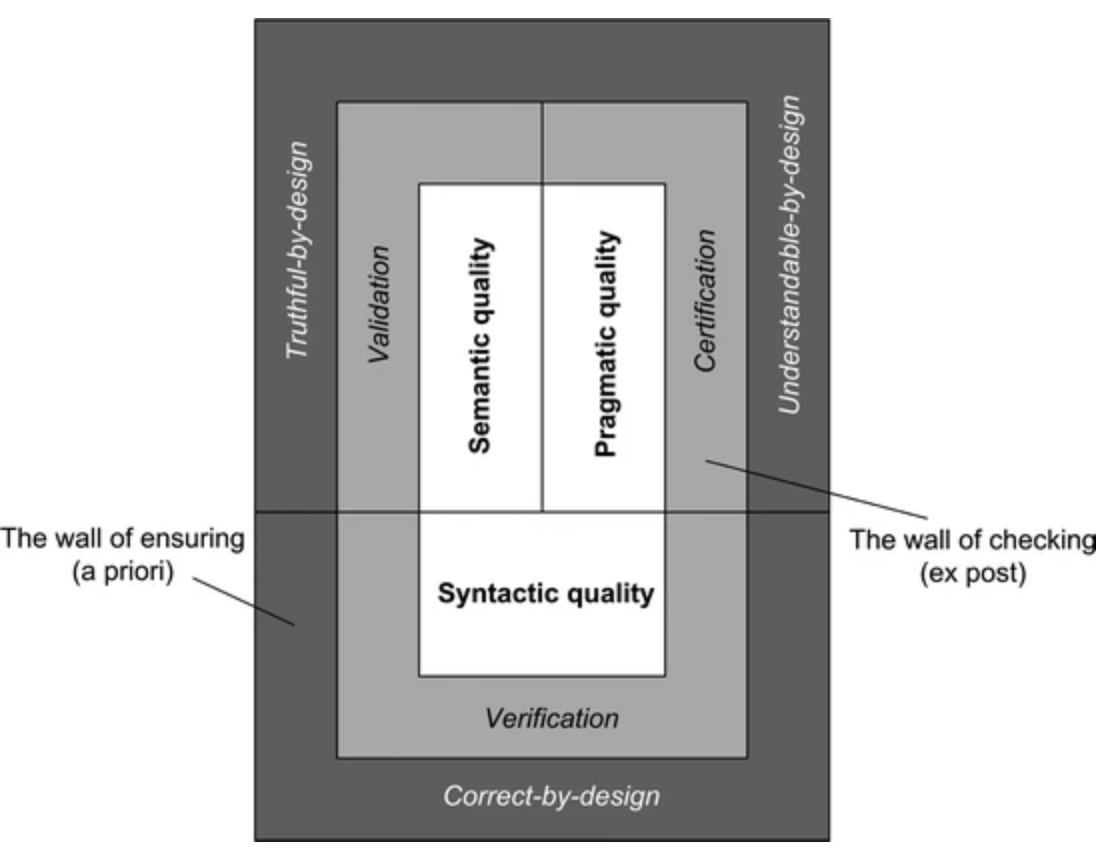}
    \caption{SIQ framework \cite{Reijers_Bsuiness_2010}.}
    \label{fig:SIQ_framework}
\end{figure}

The SIQ framework, as shown in \autoref{fig:SIQ_framework} is based on syntactic, pragmatic, and semantic quality aspects of process model quality displayed in its center. The surrounding "wall" concerns the establishment of the validity of the solution. For each of the three approaches, validation, verification, and certification, different methods can be used to inspect the degree of validity. Lastly, the "wall of ensuring" introduces the concepts of \enquote{truth-by-design}, \enquote{understandable-by-design}, and \enquote{correct-by-design}, which aim to prevent threats to the correctness of the resulting process model.

\subsection{Large Language Models}
We refer to an LLM as a language model, pretrained on large amounts of text, that exhibits strong knowledge about natural language and world facts \cite{Jurafsky_speech_2024}. It processes and generates natural language, making it a generative machine learning model. LLMs are neural networks with (typically) billions of trainable parameters, comprising stacked transformer blocks \cite{Vaswani_attention_2023} consisting of multi-head self-attention and position-wise feed-forward layers. Such LLMs can be prompted, i.e., they generate answers based on the instruction in the prompt and the input received. This makes them particularly useful for solving a variety of tasks.

\autoref{tab:LLM_overview_technical_data} provides an overview of open-source LLM families that will be considered in this study, characterized by their parameter count, i.e., the number of trainable parameters in the LLM, indicating their size, and their context length. The context length defines the maximum number of tokens the LLM can process at once, limiting how much information it can take into account when generating the output. 
Another important parameter for LLMs is the temperature, which rescales the output probability distribution before sampling, influencing randomness of the outputs \cite{Peeperkorn_Temperature_2024} and is often described as the parameter that enables creativity in LLMs.

\begin{table}[ht]
    \centering
    \tabcolsep=5pt

    \resizebox{\textwidth}{!}{
    \begin{tabular}{lrrrrr}
    \toprule
    \textbf{LLM family} &\textbf{ Release Date} & \textbf{Size} & \textbf{Context Length} \\
    \midrule
     Llama3 & April - December 2024 & 8B - 405B & 128K  \\
     Qwen2.5 & September 2024 & 0.5b - 72B & 32K \& 128K  \\
     Falcon3 & December 2024 & 1B -10B  & 8K - 32K \\
     Phi4 & December 2024 & 14B & 16K\\
     %Llama 4 & April 2025 & 17B - 288B 1M - 10M \\
     Deepseek-R1 & Januar 2025 & 1.5B - 70B & 128K \\
     Qwen3 & April 2025 & 0.6B - 235B &  40K \\
     \bottomrule
     \end{tabular}}
     \caption{LLM families and their main characteristics used in this study.}
     \label{tab:LLM_overview_technical_data}
\end{table}

Training LLMs is a multi-step procedure. First, the LLM is pretrained like other language models in a self-supervised manner by predicting the next token in a sentence using corpora of billions of tokens \cite{Jurafsky_speech_2024}. Post-training further trains the LLM to follow instructions, i.e., prompts, and to ensure that the LLM's responses align with human needs. Therefore, during instruction finetuning, the LLM is given instruction-response pairs, and it is trained to generate the response. Although the LLM generates the answer by predicting token-by-token as during pretraining, this training step is supervised since the responses are curated with human supervision \cite{Jurafsky_speech_2024}. As a result, the LLM learns to map natural language instructions to appropriate responses. Finally, to ensure that the LLM responses align with human values and needs (and to prevent harmful or toxic responses), the LLM is finetuned to generate responses from human feedback, often involving reinforcement learning techniques. Note that some LLMs are not instruction-tuned, which usually results in lower performance at following instructions in prompts. 

In order to reduce the computational and memory costs of LLMs, a technique called quantization is often applied \cite{Dettmers_GPT3_2022}. The idea is to reduce the representation of weights and activations from high-precision data types like 32-bit floating-point to lower precision ones like 8-bit integers, which can be processed faster by processors. Quantization may introduce degradation in performance, particularly at very low bit widths. The lower the target data type, the greater the precision loss. One of the most common quantization cases is going from 32-bit float to 8-bit integer, which typically offers a good balance between computational costs and precision \cite{Dettmers_GPT3_2022}.

\section{Related Work}
\label{chap:related_work}
This section introduces relevant work regarding process model quality assessment in \autoref{chap:related_work_pm_quality}, LLMs for BPM tasks in \autoref{chap:related_work_llms_bpmn}, and LLMs benchmarks, including benchmarks for LLM-based process modeling in \autoref{chap:related_work_llms_benchmark}.

\subsection{Process Model Quality Frameworks and Assessment}
\label{chap:related_work_pm_quality}

\begin{figure}[htbp!]
    \centering
    \includegraphics[width=.85\linewidth]{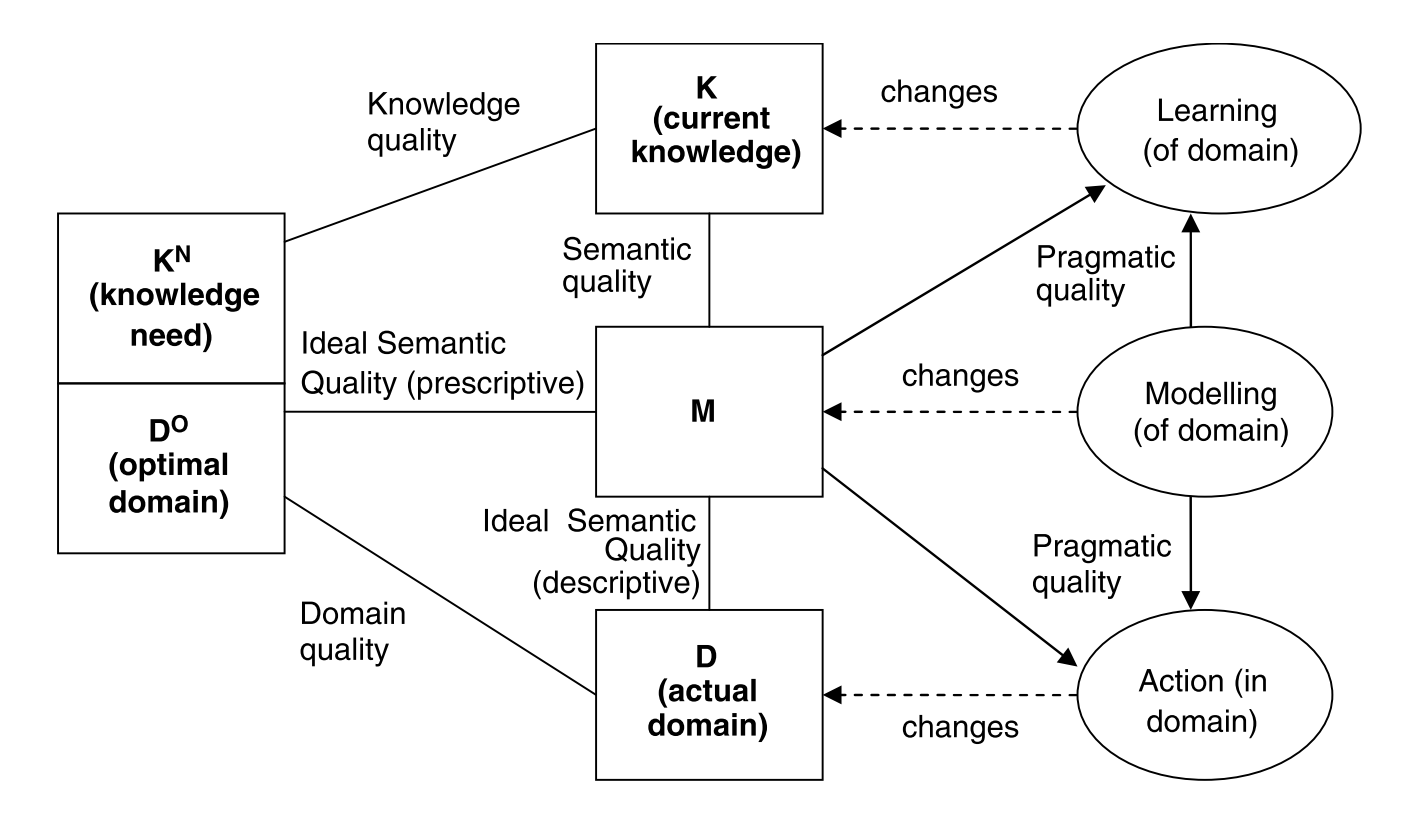}
    \caption{SEQUAL framework \cite{Krogstie_sequal_2006_old}.}
    \label{fig:sequal}
\end{figure}

This subsection summarizes related work on process quality frameworks and process model quality assessment. 
In recent work, different frameworks defining process model quality have been presented. SIQ \cite{Reijers_Bsuiness_2010}, on which BEF4LLM builds upon as introduced in \autoref{subsec:PMQ}, is one example of a process model quality framework.
Next to the SIQ framework, SEQUAL \cite{Krogstie_sequal_1995, Krogstie_sequal_2006_old} is another common framework for defining process model quality. It was first established for conceptual modeling \cite{Krogstie_sequal_1995}, but later revised to be more dynamic and suitable for assessing the quality of process models \cite{Krogstie_sequal_2006_old}. It not only focuses on a process model as knowledge, but also on it as a contribution to knowledge when it is interpreted by a human or another intelligent agent. As \autoref{fig:sequal} shows, there are six quality dimensions, which can be derived from the relation of the different requirements needed for interactive process models. In \cite{Krogstie_sequal_BPM_2016}, the SEQUAL model is specifically tailored for business process modeling, including aspects specific to business processes.

Other frameworks include the process modeling guidelines like Guidelines of Modeling (GoM) \cite{Becker_GOM_2000} or the 7 process modeling guidelines (G7) \cite{Mendling_7PMG_2010}. They are meant to be used by non-experts when creating a process model to ensure the resulting process model has decent quality.

Process model quality assessment methods provide ways to measure the quality of process models, e.g., through various process model metrics.  Since process modeling has so far been conducted by human experts, most work centers around quality assessments from a human perspective. The assessments aggregate existing metrics that measure specific characteristics of a process model. Often, the metrics are mapped to process quality dimensions or perspectives of the process, to allow a better interpretation of the scores. 

The work by \cite{Kopp_prag_quality_assess_2021} aims at measuring the pragmatic quality of a process model using various pragmatic quality metrics. To do so, they use the metrics to measure the degree of fulfillment of the seven modeling guidelines G7. However, a weakness of such works is that they focus on only one dimension, whereas a detailed analysis requires evaluating a process model from multiple quality perspectives. 

In \cite{Makni_Modeling_2010}, a tool is presented that evaluates the quality of business process models based on different metrics and process perspectives. In a first step, the quality metric framework for BPMN models is presented, which is based on the four process perspectives: functional, behavioral, organizational, and informational. For each dimension, multiple metrics are collected, which are later used to measure the quality. All in all, the paper used 9 metrics, mostly focusing on pragmatic quality aspects. After that, they present the implemented tool, allowing for the assessment of the metric values for a given BPMN model. For each process perspective, each metric is mapped to at least one of the four quality dimensions, namely maintainability, comprehension, reuse, and redesign. Further, for each metric, the user can provide thresholds, which are considered optimal values for the considered metric, along with a priority value. Based on this information, an assessment for each quality dimension in the chosen perspective is computed. As the thresholds are not fixed, the evaluations performed by different individuals are not comparable, as they most likely use different thresholds. 
The BPMIMA framework \cite{Sanchez-Gonzlez_BPMIMA_2017} is a framework used to support the improvement of business process models. The framework has three stages. It starts by \textit{measuring} the process model's quality using empirically proven metrics. Based on this, the evaluation of the measurements starts. As the pure numerical results for pragmatic quality metrics are only informative for comparative purposes, they employ thresholds for the \textit{evaluation} to indicate at which value the process quality starts to decline. Contrary to \cite{Makni_Modeling_2010}, empirically proven thresholds were used, rather than user-chosen ones, enabling objective interpretation of the metric scores. Finally, based on the evaluation results, \textit{redesign} actions are indicated, aiming at increasing the process model quality. 

The drawback of the approaches shown in \cite{Makni_Modeling_2010} and \cite{Sanchez-Gonzlez_BPMIMA_2017} is that only a part of the process quality is measured, as certain aspects are disregarded. Both are neglecting the semantic quality aspects. While \cite{Makni_Modeling_2010} omitted syntactic quality aspects entirely, \cite{Sanchez-Gonzlez_BPMIMA_2017} included them partially through the correctness dimension. However, this dimension only measures the likelihood of an error occurring, based on different pragmatic quality measures. But it is not examined whether the process model actually contains errors.

In \cite{Ullrich_kompetenzorientiertes_2024}, an e-assessment platform is presented that automates the evaluation of various graphical modeling tasks by covering all three SIQ quality dimensions, i.e., syntactic, pragmatic, and semantic quality.
The tasks can be based on different modeling languages such as BPMN, EPC, Entity-Relationship (ER) models, UML, and Petri nets. The platform is to be used in the context of learning and improving the competencies needed for process modeling, e.g., for students. The assessment, as presented in \cite{Ullrich_kompetenzorientiertes_2024}, mainly focused on the usage of Petri nets. Four checkers were implemented, which assess the solution of, e.g., a student, for tasks related to graphical process modeling. For each quality dimension of SIQ, a separate checker was implemented: a syntactic checker, a pragmatic checker, and a semantic checker, each checking different aspects of the quality dimension using several metrics. Moreover, they implemented a reachability checker that focused on the specific characteristics a Petri net, as a workflow net, should possess. A score for each checker is calculated and displayed, and the issues in the process model are marked and explained textually. 

\subsection{LLMs in BPMN Modeling}
\label{chap:related_work_llms_bpmn}

The application of LLMs to BPM tasks has expanded rapidly, with process modeling emerging as the most visible and mature application area \cite{grohs_large_2023,Norouzifar_Discovering_2024}. Prior work shows that LLMs can translate natural-language descriptions into formal process representations (imperative and declarative), supporting the extraction of activities, control-flow relations, and constraints from text. \cite{grohs_large_2023,Norouzifar_Discovering_2024}

Recent research increasingly operationalizes these capabilities in interactive modeling assistants. \cite{kourani_2024_promoai} proposed ProMoAI, a tool that leverages LLMs to generate process models from textual descriptions and supports iterative refinement via user feedback \cite{kourani_2024_promoai}.
In addition, their broader framework evaluation reports that GPT-4 performs strongly in process model generation, resolves encountered errors effectively, and integrates user feedback efficiently, while another evaluated LLM (Gemini) shows weaker results in the same setting \cite{kourani_2024_promoai}. Industry adoption further underscores the relevance of LLM-assisted process modeling: commercial tools such as Camunda's BPMN Copilot integrate LLM capabilities into established BPMN modeling environments, enabling the generation of BPMN diagrams from natural-language descriptions and supporting follow-up modifications \cite{Camunda_2026_BPMNCopilotDocs}.

Alongside academic prototypes, several systems explore alternative interaction paradigms to better align LLM assistance with established modeling practices. HyperMod \cite{Guiterrez_2024_DirectManipulationInterfaceLLMProcessModeling} exemplifies this direction by coupling direct user actions on the diagram with LLM support to enable more controllable, mixed-initiative process model construction and revision.

Tool-oriented work also investigates how to embed LLM-based process modeling into conversational user experiences. \cite{KopkeSafan2024EfficientConversationalProcessModeling} introduces the BPMN Chatbot, an interactive assistant for generating and refining BPMN models. Their evaluation reports a higher average correctness while using up to 94\% fewer tokens than an alternative tool, and includes an initial technology acceptance assessment. 
\cite{Larcardo_2025_BPMNAssistant_arXiv2509_24592} presents BPMN Assistant, which demonstrates that JSON-based structured representations can be effective for process model generation and, in particular, for manipulating and modifying models through LLM interaction. In \cite{Hoerner2026AutomaticallyGeneratingBPMN}, BPMNGen is introduced as a conversational framework for process modeling that supports iterative refinement of generated models. Two expert studies evaluated the approach with respect to semantic quality and the comprehensibility of the produced process models, reporting that LLM-generated models were perceived as equally understandable as manually created ones, and that semantic quality was at least comparable for smaller models but inferior for more complex models.

Beyond process modeling, LLMs are also explored for other BPM tasks, including support for process mining and automation-related activities, as well as conversational interfaces for AI-augmented BPM systems that aim to improve accessibility and explainability of BPM functionality \cite{Berti_Leveraging_2023,Casciani_AIAugmented_2024}.
Complementary work uses LLMs to enhance process model comprehension by enabling question answering and explanation over existing process models, thereby improving the interpretability and accessibility of complex BPMN artifacts for broader stakeholder groups \cite{Kourani_2024_EnhancedProcessModelComprehension}.

\subsection{LLM Benchmarks}
\label{chap:related_work_llms_benchmark}
The rapid advancement of LLMs has led to significant progress across many different tasks and domains. To systematically evaluate these LLMs’ capabilities, a plethora of benchmarks have been proposed — each designed to target different linguistic, reasoning, and domain-specific challenges. One prominent example is LiveBench \cite{livebench}, which evaluates LLMs across 21 diverse tasks in 7 domains, such as math, coding, and reasoning. The LLMs' performance is evaluated based on their responses to an extensive set of over 1000 questions, which are continually updated to reflect new information and challenges. The LLM-generated answers are evaluated against an objective ground truth, which serves as the basis for computing the LLM's score. The format of the answers and the scoring methodology are tailored to the specific domain being assessed.

However, the diversity and specialization of LLM benchmarks have also revealed gaps in coverage for certain domains, such as BPM, where the unique requirements and complexities demand tailored evaluation metrics and datasets. Recognizing this, the research community has been making increasing efforts to establish dedicated benchmarks for BPM-related tasks. 

\subsubsection{LLM Benchmarks for BPM Tasks}
Currently, significant efforts are being made to establish comprehensive benchmarks for various BPM tasks, evaluating on single or multiple tasks. Further, the evaluation of LLMs for BPM tasks can be conducted using different approaches, as shown by \cite{Berti_Benchmark_strategies_2024}. The first approach is automated, utilizing either an LLM-as-a-judge or a metric-based method. Alternatively, evaluations can be performed by human judges or by the evaluated LLM through self-evaluation. In the following, benchmarks corresponding to different evaluation strategies in the context of BPM will be presented.

In \cite{Fournier_Benchmark_Causal_reasoning_2025}, a benchmark is presented testing the ability of an LLM to generate sound answers to questions about a business process model. For instance, on the relationship and order between tasks in the process, or if the execution of one task causes the execution of another. These questions can be answered with "yes" or "no", allowing the calculation of a score for the correctly answered questions.

Further, multiple benchmarks have been established that not only focus on a single task in BPM but also benchmark multiple tasks together. The benchmark by \cite{Busch_Benchmark_BPM_2024} tests 4 different BPM tasks: activity recommendation, identification of RPA candidates, process question answering, and mining of declarative process models. The benchmark compares open-source and commercial LLMs to highlight task-specific performance differences. 
WONDERBREAD \cite{Wornow_Wonderbread_2024} is the first benchmark for evaluating multimodal foundation models (including LLMs) on BPM tasks beyond automation. It covers different tasks like workflow documentation generation, knowledge transfer and process improvement, or the validation of workflow completion. 

The human-centered perspective is covered by \cite{Hoerner2026AutomaticallyGeneratingBPMN}, employing two studies with experts to uncover the comprehensibility and semantic quality of the LLM-generated process models. 

\subsubsection{LLM Benchmarks for Process Modeling}
In this part, we focus on benchmarks that evaluate LLMs' ability to generate imperative or declarative process models.
The chatbot, called PRODIGY, deployed at the Hilti Group \cite{Ziche2024}, was evaluated based on human judgment. To do so, PRODIGY users had to provide feedback via a 90-minute semi-structured interview. Based on the feedback, the quality of PRODIGY was assessed. The evaluation is limited to 9 participants. 

%automatic evaluation
Most of the benchmarks previously mentioned used an automatic evaluation procedure. 
The PM-LLM-Benchmark by \cite{Berti_Benchmark_evaluation_2024} used an LLM-as-a-judge approach. The LLM-generated process models, such as Petri nets or BPMN models, are evaluated by an LLM that ranks them on a scale from 1 to 10. As no ground truth is provided for the ranking, the scoring of the LLM is rather subject to the preferences of the LLM. Further, there is a risk that the LLM fails to recognize issues in the process model, especially when the same LLM is used for both evaluation and generation. The benchmark compares the LLMs on fewer than a dozen samples.
Other benchmarks implemented a metric-based approach, so the scores for the LLMs are based on different metrics that calculate the quality of the LLM-generated BPMN model. In \cite{Busch_Benchmark_BPM_2024}, the generation of declarative process models by LLMs is evaluated based on the F1-score, with the focus on the modeling language DECLARE. To do so, they provided a ground truth model for each textual description. As declarative process models consist of constraints of different types, the true positives, false positives, and false negatives for each constraint type are assessed to calculate precision and recall for each type of constraint, as well as across all constraint types. Based on those metrics, the F1 score, the harmonic mean of precision and recall, is again calculated for each type of constraint, as well as across all constraint types. The F1 score is used as the final score for each LLM to be compared to other LLMs. 

There is already a benchmark focused solely on the generation of process models. In \cite{Kourani_Benchmark_2024}, the ability of LLMs to generate process models in the modeling language POWL is evaluated. The evaluation of the outputs is done by simulating event logs from the generated process models and comparing them, using conformance checking, against a ground event log generated from a ground truth process model. Additionally, they considered the time efficiency of the generation in the evaluation. The evaluation dataset consisted of 20 business process models.

Contrary to the LLM-as-a-judge approach as presented in \cite{Berti_Benchmark_evaluation_2024}, the metric-based approach allows a more detailed analysis of the abilities of LLM. This requires the use of a large number of metrics for evaluation. However, the previously mentioned metric-based approaches only employ a small number of metrics, which does not allow for a detailed analysis. Further, to provide a detailed picture, metrics for all quality dimensions (syntactic, semantic, and pragmatic quality) must be included in the analysis. This is a limitation of the approaches presented by \cite{Busch_Benchmark_BPM_2024} and \cite{Kourani_Benchmark_2024}, which do not cover all quality dimensions and, e.g., miss syntactic quality.

Human-centered evaluations have also been conducted. In particular, \cite{Hoerner2026AutomaticallyGeneratingBPMN} introduces BPMNGen, a conversational framework that was evaluated in two expert studies.
The first study assessed process model comprehensibility relative to manually modeled BPMN diagrams using measures such as cognitive load and level of acceptability. The second study examined semantic quality by asking experts to choose whether the BPMNGen-generated model or the manually created model was more suitable. However those do not allow for a scalable, automated approach. 

\section{BEF4LLM - BPMN Evaluation Framework for LLMs} \label{chap:framework}
In this section, the BEF4LLM framework for evaluating LLMs' ability to generate BPMN models is presented. First, the motivation and goal of the framework are shown. Following this, the elements of the framework, along with the calculation of scores, are explained in greater detail.

%Motivation
First experiments with LLMs showed promising results in BPMN modeling \cite{kourani_promoai_2024, grohs_large_2023}, with differences across LLM families \cite{Berti_Benchmark_evaluation_2024}. However, previous work lacks a detailed assessment of the generated BPMN models based on established process model quality frameworks and metrics, such as size, density or semantic label similarity. To fill this gap, the BEF4LLM framework has been developed to enable a detailed and automated evaluation of LLM-generated BPMN models, highlighting the strengths and weaknesses of different LLMs. Similar to \cite{Ullrich_kompetenzorientiertes_2024}, we extend the SIQ framework by mapping multiple process model metrics to each quality dimension, focusing on BPMN models. We preferred SIQ over SEQUAL because the latter includes social factors beyond the scope of BPMN model quality assessment. To ensure a detailed picture, it is important that each metric captures distinct aspects of the quality dimension. Other frameworks proposed in \cite{Makni_Modeling_2010} and \cite{Sanchez-Gonzlez_BPMIMA_2017} neglect, e.g., syntactical errors, omitting important quality aspects, or rely on human judgement \cite{Hoerner2026AutomaticallyGeneratingBPMN}.
%incorporates human-centered perspectives, including semantic quality and pragmatic aspects (e.g., comprehensibility). 
Reliance on human experts limits both the scalability and the degree of automation achievable with this evaluation approach. Moreover, the evaluation does not provide a dedicated assessment of syntactic quality while including semantic quality and pragmatic aspects (e.g., comprehensibility).

\begin{figure}[htbp]
    \centering
    \includegraphics[width=\linewidth]{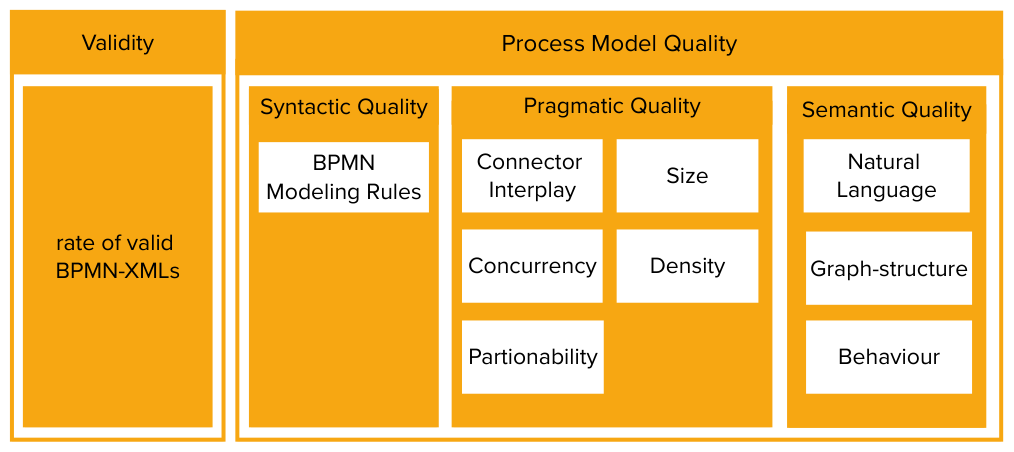}
    \caption{BEF4LLM - BPMN Evaluation Framework for LLMs.}
    \label{fig:framework_overview}
\end{figure}

% Warum BEF4LLM? Abgrenzung zu anderen Frameworks, USP
In contrast, the BEF4LLM framework, displayed in \autoref{fig:framework_overview}, enables automated assessment of BPMN model quality across four dimensions - syntactic quality, pragmatic quality, semantic quality, as well as validity of the BPMN XML file. The latter is required because LLMs are generative models, which means generated BPMN XML files may not comply with the BPMN 2.0 file format definition \cite{BPMN_rule_set}. In contrast, the validity of human-generated process model files is typically ensured by the tool used for process modeling. For instance, a graphical modeling editor always generates a valid BPMN XML file. While a validity check is therefore not required for human-generated BPMN models, it is necessary for LLM-generated ones. Validity is an important aspect since only a valid BPMN XML file can be displayed and used by subsequent software tools. It is a single measure that evaluates whether the generated XML file is a valid BPMN 2.0 file according to the XSD schema.
%In contrast, BPMN modeling editors used by humans for process modeling ensure the validity of the generated process model. 

% Was ist BEF4LLM?
The syntactic dimension focuses on syntactic rules that should be obeyed when modeling a BPMN model. Thereby, universal as well as modeling language-specific rules are included. Pragmatic quality measures the understandability of the process model for humans using metrics such as size, density, connector interplay, concurrency, or partitionability. For example, there are large process models that are precise in terms of semantic quality and comply with all modeling rules. But due to the high number of tasks, gateways, and connections, it is nearly impossible for a human to interpret them. However, a process model is only useful when humans can read and interpret it, so process modelers strive for high pragmatic quality. The semantic quality aspects are assessed based on the similarity between the generated BPMN models and the ground truth BPMN models. This is evaluated across three categories: natural language, graph structure, and behavior.

In total, the BEF4LLM framework comprises 39 individual metrics: One for validity, 16 for syntactic quality, 15 for pragmatic quality, and seven for semantic quality. Each metric takes the generated BPMN model as input (along with the ground-truth BPMN model for semantic quality) and returns a value between 0 (worst) and 1 (best).
Each generated BPMN model is assessed in two steps. First, the BPMN model's validity is checked. Only valid BPMN-XML files are subsequently evaluated with respect to syntactic, pragmatic, and semantic quality. This is necessary because these metrics can only be computed reliably when the underlying BPMN-XML is free of invalid elements. Per quality dimension, the metrics are aggregated to a single score, i.e., $Q_{\text{syn}}$ for the syntactic quality, $Q_{\text{prag}}$ for the pragmatic quality, and $Q_{\text{sem}}$ for the semantic quality. Therefore, normalization is necessary to ensure that all metrics are measured on the same scale, ranging from 0 to 1. As each dimension has different types of metrics, each uses a different normalization strategy, as explained in the corresponding section. 

Further, we decided to introduce two aggregated measures $Q_{\text{qual}}$ and $Q_{\text{total}}$. While $Q_{\text{qual}}$ only aggregates the three quality scores of the process quality dimensions, i.e., the syntactic, pragmatic, and semantic quality, the total score $Q_{\text{total}}$ additionally includes the validity. $Q_{\text{qual}}$ is intended to facilitate comparisons with human modeling competencies, because validity is not a relevant criterion in this setting. Both scores are intended to enable quick and easy comparison via a single aggregate measure, but should not be interpreted as general-purpose quality indicators. 

We conducted a targeted literature search to identify established metrics associated with each dimension. Because the framework is intended to support automated evaluation, each selected metric must be computable fully automatically from a generated BPMN-XML file and a corresponding ground-truth process model (for semantic quality). This implies that metrics that require manual human judgment are excluded from the framework, as they would hinder automated, scalable evaluation. To ensure reproducibility, we considered only metrics with clear operational definitions established in prior BPMN or process modeling research.
Moreover, because the framework is applied across heterogeneous process models, the selected metrics must be sufficiently robust to retain meaning across varying process model structures and sizes. For pragmatic quality and, in part, for semantic quality, we relied on existing literature reviews that compile and categorize relevant measures in order to select metrics suitable for our setting. For syntactic quality, we prioritized metrics that reflect constraints defined by the modeling language (BPMN in our case) as well as general well-formedness rules that apply across multiple process modeling languages.

The framework provides a standardized procedure for evaluating LLM capabilities in BPMN modeling using established metrics. Thereby, a systematic analysis per dimension and between LLMs can be made. However, note that the layout of the BPMN model is not part of the assessment and therefore not checked. The layout can be added algorithmically to a BPMN XML file, omitting the requirement to generate it with an LLM. Further, certain BPMN elements are not covered by this framework because our selected metrics primarily focus on the process flow and, in particular, do not account for BPMN-specific constructs. Specifically, we exclude artifacts, which we define as BPMN elements that provide supplementary information but are not part of the process flow (e.g., groups, data objects, and text annotations). So are data objects omitted, as the data perspective is not assessed within the chosen metrics and therefore outside the scope of our framework. The same applies to groupings and subprocesses. Although they can influence understandability and thus pragmatic quality, they introduce hierarchical structures that complicate metrics such as size due to the need to aggregate across nested graphs.

\subsection{Syntactic Quality}
Syntactic quality concerns whether a process model follows the rules of the modeling language. 
Therefore, validation metrics have been derived from BPMN modeling rules based on \cite{Dijkman_2007Formal_SA}, \cite{BPMN_rule_set}, and \cite{Wong_2008_Process}. They are summarized in \autoref{Tab:Criteria_syntactic_quality} and aggregated to a score $Q_{\text{syn}}$. We use two types of metrics, where one is a Boolean measure that checks whether all necessary elements exist or all mandatory regulations are followed. The other measures the percentage of how often a certain rule is followed. 

\begin{table}[htbp]
\footnotesize
\centering
\begin{tabular}{clc}
\toprule
 & \textbf{Metric description} & \textbf{Ref.}\\
\midrule
1 & Existence of a start event &
\cite{Dijkman_2007Formal_SA}\\
2 & Existence of an end event &
\cite{Dijkman_2007Formal_SA}\\
3 & One start event per process &
\cite{Dijkman_2007Formal_SA}\\
4 & One end event per process &
\cite{Dijkman_2007Formal_SA}\\
5 & Sequence‑flow connection rules &
\cite{BPMN_rule_set}\\
6 & Message‑flow connection rules &
\cite{BPMN_rule_set}\\
7 & Start event: $in=0$, $out=1$ &
\cite{Dijkman_2007Formal_SA}\\
8 & End event: $in=1$, $out=0$ &
\cite{Dijkman_2007Formal_SA}\\
9 & Split gateway has matching join gateway &
\cite{Dijkman_2007Formal_SA}\\
10 & Exactly one process per pool &
\cite{Wong_2008_Process}\\
11 & Each observable task has a label &
\cite{BPMN_rule_set}\\
12 & Task: $in=1$, $out=1$ &
\cite{Dijkman_2007Formal_SA}\\
13 & Non‑exception intermediate event: $in=1$, $out=1$ &
\cite{Dijkman_2007Formal_SA}\\
14 & Exception event: $in=0$, $out=1$ &
\cite{Dijkman_2007Formal_SA}\\
15 & Split gateway: $in=1$, $out>1$ &
\cite{Dijkman_2007Formal_SA}\\
16 & Join gateway: $in>1$, $out=1$ &
\cite{Dijkman_2007Formal_SA}\\
\bottomrule
\end{tabular}
%%\footnotetext{Exception events are excluded from the matching‑gateway rule.}
\caption{Metrics for syntactic quality in the \textsc{BEF4LLM} framework. A complete table listing the metric formulas is provided in \autoref{Tab:Criteria_syntactic_quality_long}.}
\label{Tab:Criteria_syntactic_quality}
\end{table}

The metrics include general modeling rules, such as the existence of a start node, as well as modeling-language-specific metrics, such as sequence flow connection rules. Boolean measures either evaluate to 0 (non-existent) or 1 (existent), e.g., the existence of a start node. A BPMN model with no start event would receive a score of 0 for the "existence of a start node" metric. Counting metrics are computed by dividing the number of elements not following the rule by the total number of elements covered by this rule. Therefore, the scores are normalized to a range of 0 to 1. Given a BPMN model with 8 labeled activities out of 10, the "observable task" metric would evaluate to $0.8$.

\subsection{Pragmatic Quality}
The pragmatic quality focuses on whether a process model can be understood by a human. Thus, pragmatic quality is connected to the way the process is modeled, but not by its content. %The certification in the SIQ framework refers to checking the pragmatic quality. 
\autoref{Tab:Criteria_pragmatic_quality} provides an overview of the criteria that are used in our framework to measure the pragmatic quality of a process model, aggregated to the quality score $Q_{\text{prag}}$. The measurements can be categorized into six types \cite{Mendling_Metrics_2008}, describing different characteristics of the process model influencing its pragmatic quality: (i) size (metrics of this type assess the overall size of the process model), (ii) density (includes metrics relating the number of nodes to the number of arcs in the process model), (iii) connector interplay (focuses on the gateways and their interplay), (iv) partitionability (measures the relation between the subcomponents within a process model), (v) cyclicity (refers to the presence of cycles or loops in process models), and (vi) concurrency (metrics examine the concurrent paths within the process model).

Normalization to scores between 0 and 1 in this dimension is done using four empirically validated thresholds (\cite{Snachez_2011_BPMImprovementMeasures, Sanchez_2015_CaseStudyThresholds, boomsma_2009_evaluation, ekstedt_2015_quality}) per metric that separate the values into five distinct groups. The thresholds for each metric are given in \autoref{Tab:thresholds} in the appendix. For some metrics, e.g., token split or connectivity coefficient, greater is better, and the function $norm_{asc}(x)$ \autoref{eq:grouping_equation_asc} is used. For other metrics like the total number of nodes or density, where lower is better, the function $norm_{desc}(x)$ \autoref{eq:grouping_equation_desc} is used for normalization.

\vspace{1em}
%\begin{equation*}
\noindent
\begin{minipage}{0.48\textwidth}
\footnotesize
    %\[
    \begin{equation}
    \text{norm}_{desc}(x) =
    \begin{cases}
        1.0,    & \text{if } x < t_1\\
        0.75, & \text{if } t_1 \leq x < t_2 \\
        0.5,  & \text{if } t_2 \leq x < t_3 \\
        0.25, & \text{if } t_3 \leq x < t_4 \\
        0,  & \text{if } x \geq t_4 \\
    \end{cases}
    \label{eq:grouping_equation_desc}
    \end{equation}
    %\]
\end{minipage}\hfill
\begin{minipage}{0.48\textwidth}
    \footnotesize
    \begin{equation}
    %\[
    \text{norm}_{asc}(x) =
    \begin{cases}
        1.0,    & \text{if } x < t_1\\
        0.75, & \text{if } t_1 \geq x > t_2 \\
        0.5,  & \text{if } t_2 \geq x > t_3 \\
        0.25, & \text{if } t_3 \geq x > t_4 \\
        0,  & \text{if } x \geq t_4 \\
    \end{cases}
    %\]
    \label{eq:grouping_equation_asc}
    \end{equation}
\end{minipage}
%\end{equation*}
\vspace{1em}

For example, consider a BPMN model with 45 nodes, for which the TNN (total number of nodes) metric is computed. The thresholds $t_1$ to $t_4$ for this metric are given by \cite{Snachez_2011_BPMImprovementMeasures} as follows: $t_1 = 29.9$, $t_2 = 43.7$, $t_3 = 58.1$, and $t_4 = 81.1$. For this metric, a lower number of nodes makes the BPMN model easier to understand, which is why the descending normalization function $\text{norm}_{desc}(x)$ is used. Since 45 $\ge$ 43.7 and 45 $<$ 58.1, the score $0.5$ is assigned, i.e., group 3.

\begin{table}[t]
\footnotesize
\centering
\begin{tabular}{clc}
\toprule
 & \textbf{Metric} & \textbf{Ref.} \\ 
\midrule
\multicolumn{3}{c}{\textbf{Size}} \\ \midrule
1 & TNN (total number of nodes) &
\cite{Mendling_Metrics_2008}\\
2 & TNG (total number of gateways) &
\cite{Rolon_2008_Evaluation}\\
3 & TNSF (total number of sequence flows) &
\cite{Rolon_2008_Evaluation}\\
4 & TNMF (total number of message flows) &
\cite{Rolon_2008_Evaluation}\\
5 & Diameter &
\cite{Mendling_Metrics_2008}\\
\midrule
\multicolumn{3}{c}{\textbf{Density}} \\ \midrule
6 & Density &
\cite{Mendling_Metrics_2008}\\
7 & AGD (average gateway degree) &
\cite{Mendling_Metrics_2008}\\
8 & CNC (connectivity coefficient) &
\cite{Mendling_Metrics_2008}\\
\midrule
\multicolumn{3}{c}{\textbf{Connector interplay}} \\ \midrule
9 & GH (gateway heterogeneity) &
\cite{Mendling_Metrics_2008}\\
10 & CFC (control‑flow complexity) &
\cite{Mendling_Metrics_2008}\\
11 & CC (cross‑connectivity) &
\cite{Vanderfeesten_crossconnectivity_2008}\\
\midrule
\multicolumn{3}{c}{\textbf{Partitionability}} \\ \midrule
12 & Sequentiality &
\cite{Mendling_Metrics_2008}\\
13 & Separability &
\cite{Mendling_Metrics_2008}\\
14 & Depth &
\cite{Mendling_Metrics_2008}\\
\midrule
\multicolumn{3}{c}{\textbf{Concurrency}} \\ \midrule
15 & TS (token split) &
\cite{Mendling_Metrics_2008}\\
\bottomrule
\end{tabular}
\caption{Metric set for pragmatic quality in the \textsc{BEF4LLM} framework. A complete table listing the metric formulas is provided in \autoref{Tab:Criteria_pragmatic_quality_long}.}
\label{Tab:Criteria_pragmatic_quality}
\end{table}

Cyclicity, frequently employed as a pragmatic measure, is not included in the BEF4LLM framework because existing research does not provide multiple thresholds for cyclicity metrics, which prevents categorization of these metrics in a manner consistent with the other metrics used in the framework.

Various studies \cite{mendeling_thresholds_2012, Snachez_2011_BPMImprovementMeasures, ekstedt_2015_quality, boomsma_2009_evaluation} have employed threshold-based pragmatic quality measurement approaches to group metric values, utilizing different numbers of thresholds and corresponding groups. Some provide a binary classification, categorizing into the groups "easy to understand" and "difficult to understand", while others employ 4 or 5 thresholds, resulting in 5 or 6 distinct groups. Since a classification based on 4 thresholds was available for more metrics than one based on 5 thresholds, we opted to use 4 thresholds per metric, resulting in a fine-grained assessment into 5 distinct groups. However, this decision also affected the selection of the metrics for this dimension. Although additional metrics appeared reasonable to include (e.g., cyclicity), we excluded them because no consistent assignment to our pragmatic quality categorization was available.

Because pragmatic quality is computed from metrics capturing model size and structural complexity, its score decreases as a process model becomes larger and more complex. However, “simpler” is not always better, since some process descriptions inherently require larger and more complex models. Pragmatic quality should therefore be interpreted in conjunction with the other BEF4LLM dimensions—especially semantic quality—rather than in isolation, as it primarily reflects the potential comprehension costs of a generated model. In this setting, pragmatic quality enables the analysis of which LLMs can represent complex processes at comparatively lower comprehension costs. To better understand the drivers of a low pragmatic score, it is useful to inspect not only the aggregated pragmatic score but also the subgroup scores (e.g., size and density). This helps identify whether specific metric groups, such as size-related metrics, disproportionately contribute to the decline in pragmatic quality.

\subsection{Semantic Quality}
Semantic quality addresses what a process model says about reality: every true statement about the target process must be present (completeness) and every statement contained in the process model must, in fact, be true (validity). Directly verifying these properties against the real world is impractical, so we compare a candidate process model $M_{c}=(N_{c}, E_{c},\tau_{c})$ with a ground‑truth process model $M_{g}=(N_{g}, E_{g},\tau_{g})$ that is assumed to be both complete and valid. The closer $M_{c}$ resembles $M_{g}$, the higher its semantic quality.

We employ automated similarity measures from three groups identified by \cite{Schoknecht_SimilaritySOTA_2017}: (i) natural‑language similarity, which judges how well node labels match on syntactic, semantic, and neighbourhood‑context levels; (ii) graph‑structure similarity, which compares the models’ topology via graph‑edit distance and the share of common nodes and edges; and (iii) behavioural similarity, which contrasts their execution semantics using causal footprints and dependency graphs. Human‑estimation and “other” measures are omitted because the evaluation must run without manual intervention.

\begin{table}[htbp!]
\footnotesize
\centering
\begin{tabular}{C{0.01\linewidth}L{0.5\linewidth}C{0.1\linewidth}}
\toprule
&\textbf{Metric} & \textbf{Ref.} \\
\midrule
\multicolumn{3}{c}{\textbf{Natural‑language similarity}}\\\midrule
1 & Syntactic label similarity &
\cite{Dijkman_Similarity_2011}\\

2 & Semantic label similarity &
\cite{Dijkman_Similarity_2011}\\

3 & Context similarity &
\cite{Dijkman_Similarity_2011}\\
\midrule
\multicolumn{3}{c}{\textbf{Graph‑structure similarity}}\\\midrule
4 & Graph‑edit distance &
\cite{Dijkman_Similarity_2011}\\

5 & Common nodes and edges &
\cite{Becker_comparative_2012}\\
\midrule
\multicolumn{3}{c}{\textbf{Behavioural similarity}}\\\midrule
6 & Causal‑footprint overlap &
\cite{Dijkman_Similarity_2011,vanDongen_Measuring_2013}\\

7 & Dependency‑graph overlap &
\cite{Dijkman_Similarity_2011,vanDongen_Measuring_2013}\\
\bottomrule
\end{tabular}
\caption{Metric set for semantic quality in the \textsc{BEF4LLM} framework. A complete table listing the metric formulas is provided in \autoref{Tab:Criteria_semantic_quality_long}.}
\label{Tab:Criteria_semantic_quality}
\end{table}

The measurements in this quality dimension also have inherent limitations. Semantic quality is assessed via similarity to a ground-truth model, which constrains the evaluation to what is expressed in that reference. Label-based similarity metrics further restrict the assessment by approximating equivalence through one-to-one matching, even when employing synonym handling and related techniques, and may therefore penalize valid alternative phrasings or differences in granularity.

\subsection{Validity}
Validity is a single measure that checks whether the BPMN XML file is parsable, allowing the process model to be displayed and processed by subsequent software. This is important because BPMN models are intended to be read and used by humans, for which visualization is crucial. Although validity could be considered part of syntactic quality, because errors in the BPMN XML file can be interpreted as syntactic errors, we decided to decouple this measure. The rationale is that validity serves as a gatekeeping criterion as the remaining metrics can only be computed if a valid BPMN XML file is available. Moreover, validity and syntactic quality operate at different layers. While syntactic quality captures conformance to BPMN modeling rules, validity captures conformance to the BPMN XML schema.

We check the validity $Q_{\text{val}}$ of the BPMN XML based on the XSD schema\footnote{\url{https://github.com/bpmn-io/bpmn-moddle/tree/main/resources/bpmn/xsd}}, containing all structural and formal constraints that the BPMN XML must adhere to according to the BPMN 2.0 convention using the following formula \autoref{eq:q_val}.

\begin{equation}
    Q_{\text{val}} =
        \begin{cases}
        1.0, & \text{if } \text{BPMN XML is valid} \\
        0.0, &  \text{else} \\
        \end{cases}
    \label{eq:q_val}
\end{equation}

\subsection{Aggregation}
We aggregate the metrics first per dimension, followed by aggregating the dimension scores into overall scores.
Since all metrics within a given quality dimension are normalized using the same strategy, the resulting metric scores are directly comparable and can be aggregated at the dimension level.
%At the same time, we note that normalization alone does not imply that all metrics are fully comparable, nor that their aggregation constitutes a theoretically validated quality construct. Hence, the aggregated scores should be interpreted as descriptive summary indices for benchmarking and comparison purposes.
Normalization alone does not guarantee that all metrics are fully comparable or that their aggregation forms a theoretically validated quality construct, so the aggregated scores are best understood as descriptive summary indices for benchmarking and comparison purposes. As we currently lack empirical evidence that any metric should receive a higher weight than the others, we aggregate the metrics by taking their arithmetic mean, thereby avoiding favoring an individual metric. This choice is therefore a transparent baseline aggregation strategy rather than a claim of empirically grounded weighting or linear contribution of all metrics.

For example, to compute the quality score for the syntactic quality dimension $Q_{\text{syn}}$, we sum the individual metric scores and divide by the number of metrics applied. \autoref{eq:aggregation_syn} gives the formula for that dimension where $score(m)$ gives the score of a certain metric $m$.

\begin{equation}
\label{eq:aggregation_syn}
     Q_{syn} = \frac{\sum_{m \in metrics_{syn}} score(m) }{|metrics_{syn}|} =  \frac{\sum_{m \in metrics_{syn}} score(m)}{16} 
\end{equation}

For easier comparison, we introduce two more measures that aggregate the dimension quality scores. Similar to the aggregation within the quality dimensions, we found no empirical evidence that any quality dimension should be considered more important than the others.
Therefore, we adopted equal weighting across dimensions. \autoref{eq:q_qual} shows $Q_{\text{qual}}$ as the average of the three quality dimension scores, while $Q_{\text{total}}$, shown in \autoref{eq:q_total}, is the average over the three quality dimension scores and the validity score. Note that those two scores are therefore also in the range between 0 (worst) and 1 (best).
Since both lack empirical validation, they should not be interpreted as a general quality indicator; they are simply aggregations of the individual quality dimensions.

\begin{equation}
     Q_{\text{qual}} = \frac{Q_{syn} + Q_{prag} + Q_{sem}}{3}
     \label{eq:q_qual}
\end{equation}

\begin{equation}
     Q_{\text{total}} = \frac{Q_{\text{syn}} + Q_{\text{prag}} + Q_{\text{sem}} + Q_{\text{val}}}{4}
     \label{eq:q_total}
\end{equation}

\section{Experiment settings} 
\label{chap:exp_settings}
In the following, we explain how we conducted the benchmark, evaluating the ability of several open-source LLMs to generate BPMN models based on textual descriptions. With the help of the benchmark, we want to answer the following questions: 
\begin{enumerate}
    \item Which LLMs are the best to use for process modeling based on a textual description?
    \item How far does the size of the LLM influence the quality of the generated process models? 
    \item What is the difference in the quality of a process model generated by an LLM in contrast to a process model modeled by a human? 
\end{enumerate}

To answer those questions, the BEF4LLM framework and the different quality scores are used. The individual scores provide a detailed picture of the process modeling abilities of different LLMS, while the $Q_{\text{qual}}$ score allows for comparing the quality of human-generated BPMN models with that of LLM-generated ones. Next, the concrete procedure, including prompting, LLM choices, and the implementation details, is shown. This setting is largely similar to the one presented in \cite{Lauer_Generating_2025}.

With this experiment, we aim to establish a baseline evaluation.
Accordingly, the research questions stated above are addressed using a basic prompting setup, without additional adjustments such as training, intermediate representations, or advanced prompting strategies. The implementation of the framework, LLM experiments, and the datasets used are available at \url{https://gitlab-iwi.dfki.de/lauer/bef4llm}.

\subsection{Procedure}
\begin{figure}[ht!]
    \centering
    \includegraphics[width=.7\linewidth]{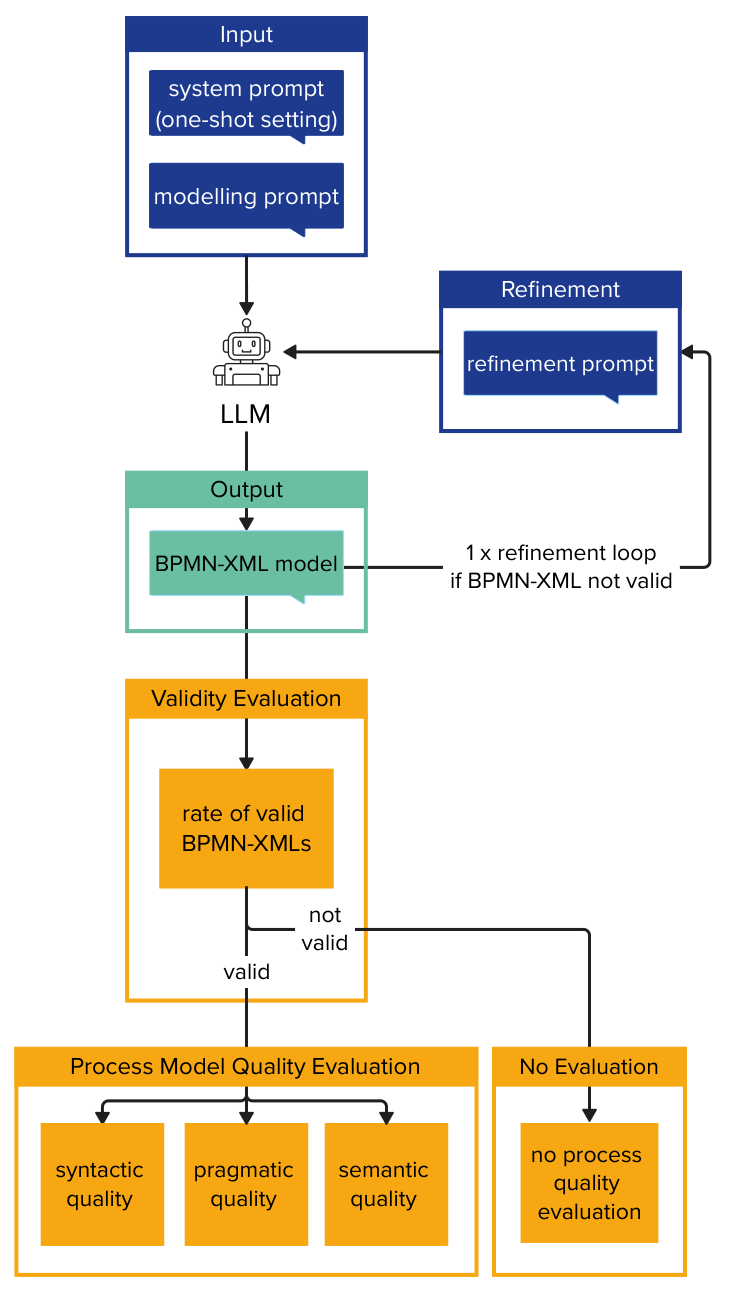}
    \caption{Complete procedure to conduct the experiment for one LLM.}
    \label{fig:scenario_4}
\end{figure}

\autoref{fig:scenario_4} illustrates the complete procedure, including the input and prompts for the LLM, the refinement loop, and the evaluation steps.
Initially, the LLM receives a system prompt, explaining the general setting with a single example, i.e., a one-shot prompt\footnote{For more details, see \autoref{subsec:prompting}}. The system prompt also explains that no layout should be generated by the LLM.
After that, a modeling prompt containing the textual description of the BPMN model to generate is sent to the LLM. The LLM then returns its response, which should be the BPMN model in XML. If the BPMN XML file returned by the LLM is invalid, a refinement loop is started. Here, a refinement prompt is sent, stating which errors the BPMN XML file contains and that it should be fixed. Note that only one refinement per description is done. Next, the BPMN XML file is checked for validity; if it is valid, the process model quality scores $Q_{\text{syn}}$, $Q_{\text{prag}}$, and $Q_{\text{sem}}$ are calculated. If the validity evaluation fails, no quality scores are computed but only $Q_{\text{val}}$. 

\subsection{LLM Selection and Configuration}
The goal in this experiment is to compare the process modeling abilities of open-source LLMs of different sizes and architectures. We included different types of LLMs, such as non-thinking LLMs (e.g., Qwen2.5), Mixture-of-Expert (MoE) based ones (e.g., Qwen3), and thinking LLMs like Deepseek. 
To measure the influence of the parameter size for the LLM version, at least two different parameter sizes for each LLM family were tested, if possible. \autoref{tab:Experiment_LLM} summarizes the LLMs that were tested. The LLMs are grouped by parameter count: smaller than 7b as small LLMs, between 7b and 14b as mid-sized LLMs, and over 14b as large LLMs. 
%We used q8 quantized versions of all LLMs.

\begin{table}[ht]
    \centering
    \tabcolsep=5pt

    %\vspace*{1mm}
    \resizebox{\textwidth}{!}{
    \begin{tabular}{llll}
    \toprule
      \textbf{LLM Familiy }& \textbf{Small} ($\leq 7B$) & \textbf{Medium} ($> 7B \textbf{ \&} \leq 14B$) & \textbf{Large} ($> 14B$) \\ \midrule
      Llama3 & llama3.3:1b-instruct & llama3.3:8b-instruct & llama3.3:70b-instruct \\
      Qwen2.5 & qwen2.5:1.5b-instruct & qwen2.5:14b-instruct & qwen2.5:32b-instruct \\
      Qwen3 & qwen3:1.7b & qwen3:14b, qwen3:30b-a3b & qwen3:235b-a22b \\
      Deepseek (Llama) & &  deepseek-r1:8b-llama-distill & deepseek-r1:70b-llama-distill \\
      Deepseek (Qwen) & deepseek-r1:1.5b-qwen-distill &  deepseek-r:14b-qwen-distill & \\
      Phi4 & & phi4:14b& \\
      Falcon & falcon3:1b-instruct& falcon3:10b-instruct& \\
    \bottomrule
    \end{tabular}}
    \caption{LLMs used for the experiments with their respective Ollama tag.}
    \label{tab:Experiment_LLM}
\end{table}

The \textit{temperature} was set to 0.1 to encourage strict adherence to the prompt instructions, in particular, the BPMN 2.0 constraints required to produce a valid BPMN-XML file. We did not use a \textit{temperature} of 0, which would result in greedy decoding (selecting the highest-probability next token), as it can still yield non-deterministic behavior in practice and provides no controlled flexibility for interpreting ambiguous process descriptions. At a temperature of 0.1, generation remains concentrated on high-probability tokens while allowing limited stochasticity, which we found preferable for this task. This allows the LLM to produce slight variations in the expected structure, such as alternative control-flow paths or task labels, while still adhering to the given instructions. For the same reason, we refrained from using a seed.
To ensure a good balance between accuracy and memory usage, we decided to use \textit{8-bit quantized} LLMs. We preferred to use the instruction-tuned variants of the LLMs, indicated by the instruction tag in \autoref{tab:Experiment_LLM}. Those are trained to follow instructions and are therefore more likely to generate useful BPMN models. However, for some LLM families, such as Qwen3 or Deepseek-R1, no instruction-tuned variants are available, so we resorted to non-instruction-tuned variants. For the context length, we used a fixed value of 40k, ensuring that the context always captures the length of the modeling and refinement prompt. 

\subsection{Prompting}
\label{subsec:prompting}
For the experiment, a \textit{system prompt}, a \textit{modeling prompt}, and an \textit{error-handling prompt} were used.

\paragraph{System Prompt}
Before the LLM is given the textual description of the process, the system prompt outlines the tasks, the input, and the expected output. We opted for role prompting in the system prompt, explaining the LLM's role in the interaction. Further, a single output example is provided in the system prompt (without a matching input), i.e., one-shot prompting, to ensure the output matches the expected output format.

%\vspace{1em}
\begin{minipage}{0.95\textwidth}
    \footnotesize
    \label{fig:system_prompt}
    \fbox{\parbox{0.99\textwidth}{
    \#\#\# Your Role: \\
        You are an expert in modeling business processes...\\
        \#\#\# Guidelines:\\
            1. **BPMN XML Format**: ... \\
        \#\#\# Template for an BPMN XML \\
            $<$definitions$>$.... \\
        \#\#\# Output: As an output, a BPMN in XML format is needed... \\
        \#\#\# Example Output: \\
            $<$?xml version=``1.0'' encoding=``UTF-8''?$>$... }}
    \captionof{figure}{System prompt \cite{Lauer_Generating_2025}.}
    
\end{minipage}
%\vspace{1em}

\paragraph{Modeling Prompt}
The modeling prompt instructs the LLM to generate a BPMN model based on a textual description. In contrast to \cite{Lauer_Generating_2025}, where the modeling prompt is limited to a textual description, our approach utilizes a more comprehensive prompt that includes an instruction requiring the LLM to generate a BPMN XML file, followed by a detailed textual description. Note that each textual description is processed individually by one LLM call, which does not contain information from previous calls.

\vspace{1em}
\begin{minipage}{0.95\textwidth}
    \footnotesize
    \label{fig:modeling_prompt}
    \fbox{\parbox{0.99\textwidth}{
        Create a BPMN model in XML format for the following textual description of a process. Make sure to follow the BPMN 2.0 modeling guidelines. \\
        If goods shall be shipped, the secretary clarifies who will do the shipping...}}
    \captionof{figure}{Modeling prompt.}
\end{minipage} 
\vspace{1em}

\paragraph{Refinement Prompt} 
Using the refinement loop, the LLM is given the chance to correct the generated BPMN XML file once. If the XML validation fails, the refinement prompt instructs the LLM to fix the invalid BPMN XML file. To guide the LLM, a list of the most common mistakes, along with the actual mistakes found by the XML validator, is added to the prompt. Note that the validation can also fail if the LLM generated no BPMN XML file, but any other textual process description.

\vspace{1em}
\begin{minipage}{0.95\textwidth}
    \footnotesize
    \label{fig:error_handling}
    \fbox{\parbox{0.99\textwidth}{It seems that the BPMN XML format is invalid.                           Ensure that:\\
                            - the id of the elements are unique \\
                            - ... \\
                            Make sure to correct all the mistakes given in the following list:\\
                            - missing targetRef for sequenceFlow 123 \\
                            - ...}}
    \captionof{figure}{Refinement prompt \cite{Lauer_Generating_2025}.}
\end{minipage} 
\vspace{1em}

\subsection{Data Description}
To obtain reliable results, a high-quality dataset is required. Further, since the semantic quality dimension requires a ground truth BPMN model for comparison, pairs of textual descriptions and BPMN process models are needed. The ground truth BPMN model must be of high quality and, more importantly, capture all the semantic aspects of the process as defined by the textual description. Therefore, we manually searched for high-quality text-BPMN pairs in publicly available datasets. 

\autoref{tab:datasets} lists the four datasets found, with the number of textual descriptions and corresponding BPMN models per description. Each dataset is created by an expert, ensuring that the textual descriptions and the corresponding ground truth BPMN models are of high quality.

\begin{table}[ht]
    \centering
    \tabcolsep=5pt

    %\vspace*{1mm}
    \resizebox{\textwidth}{!}{
    \begin{tabular}{lrrrrrr}
    \toprule
    \textbf{Dataset} & \textbf{Textual descriptions} & \textbf{BPMNs per description} & \textbf{Language} \\ \midrule
    Camunda\tablefootnote{\url{https://github.com/camunda/bpmn-for-research}}& 8 & 1 & EN, DE & \\
    BPMN text and model \cite{mangler_textual_2023} & 24 & 6-12 &EN &\\
    Dataset for LRE original\tablefootnote{\label{fn:lre_dataset}\url{https://github.com/setzer22/alignment_model_text/tree/master/datasets}} & 48 & 1 & EN & \\
    Dataset for LRE new\footref{fn:lre_dataset}  & 25 & 1 & EN & \\
    \bottomrule
    Total number of text-model pairs & 105 & & &   \\
    \end{tabular}
    }
    \caption{Details on the datasets used \cite{Lauer_Generating_2025}.}
    \label{tab:datasets}
\end{table}

In total, 105 text-BPMN model pairs have been built, each constituting a single sample. Each sample has been verified manually for high quality. From the BPMN models in \cite{mangler_textual_2023}, which have a quality ranking from 1 to 5, only those with the highest quality ranking were chosen.

\subsection{Quality Score Calculation \& Implementation Details} 
For each sample, the $Q_{\text{val}}$ score is computed. The quality scores $Q_{\text{syn}}$, $Q_{\text{prag}}$, and $Q_{\text{sem}}$ can only be computed if the resulting BPMN is valid. Similarly, if the LLM did not return a result after six minutes, the attempt was treated as invalid, and no quality scores were calculated for this sample. This issue occurred randomly, without any pattern or relation to a specific sample. Therefore, we attribute this to the LLM generating text infinitely, which is a technical issue. 
Additionally, if the resulting BPMN model contains fewer than 2 nodes (all ground truth BPMN models have more than 2 nodes), no quality scores can be calculated, and the run is also counted as invalid. Therefore, the $Q_{\text{val}}$ can only be $1$ if and only if a valid BPMN XML file with at least two flow objects has been returned after 6 minutes. Otherwise, the score is $0$.

Since we intentionally created variation in the generated BPMN models by setting the temperature to $0.1$, five runs per sample are conducted and averaged. In total, for each LLM, 5 runs for each of the 105 samples are performed and averaged. In the results section, we will report the average values, denoted by $\overline{Q}$, for all scores introduced previously.

\paragraph{Server Details}
The experiments ran on a server equipped with two AMD EPYC 9474F CPUs (96 cores total), 1.2 TB DDR5 RAM, NVIDIA H200 NVL GPU (141 GB VRAM), PCIe Gen5 support while running Ubuntu 24.04 and CUDA 12.4. For the inference framework, ollama 0.9.6 was used.
The GPU has sufficient capacity to handle the required context lengths and processes one request at a time. The runtimes per LLM differed significantly. The whole experiment, with a total of $8,925$ BPMN model generations\footnote{17 LLMs, each being tested on 105 text BPMN model pairs 5 times, results in $8,925$ runs.}, took around 5 days to finish.

\subsection{Statistical Evaluation}

To determine if the observed performance variations between LLMs were statistically significant, we adopted a robust, non-parametric evaluation framework. This approach was chosen because our quality scores are ordinal (rank-based) and our dataset has missing values whenever an LLM fails to produce a valid BPMN XML file, making standard parametric tests unsuitable.

Our analysis began with a global test using the Skillings-Mack test \cite{Chatfleld_Skillings_Mack_2009}, selected for its unique ability to handle missing data when other methods like the Friedman test would fail. This test ranks the valid LLM outputs for each prompt to determine whether significant performance variations exist across the entire group, yielding a test statistic $T$ that is compared against a $\chi^2$ distribution. Once a global effect was confirmed, we described each LLM's performance by calculating the mean score ($\bar{y}$) of its valid outputs.

Finally, to isolate specific performance gaps, we conducted head-to-head comparisons using the Wilcoxon signed-rank test \cite{gibbons1993nonparametric}, a powerful method chosen because our data consist of paired samples (i.e., two LLMs evaluated on the same textual description) and does not assume a normal distribution. To ensure the integrity of these multiple comparisons, we applied a strict Bonferroni correction, a conservative method designed to rigorously prevent false positives. This technique adjusts each raw p-value $p$ from the $m$ tests to $p_{\text{adj}} = \min(mp, 1)$, and a difference was only deemed significant if this adjusted value fell below 0.05.

\section{Results}
\label{chap:results}

This section summarizes the empirical findings and is structured around three research questions. 
Subsection \ref{sec:rq1} compares the quality of BPMN models generated by the LLMs. 
Subsection \ref{sec:size_effects} investigates how the size of LLMs influences process model quality.
Subsection \ref{sec:llm_vs_human} benchmarks the best‑performing LLMs against human modelers. 
Together, these results describe the current strengths and limitations of LLM‑based process model generation and set the stage for the following discussion.

\paragraph{Exclusion of LLMs}
During the experiments, we observed that some LLMs, especially small ones with fewer than 7 B parameters, often generated invalid BPMN XML files. For instance, \textit{deepseek-r1:1.5b-qwen-distill} generated not a single valid BPMN XML file, and \textit{llama3.2:1b} only an average of 1.4 BPMN XML files across the 5 runs\footnote{Results for these LLMs can be found in the appendix in \autoref{tab:results_full}}.

Even though a dedicated metric, $Q_{\text{val}}$, exists to assess performance in generating a valid BPMN XML file, the underperformance of these LLMs would distort the overall interpretation of the results and the statistical tests. Therefore, we considered only LLMs in the results that, on average across all 105 samples and five runs, produce at least 30 valid BPMN XML files. We used the formula \autoref{eq:avbm} to compute the average number of valid BPMN XML files across all five runs, where $\text{VBMPR}_i$ is the number of valid BPMN XML files in run $i$. 

\begin{equation}
     \text{AVBM} = \frac{\sum_{i=1}^{5} \text{VBMPR}_i}{5}
     \label{eq:avbm}
\end{equation}

If an LLM was unable to generate 30 valid BPMN XML files on average, i.e., $\text{AVBM} < 30$, it was not considered. Six LLMs were not able to generate a sufficient number of valid BPMN XML files, which reduces the number of LLMs analyzed in the following to eleven\footnote{In the end, the following LLMs were dropped: \textit{deepseek-r1:1.5b-qwen-distill, deepseek-r1:8b-llama-distill, falcon3:3b-instruct, llama3.2:1b-instruct, qwen2.5:1.5b-instruct, qwen3:1.7b.}}.

\subsection{RQ 1 – Comparative Quality of LLM‑Generated BPMN Models}\label{sec:rq1}

This subsection answers \textit{RQ1} by examining how the eleven evaluated LLMs differ in their capacity to transform textual descriptions into high‑quality BPMN models. Subsection \ref{sec:global_eq_test_across_models} first tests for overall performance variation across LLMs. Subsection \ref{sec:descriptive_performance} then profiles individual strengths and weaknesses statistically. Lastly, subsection \ref{sec:pairwise_performance_contrast} pinpoints pairwise differences with rigorous statistical contrasts. Together, these analyses establish a clear ranking of current LLM families and provide the empirical basis for the following analysis.

\subsubsection{Global Equality Tests Across LLMs}
\label{sec:global_eq_test_across_models}

Before examining individual LLMs, we verify whether genuine variation exists across the quality dimensions. We only test the three criteria, syntactic, pragmatic, and semantic quality. Validity is intrinsically handled by the Skillings–Mack design matrix through structurally missing values and therefore requires no separate test. Formally, we tested the following global null hypothesis:

\begin{quote}
\textbf{H\textsubscript{0}}: \textit{All eleven LLMs generate BPMN models of equal pragmatic, semantic, and syntactic quality.}
\end{quote}

This preliminary step ensures that subsequent descriptive and pairwise analyses are statistically meaningful. The results of the global Skillings-Mack tests are summarized in Table~\ref{tab:global_skillingsmack_results}.

\begin{table}[ht!]
    \centering

    \resizebox{\linewidth}{!}{%
    \begin{tabular}{lcccc}
        \toprule
        \textbf{Quality Dimension} & \textbf{Test Statistic (\textit{T})} & \textbf{df} & \textbf{\textit{p}-value} & \textbf{Conclusion} \\
        \midrule
        Syntactic & 313.78 & 10 & $<10^{-15}$ & Reject H$_0$: LLMs differ significantly \\
        Pragmatic & 170.34 & 10 & $<0.001$ & Reject H$_0$: LLMs differ significantly \\[2pt]
        Semantic & 310.92 & 10 & $<0.001$ & Reject H$_0$: LLms differ significantly \\[2pt]
        \bottomrule
    \end{tabular}}
    \caption{Global Skillings–Mack test results across quality dimensions.}
    \label{tab:global_skillingsmack_results}
\end{table}

Three key insights follow from these results. First, the global null hypothesis of equal performance across the eleven LLMs is decisively rejected for all three dimensions, confirming the existence of genuine quality differences. Second, the magnitude of the test statistic $T$ approximately doubles from pragmatic to semantic quality and slightly increases further for syntactic quality. This pattern indicates increasing differentiation in performance as the evaluation criteria shift from subjective readability towards more objective, rule-based correctness. Third, despite substantial data sparsity, all tests retained a full-rank covariance matrix. This provides assurance that the observed data are sufficiently interconnected to support robust statistical conclusions. Having confirmed significant global differences, we proceed in the following subsections to detailed descriptive statistics (\autoref{sec:descriptive_performance}) and controlled pairwise comparisons (\autoref{sec:pairwise_performance_contrast}), aimed at precisely identifying the nature and extent of these differences.

\subsubsection{Descriptive Performance Profiles by LLM}
\label{sec:descriptive_performance}
\autoref{tab:results_temp_01} shows that no single LLM achieves the highest score in all quality dimensions.
\begin{table}[ht]
    \centering
    \tabcolsep=5pt

    \resizebox{\textwidth}{!}{
    \begin{tabular}{lccccccc}
    \toprule
        & \multicolumn{2}{c}{\textbf{Validity}} & \multicolumn{3}{c}{\textbf{Process Model Quality}} & \multicolumn{2}{c}{\textbf{Total Scores}} \\
        \cmidrule(lr){2-3} \cmidrule(lr){4-6}  \cmidrule(lr){7-8}
        \textbf{LLM} & $\overline{Q_{\text{val}}}$ & AVBM & $\overline{Q_{\text{syn}}}$ & $\overline{Q_{\text{prag}}}$ & $\overline{Q_{\text{sem}}}$ & $\overline{Q_{\text{qual}}}$ & $\overline{Q_{\text{total}}}$  \\ \midrule
        deepseek-r1:14b-qwen-distill & 0.6362 & 66.8 & 0.8548 & 0.8841 & 0.5270 & 0.7553 & 0.7114 \\
        deepseek-r1:70b-llama-distill & 0.7905  & 83 & 0.8708 & 0.8636 & 0.5609 & 0.7651 &0.7560 \\ \midrule
        falcon3:10b-instruct & 0.3067 & 32.2 & \textbf{0.9082 }& 0.8837 & 0.5544 & \textbf{0.7821} &0.6475   \\ \midrule
        llama3.1:8b-instruct & 0.7067 & 74.2 & 0.8597 & 0.8954 & 0.5389 & 0.7647 & 0.7347 \\
        llama3.3:70b-instruct & \textbf{ 0.9733} & \textbf{102.2 }& 0.8955 & 0.8721 & 0.5747 & 0.7808 & \textbf{0.8143}   \\ \midrule
        phi4:14b & 0.5848 &61.4& 0.8580 & 0.8710 & 0.5666 & 0.7652 & 0.7056   \\ \midrule
        qwen2.5:14b-instruct & 0.5467 &57.4& 0.8076 &\textbf{ 0.8907 }& 0.5521 & 0.7501 & 0.6873 \\
        qwen2.5:32b-instruct & 0.5105 &53.6 & 0.8782 & 0.8834 & \textbf{0.5768} & 0.7795 &0.6984\\ \midrule
        qwen3:14b & 0.6533 &68.6 & 0.8643 & 0.8995 & 0.5720 & 0.7678 &0.7250 \\
        qwen3:30b-a3b & 0.4895 & 51.4 & 0.8792 & 0.8877 & 0.5485 & 0.7718 &0.6862 \\
        qwen3:235b-a22b & 0.5771 &60.6 & 0.8790 & 0.8537 & 0.5620 & 0.7649 &0.7036\\
     \bottomrule
     \end{tabular}
     }
     \caption{Results of the first experiment.}
     \label{tab:results_temp_01}
\end{table}
\textit{falcon3:10b-instruct} achieved the highest syntactic quality score (0.9082), reflecting its superior adherence to BPMN syntax rules. Pragmatic quality was highest in \textit{qwen2.5:14b-instruct} (0.8907), indicating that the generated BPMN models are easy to understand. In terms of semantic quality, \textit{qwen2.5:32b-instruct} showed the highest score (0.5768). \textit{llama3.3:70b-instruct} excelled significantly in validity, producing valid XML structures in 97.33\% of cases. Although \textit{falcon3:10b-instruct} had the highest combined process score $Q_{\text{qual}}$ (0.7821), when validity is not considered, \textit{llama3.3:70b-instruct} emerged as the overall best-performing LLM with a $Q_{\text{qual}}$ score of 0.8143.
A detailed analysis reveals consistent strengths and weaknesses across the evaluated LLMs. Syntactic quality was robust across all LLMs (scores consistently above 0.75), and the pragmatic quality remained high (scores exceeding 0.8). However, semantic quality consistently lagged behind.

\subsubsection{Pairwise Performance Contrasts}
\label{sec:pairwise_performance_contrast}

To localize the quality gaps identified in the global and descriptive analyses, every admissible pair of checkpoints was compared with a Wilcoxon signed–rank test, applied independently to syntactic, pragmatic, and semantic scores. Each test used only those samples for which both LLMs produced a valid BPMN XML file. Raw probabilities were Bonferroni‑adjusted to hold the family–wise error rate at $\alpha = 0.05$. All results can be found in \autoref{tab:syntax_pairwise_full}, \autoref{tab:prag_pairwise_full}, and \autoref{tab:sem_pairwise_full}. Rows labeled “n.s.” indicate that the adjusted probability exceeded the significance threshold.

Overall, significant contrasts are largely driven by weaker LLMs, most notably \textit{qwen3:235b-a22b}, \textit{deepseek-r1:70b-llama-distill}, and \textit{falcon3:10b-instruct}. Once these LLMs are set aside, pragmatic quality differences among the remaining systems vanish beneath the detection limit, whereas syntactic and semantic quality still display measurable separations. A clear hierarchy emerges in the syntactic and semantic dimensions, with \textit{llama3.3:70b-instruct} consistently positioned as a top performer. 

Syntactic quality shows a notable trend, with 25 of 55 contrasts yielding statistically significant results. (see Table~\ref{tab:syntax_pairwise_full}). The results reveal a clear performance stratification. \textit{llama3.3:70b-instruct} and \textit{qwen2.5:14b-instruct} are the most structurally robust LLMs, with the former achieving the single most decisive victory against any other LLM in the entire analysis. Conversely, \textit{falcon3:10b-instruct} consistently produces less syntactically correct BPMN models than its peers. While several high-performing LLMs show no statistically significant differences among themselves, the hierarchy confirms that rule conformance has not yet reached parity across the full distribution.

\begin{footnotesize}
\begin{longtable}{L{0.33\textwidth}L{0.33\textwidth}C{0.14\textwidth}C{0.07\textwidth}}
\label{tab:syntax_pairwise_full}\\
\toprule
\textbf{Better LLM} & \textbf{Worse LLM} & $p_{\text{adj}}$ & \textbf{Paired cases}\\
\midrule
\endfirsthead
\multicolumn{4}{l}{\textit{Table~\ref{tab:syntax_pairwise_full} (continued).}}\\
\toprule
\textbf{Better LLM} & \textbf{Worse LLM} & $p_{\text{adj}}$ & \textbf{Paired cases}\\
\midrule
\endhead
\midrule
\multicolumn{4}{r}{\textit{Continued on next page}}\\
\endfoot
\bottomrule
\addlinespace
\caption{Wilcoxon contrasts for the syntactic quality (Bonferroni‑adjusted). “n.s.” denotes $p_{\text{adj}}>0.05$.}
\endlastfoot
llama3.3:70b-instruct & qwen2.5:14b-instruct & $4.39\times10^{-31}$ & 283\\
qwen2.5:14b-instruct & deepseek-r1:70b-llama-distill & $1.69\times10^{-19}$ & 233\\
qwen2.5:14b-instruct & qwen3:235b-a22b & $4.54\times10^{-14}$ & 195\\
qwen2.5:14b-instruct & qwen3:30b-a3b & $1.54\times10^{-13}$ & 145\\
llama3.3:70b-instruct & phi4:14b & $2.55\times10^{-13}$ & 299\\
qwen2.5:14b-instruct & qwen2.5:32b-instruct & $6.82\times10^{-12}$ & 141\\
qwen2.5:14b-instruct & falcon3:10b-instruct & $9.96\times10^{-12}$ & 96\\
llama3.3:70b-instruct & deepseek-r1:14b-qwen-distill & $2.03\times10^{-11}$ & 324\\
llama3.3:70b-instruct & llama3.1:8b-instruct & $3.48\times10^{-10}$ & 364\\
qwen3:14b & qwen2.5:14b-instruct & $1.84\times10^{-09}$ & 186\\
deepseek-r1:14b-qwen-distill & falcon3:10b-instruct & $1.50\times10^{-07}$ & 108\\
llama3.3:70b-instruct & deepseek-r1:70b-llama-distill & $2.86\times10^{-06}$ & 405\\
qwen2.5:14b-instruct & phi4:14b & $4.58\times10^{-06}$ & 179\\
llama3.1:8b-instruct & qwen2.5:14b-instruct & $8.31\times10^{-06}$ & 210\\
qwen2.5:14b-instruct & deepseek-r1:14b-qwen-distill & $1.11\times10^{-05}$ & 191\\
qwen3:14b & llama3.3:70b-instruct & $1.44\times10^{-05}$ & 308\\
llama3.1:8b-instruct & falcon3:10b-instruct & $7.54\times10^{-05}$ & 121\\
phi4:14b & falcon3:10b-instruct & $9.68\times10^{-05}$ & 101\\
deepseek-r1:70b-llama-distill & falcon3:10b-instruct & $5.21\times10^{-04}$ & 134\\
deepseek-r1:14b-qwen-distill & qwen3:30b-a3b & $6.66\times10^{-04}$ & 147\\
phi4:14b & qwen3:235b-a22b & $8.57\times10^{-04}$ & 204\\
qwen3:14b & falcon3:10b-instruct & $7.18\times10^{-03}$ & 100\\
deepseek-r1:14b-qwen-distill & qwen3:235b-a22b & $7.20\times10^{-03}$ & 228\\
llama3.3:70b-instruct & qwen2.5:32b-instruct & $3.93\times10^{-02}$ & 262\\
llama3.1:8b-instruct & qwen3:235b-a22b & $4.72\times10^{-02}$ & 25\\
\multicolumn{4}{c}{\textit{All remaining 30 pairs: n.s.}}\\
\end{longtable}
\end{footnotesize}

Across the pragmatic quality matrix, 27 of the 55 possible pairs remain significant after adjustment, a figure that represents just under one half of the admissible comparisons (see ~\autoref{tab:prag_pairwise_full}). Every highly significant contrast places either \textit{Deepseek-r1 70b} or \textit{Qwen3 235b-a22b} on the lower side. No comparison between two of the strongest LLM approaches is significant. The pragmatic quality has therefore reached a practical ceiling: once the median score approaches 0.90, residual differences become too small to detect under the stringent Bonferroni correction.

\begin{footnotesize}
\begin{longtable}{L{0.33\textwidth} L{0.33\textwidth} C{0.14\textwidth} C{0.07\textwidth}}
\label{tab:prag_pairwise_full}\\
\toprule
\textbf{Better LLM} & \textbf{Worse LLM} & $p_{\text{adj}}$ & \textbf{Paired cases}\\
\midrule
\endfirsthead
\multicolumn{4}{l}{\textit{Table~\ref{tab:prag_pairwise_full} (continued).}}\\
\toprule
\textbf{Better LLM} & \textbf{Worse LLM} & $p_{\text{adj}}$ & \textbf{Paired cases}\\
\midrule
\endhead
\midrule
\multicolumn{4}{r}{\textit{Continued on next page}}\\
\endfoot
\bottomrule
\addlinespace
\caption{Wilcoxon contrasts for the pragmatic quality (Bonferroni‑adjusted). “n.s.”~denotes $p_{\text{adj}}>0.05$.}
\endlastfoot
llama3.18b-instruct & qwen3:235b-a22b & $7.95\times10^{-16}$ & 251\\
llama3.18b-instruct & deepseek-r1:70b-llama-distill & $2.54\times10^{-13}$ & 300\\
qwen2.5:14b-instruct & qwen3:235b-a22b & $2.77\times10^{-12}$ & 195\\
qwen2.5:14b-instruct & deepseek-r1:70b-llama-distill & $1.50\times10^{-10}$ & 233\\
llama3.370b-instruct & llama3.18b-instruct & $1.52\times10^{-9}$ & 364\\
qwen3:14b & llama3.18b-instruct & $2.75\times10^{-9}$ & 235\\
qwen2.5:14b-instruct & phi4:14b & $4.34\times10^{-8}$ & 179\\
deepseek-r1:14b-qwen-distill & qwen3:235b-a22b & $4.63\times10^{-8}$ & 228\\
qwen3:30b-a3b & qwen3:235b-a22b & $6.41\times10^{-8}$ & 189\\
llama3.370b-instruct & qwen2.5:14b-instruct & $1.68\times10^{-7}$ & 283\\
qwen3:14b & qwen2.5:14b-instruct & $2.54\times10^{-7}$ & 186\\
qwen2.5:32b-instruct & qwen3:235b-a22b & $1.29\times10^{-6}$ & 191\\
deepseek-r1:14b-qwen-distill & deepseek-r1:70b-llama-distill & $2.04\times10^{-6}$ & 267\\
llama3.18b-instruct & phi4:14b & $6.10\times10^{-5}$ & 230\\
qwen2.5:14b-instruct & qwen3:30b-a3b & $2.45\times10^{-4}$ & 145\\
qwen3:14b & qwen2.5:32b-instruct & $1.48\times10^{-3}$ & 166\\
llama3.18b-instruct & qwen3:30b-a3b & $1.55\times10^{-3}$ & 183\\
qwen2.5:14b-instruct & qwen2.5:32b-instruct & $1.81\times10^{-3}$ & 141\\
llama3.370b-instruct & qwen3:235b-a22b & $2.09\times10^{-3}$ & 335\\
qwen3:14b & deepseek-r1:14b-qwen-distill & $3.79\times10^{-3}$ & 217\\
qwen2.5:14b-instruct & falcon3:10b-instruct & $5.41\times10^{-3}$ & 96\\
llama3.18b-instruct & falcon3:10b-instruct & $1.20\times10^{-2}$ & 121\\
qwen3:14b & qwen3:30b-a3b & $1.53\times10^{-2}$ & 170\\
qwen2.5:32b-instruct & deepseek-r1:70b-llama-distill & $2.06\times10^{-2}$ & 204\\
phi4:14b & qwen3:30b-a3b & $2.59\times10^{-2}$ & 147\\
phi4:14b & qwen3:235b-a22b & $3.84\times10^{-2}$ & 204\\
llama3.370b-instruct & deepseek-r1:70b-llama-distill & $5.00\times10^{-2}$ & 405\\
\multicolumn{4}{c}{\textit{All remaining 28 pairs: n.s.}}\\ 
\end{longtable}
\end{footnotesize}

Semantic quality displays the clearest hierarchy, with twenty-seven of the fifty-three admissible pairs remaining significant (see Table~\ref{tab:sem_pairwise_full}). \textit{llama3.3:70b-instruct} establishes itself as the front-runner, showing statistically reliable separation from nearly all other LLMs. The most pronounced gaps appear between this top performer and the weakest LLMs for this task, namely \textit{deepseek-r1:14b-qwen-distill} and \textit{qwen3:235b-a22b}. Semantic quality, therefore, remains a key dimension in which meaningful differences can be observed among the best open-source LLMs.

\begin{footnotesize}
\begin{longtable}{L{0.33\textwidth} L{0.33\textwidth} C{0.14\textwidth} C{0.07\textwidth}}
\label{tab:sem_pairwise_full}\\
\toprule
\textbf{Better LLM }& \textbf{Worse LLM} & $p_{\text{adj}}$ & \textbf{Paired cases}\\
\midrule
\endfirsthead
\multicolumn{4}{l}{\textit{Table~\ref{tab:sem_pairwise_full} (continued).}}\\
\toprule
\textbf{Better LLM} & \textbf{Worse LLM} & $p_{\text{adj}}$ & \textbf{Paired cases}\\
\midrule
\endhead
\midrule
\multicolumn{4}{r}{\textit{Continued on next page}}\\
\endfoot
\bottomrule
\addlinespace
\caption{Wilcoxon contrasts for the semantic quality (Bonferroni‑adjusted). “n.s.” denotes $p_{\text{adj}}>0.05$.}
\endlastfoot
llama3.370b-instruct & deepseek-r1:14b-qwen-distill &$3.65\times10^{-19}$ & 324\\
llama3.370b-instruct & llama3.18b-instruct &$6.29\times10^{-16}$ & 364\\
deepseek-r1:14b-qwen-distill &qwen3:235b-a22b &$1.02\times10^{-14}$ & 228\\
deepseek-r1:14b-qwen-distill &phi4:14b &$2.71\times10^{-12}$ & 201\\
llama3.18b-instruct &qwen3:235b-a22b &$5.88\times10^{-12}$ & 251\\
qwen3:14b &deepseek-r1:14b-qwen-distill &$6.89\times10^{-12}$ & 217\\
deepseek-r1:14b-qwen-distill &qwen2.5:32b-instruct &$4.92\times10^{-10}$ & 175\\
llama3.370b-instruct &qwen2.5:14b-instruct &$1.85\times10^{-09}$ & 283\\
deepseek-r1:14b-qwen-distill &deepseek-r1:70b-llama-distill &$2.43\times10^{-08}$ & 267\\
llama3.18b-instruct &phi4:14b &$2.89\times10^{-08}$ & 230\\
deepseek-r1:14b-qwen-distill &qwen3:30b-a3b &$1.60\times10^{-07}$ & 147\\
llama3.18b-instruct &deepseek-r1:70b-llama-distill &$1.39\times10^{-06}$ & 300\\
qwen3:14b &llama3.18b-instruct &$6.83\times10^{-06}$ & 235\\
llama3.18b-instruct &qwen2.5:32b-instruct &$9.84\times10^{-06}$ & 194\\
qwen3:30b-a3b &qwen3:235b-a22b &$3.84\times10^{-05}$ & 189\\
llama3.370b-instruct &qwen3:30b-a3b &$4.73\times10^{-05}$ & 251\\
qwen2.5:14b-instruct &phi4:14b &$5.62\times10^{-05}$ & 179\\
qwen3:14b &qwen2.5:14b-instruct &$6.23\times10^{-05}$ & 186\\
phi4:14b &falcon3:10b-instruct &$9.09\times10^{-05}$ & 101\\
llama3.370b-instruct &falcon3:10b-instruct &$1.12\times10^{-04}$ & 156\\
llama3.370b-instruct &deepseek-r1:70b-llama-distill &$1.12\times10^{-03}$ & 405\\
qwen3:14b &qwen3:30b-a3b &$4.86\times10^{-03}$ & 170\\
qwen2.5:14b-instruct &deepseek-r1:70b-llama-distill &$1.24\times10^{-02}$ & 233\\
qwen2.5:32b-instruct &falcon3:10b-instruct &$1.53\times10^{-02}$ & 87\\
falcon3:10b-instruct &qwen3:235b-a22b &$1.82\times10^{-02}$ & 120\\
qwen2.5:14b-instruct &qwen2.5:32b-instruct &$2.01\times10^{-02}$ & 141\\
qwen2.5:14b-instruct &qwen3:235b-a22b &$3.25\times10^{-02}$ & 195\\
\multicolumn{4}{c}{\textit{All remaining 26 pairs: n.s.}}\\
\end{longtable}
\end{footnotesize}

\subsection{RQ 2 – Influence of LLM Model Size on BPMN Quality}
\label{sec:size_effects}
Scaling laws derived from open‑domain language modeling often suggest that adding parameters monotonically improves performance.  
Our evaluation of four open‑weight families, Llama3, Qwen2.5, Qwen3, and Deepseek‑R1, shows that this principle does \emph{not} transfer wholesale to BPMN diagram generation.
For each prompt where both checkpoints within a family produced valid BPMN models, we conducted paired Wilcoxon signed-rank tests. We controlled the family-wise error rate at 5\% using a Bonferroni adjustment.

\begin{table}[htbp]
\centering
\resizebox{0.7\textwidth}{!}{
\begin{tabular}{lcccc}
\toprule
\textbf{LLM Family} & \textbf{Parameters} & $\overline{Q_{\text{syn}}}$ & $\overline{Q_{\text{prag}}}$ & $\overline{Q_{\text{sem}}}$ \\
\midrule
Llama3   & 8\,B   & 0.860 & \textbf{0.895} & 0.539\\
          & 70\,B  & \textbf{0.896} & 0.872 & \textbf{0.575}\\ \midrule
Qwen2.5  & 14\,B  & 0.808 & \textbf{0.891} & 0.553\\
          & 32\,B  & \textbf{0.877} & 0.883 & \textbf{0.576}\\ \midrule
Qwen3   & 14\,B  & 0.865 & 0.867 & \textbf{0.572}\\
          & 30\,B  & \textbf{0.879} & \textbf{0.883} & 0.551\\
          & 235\,B & \textbf{0.878} & 0.859 & \textbf{0.573}\\ \midrule
Deepseek‑R1 & 14\,B & 0.854 & \textbf{0.884} & 0.526\\
           & 70\,B & \textbf{0.870} & 0.864 & \textbf{0.561}\\
\bottomrule
\end{tabular}}
\caption{Mean syntactic, pragmatic, and semantic scores (0–1 scale).}
\label{tab:means}
\end{table} 

\autoref{tab:means} reveals two broad patterns.  
First, larger LLMs generally boost semantic and syntactic means by three to five percentage points.  
Second, those same expansions frequently depress pragmatic scores by about two percentage points.
%, a practically visible drop during blind user reviews.  
Thus, parameter count alone is a poor proxy for overall BPMN quality.

The inferential evidence in \autoref{tab:wilcoxon_results} corroborates the descriptive trends.  
For syntactic and semantic quality, larger checkpoints win decisively in three families (\textit{Llama3}, \textit{Qwen2.5}, \textit{Deepseek‑R1}), with $p$‑values as small as $10^{-16}$. However, pragmatic quality consistently moves in the opposite direction, a statistically strong but negative shift.

\begin{table}[ht]
    \centering
    \tabcolsep=5pt
    \resizebox{\textwidth}{!}{
    \begin{tabular}{lcccc}
        \toprule
        \makecell{\textbf{Size}\\\textbf{Contrast}}
        & $p_{\text{syntactic}}$
        & $p_{\text{pragmatic}}$
        & $p_{\text{semantic}}$
        & \makecell{\textbf{Direction}\\(syntactic / pragmatic / semantic)} \\ \midrule
        Llama3 8B $\rightarrow$ 70B      & $3.5\times10^{-10\,***}$ & $1.5\times10^{-9\,***}$  & $6.3\times10^{-16\,***}$ & $\uparrow$70B \,/\, $\uparrow$8B  \,/\, $\uparrow$70B \\
        Qwen2.5 14B $\rightarrow$ 32B  & $6.8\times10^{-12\,***}$ & $1.8\times10^{-3\,**}$   & $2.0\times10^{-2\,*}$    & $\uparrow$32B \,/\, $\uparrow$14B \,/\, $\uparrow$32B \\
        Qwen3 14B $\rightarrow$ 30B    & n.\,s.                   & $1.5\times10^{-2\,*}$    & $4.9\times10^{-3\,**}$   & -- \,/\, $\uparrow$30B \,/\, $\uparrow$14B \\
        Qwen3 14B $\rightarrow$ 235B   & n.\,s.                   & n.\,s.                   & n.\,s.                    & -- \\
        Qwen3 30B $\rightarrow$ 235B   & n.\,s.                   & $6.4\times10^{-8\,***}$  & $3.8\times10^{-5\,***}$  & -- \,/\, $\uparrow$30B \,/\, $\uparrow$235B \\
        Deepseek 14B $\rightarrow$ 70B  & $2.8\times10^{-4\,***}$  & $2.0\times10^{-6\,***}$  & $1.5\times10^{-8\,***}$  & $\uparrow$70B \,/\, $\uparrow$14B \,/\, $\uparrow$70B \\
        \bottomrule
    \end{tabular}}
    \vspace{2pt}
    \footnotesize
    $^{\dagger}$Arrows: $\uparrow$ = larger checkpoint wins, -- = non‑significant.  
    Significance: *$p<0.05$, **$p<0.01$, ***$p<0.001$ (Bonferroni‑corrected); n.\,s.\ = not significant.
    \caption{Wilcoxon signed–rank tests contrasting small and large checkpoints.  
             Arrows indicate the checkpoint with the higher median when the shift is significant.}
    \label{tab:wilcoxon_results}
\end{table}

The \textit{Qwen3} family illustrates how size effects can plateau or even reverse.  
Expanding from 14 B to 30 B sacrifices semantic quality ($\Delta=-0.021$, $p=4.9\times10^{-3}$) yet improves pragmatic quality ($\Delta=+0.016$, $p=1.5\times10^{-2}$).  
A further leap to 235 B restores the lost semantic quality ($\Delta=+0.022$, $p=3.8\times10^{-5}$) but erodes pragmatic quality ($\Delta=-0.024$, $p=6.4\times10^{-8}$) without affecting syntax.  
These oscillations imply that raw parameter count is not the limiting factor. Architecture‑specific bottlenecks or training‑data saturation likely dominate beyond 30 B. Practical importance depends jointly on magnitude and consistency.

\textit{Deepseek‑R1}’s move from 14 B to 70 B yields syntactic and semantic quality gains of $+0.035$ and $+0.016$ respectively, both significant, offset by a pragmatic quality loss of $-0.020$.  

Syntactic and semantic quality tend to improve in parallel, likely because they share a reliance on extended relational memory and longer effective context windows. In contrast, pragmatic quality often deteriorates when LLMs overoptimize for frequent patterns, resulting in unnecessarily detailed outputs. This effect is analogous to the inflated responses observed in open-ended text generation.
Given these findings, a single checkpoint seldom satisfies all objectives.  
Large LLMs such as \textit{llama3.3:70b-instruct} or \textit{deepseek-r1:70b-llama-distill} are justified when semantic and syntactic quality dominate, whereas small and medium-sized LLMs (\textit{llama3.1:8b-instruct}, \textit{qwen2.5:14b-instruct}) were superior in creating BPMN models that are easy for humans to understand at lower inference cost.  
The fact that \textit{qwen3:14b} statistically matches its 235B MoE-counterpart on every metric underscores the risk of paying for parameters that yield no domain‑specific benefit.

In sum, parameter scaling improves syntactic and semantic quality but often degrades pragmatic quality, and the size of each effect is architecture dependent. However, a higher semantic score might also be associated with larger, more complex process models. Therefore, an increase in semantic quality might be accompanied by a decrease in pragmatic quality. To determine the causes of the declining pragmatic quality, a detailed analysis by a human judge is necessary, but is out of scope for this paper.
Effective BPMN model generation requires metric‑aligned checkpoint selection, grounded in both descriptive means and inferential tests, rather than an automatic preference for the largest available LLMs.

\subsection{RQ 3 – Comparison to competences of human process modeling}
\label{sec:llm_vs_human}
In addition to comparing the process model generation abilities of LLMs, it is also interesting to compare their abilities with those of humans. We decided to compare the LLMs to experts in the field of process modeling, as those represent the "gold standard" for process modeling. As the experts who participated in this experiment were not fluent in English but in German, we created a new dataset for this experiment. To achieve this, we utilized four German textual descriptions from Camunda\footnote{\url{https://github.com/camunda/bpmn-for-research}} and five additional textual descriptions created by a process consulting company. For each textual description, we had a corresponding ground truth BPMN model, with both the BPMN model and the accompanying textual description provided. 
We asked 5 consultants with several years of expertise in process modeling to create a BPMN model for each of the nine textual descriptions using publicly available BPMN modeling tools. Each expert was provided with a complete set of 9 textual descriptions, prefaced by a standardized instruction set outlining the specific requirements and tasks to be performed. The instructions demanded that they create one BPMN model after another, ensuring the same separation between each sample, as in the LLM settings. In total, 45 BPMN models have been created by the experts.

\begin{table}[t]
    \centering
    \tabcolsep=5pt

    \resizebox{\textwidth}{!}{
    \begin{tabular}{lccccccc}
    \toprule
       & \multicolumn{2}{c}{\textbf{Validity}} & \multicolumn{3}{c}{\textbf{Process Model Quality}} & \multicolumn{2}{c}{\textbf{Total Scores}} \\
        \cmidrule(lr){2-3} \cmidrule(lr){4-6}  \cmidrule(lr){7-8}
        \textbf{LLM} & $\overline{Q_{\text{val}}}$ & AVBM & $\overline{Q_{\text{syn}}}$ & $\overline{Q_{\text{prag}}}$ & $\overline{Q_{\text{sem}}}$ & $\overline{Q_{\text{qual}}}$ & $\overline{Q_{\text{total}}}$  \\ \midrule
        deepseek-r1:14b-qwen-distill & 0.8222 & 7.4 & 0.8412 & 0.8795 & 0.3337 & 0.6848 & 0.7191 \\
        deepseek-r1:70b-llama-distill & 0.7111 & 6.4 & 0.8738 & 0.8671 & 0.4242 & 0.7217 & 0.7190 \\ \midrule
        falcon3:10b-instruct & 0.3111 & 2.8 & 0.9109 & 0.8142 & 0.3906 & 0.7052 & 0.6067 \\ \midrule
        llama3.18b-instruct & 0.7556 & 6.8 & 0.8513 & \textbf{0.8856} & 0.4318 & 0.7229 & 0.7311 \\
        llama3.370b-instruct & \textbf{1.0000} &\textbf{ 9} & 0.8909 & 0.8763 & 0.5006 &\textbf{ 0.7559 }& 0.8169 \\ \midrule
        phi4:14b & 0.8000 & 7.2 & 0.8452 & 0.8563 & 0.4344 & 0.7120 & 0.7340 \\ \midrule
        qwen2.5:14b-instruct & 0.5111 & 4.6 & 0.8390 & 0.8763 & 0.3692 & 0.6949 & 0.6489 \\
        qwen2.5:32b-instruct & 0.8222 & 7.4 & 0.8724 & 0.8841 & 0.4897 & 0.7487 & 0.7671 \\ \midrule
        qwen3:14b & 0.5111 & 4.6 & 0.9104 & 0.8275 & 0.4295 & 0.7225 & 0.6696 \\
        qwen3:30b-a3b & 0.1778 & 1.6 & 0.5406 & 0.5106 & 0.2327 & 0.7132 & 0.3654 \\
        qwen3:235b-a22b & 0.6444 & 5.8 & \textbf{0.9136 }& 0.8070 & 0.4658 & 0.7288 & 0.7077 \\ \midrule
        \midrule
        human experts & \textbf{1.0}& \textbf{9} & 0.8826 & 0.7371& \textbf{0.5152}& 0.7116 & \textbf{0.7837}\\
     \bottomrule
    \end{tabular}
    }
    \caption{Results for the Comparison of Human Experts and LLM capabilities in Generating BPMN Models (on the Subset for the Human Evaluation).}
    \label{tab:comparison_human}
\end{table}

Similar to the LLM-generated BPMN models, the human-modeled BPMN models are analyzed to get the scores for the three quality dimensions $Q_{\text{syn}}$, $Q_{\text{prag}}$, and $Q_{\text{sem}}$. Results are shown in \autoref{tab:comparison_human}. 
We repeated the experiments with the LLMs on the same set of text-BPMN model pairs as used for the modeling experts.
This ensures that the basis of comparison is the same and that no differences in the complexity of the process descriptions or the number of textual descriptions distort the comparison. 

Upon examining the results, it becomes evident that the scores of the LLMs and human experts are largely comparable, falling within a similar range and having mostly similar tendencies. In some areas, the LLMs surpassed the quality scores of the experts. In terms of both pragmatic and syntactic quality, several LLMs outperformed the human experts, obtaining higher scores. However, for semantic quality, human experts achieved the best result with 0.5152. This indicates that LLMs make fewer syntactic errors, while humans reflect more on the textual descriptions and the overall semantic quality.
Nevertheless, the highest $Q_{qual}$ score, ignoring the validity dimension, was achieved by \textit{llama3.370b-instruct} with only four LLMs performing worse and seven better than the experts. When validity was taken into account, the experts performed best. 

\section{Discussion} \label{chap:discussion}
In the following, we discuss the BEF4LLM framework, the experimental results, and the limitations of our work.

\subsection{BEF4LLM Framework}
The BEF4LLM framework, presented in this work, is the first LLM-specific framework for assessing and comparing their performance in BPMN modeling, focusing on qualitative aspects of the generated process models. It provides a detailed view of the syntactic, pragmatic, and semantic quality, as well as the validity of BPMNs. Its grounding in established process model quality metrics and frameworks allows for a detailed and systematic evaluation of LLMs using 39 different metrics. Validity, which has been added as a fourth dimension to BEF4LLM due to the generative nature of LLMs, adds an important aspect to LLM-driven BPMN modeling: The generated BPMN models are only useful if they conform to the BPMN standard, which is often not the case when generating BPMN models with LLMs. Even if an LLM scores high in the three quality dimensions, it is not useful when most BPMN models are invalid and thus not parsable for subsequent software. In combination with the three quality dimensions, they allow for a detailed and extensive analysis of the strengths and weaknesses of LLMs in BPMN modeling. 

The BEF4LLM framework is adaptable in terms of the metrics it supports and the modeling languages it covers. New metrics can be added or existing ones removed based on their relevance. Further, most metrics can also be computed for other modeling languages like Petri nets or EPCs, which allows adapting the framework to those languages. Currently, all metrics and dimensions in the BEF4LLM framework are weighted equally. If there is evidence that certain metrics or dimensions are more important than others, the weights can be adjusted accordingly.

\subsection{Strengths and Weaknesses of LLMs in BPMN modeling}
The experimental results reveal several strengths and weaknesses of LLMs in BPMN modeling. In the following, we will elaborate on them.
The high pragmatic scores $Q_{\text{prag}}$, consistently above $0.8$, show a strength of LLMs in generating BPMN models that are easy to understand by humans. Similarly, the syntactic scores $Q_{\text{syn}}$ of more than $0.75$ for all LLMs hint at strong performance in following formal BPMN 2.0 syntax rules. However, deficits remain in the semantic quality of the generated BPMN models. 

Further inspection of specific quality metrics revealed common limitations. In terms of syntactic quality, many LLMs frequently violated the requirement to match join gateways to each split gateway, suggesting a systemic issue. 
Pragmatic quality metrics, particularly sequentiality and separability, were often lower (around $0.5$), indicating that LLMs often generated processes with multiple parallel paths or loops and limited block structuring. This makes the BPMN model appear unorganized. 
Within semantic quality, labeling accuracy (natural language subgroup) consistently scored lower than structural semantic quality metrics. This indicates that the LLMs have issues in generating fitting labels with regard to the automated measures. This suggests that activity naming remains challenging to match expert reference labels exactly. However, this result depends on the chosen similarity operationalization and may under-credit acceptable paraphrases.

A major weakness of LLMs is their ability to generate valid BPMN XML files; even though the layout part in the BPMN file had not to be generated, and one refinement loop was allowed. Only \textit{llama3.3:70b-instruct} reached a validity correctness of above 90\%, allowing it to be used in practice. There is substantial variability among LLMs, with $Q_{\text{val}}$ values ranging from 0.3067 for \textit{falcon3:10b-instruct} to 0.9733 for \textit{llama3.3:70b-instruct}.
Nevertheless, no specific textual description consistently led to invalid outputs across all LLMs, suggesting validity issues are primarily LLM-specific rather than input-dependent. However, generating a valid BPMN XML file can be considered a challenge in its own, given the length of a typical BPMN XML file and its strict requirements.

Additionally, operational challenges such as frequent generation timeouts, particularly in smaller LLMs, were identified. Further, the limited context length, especially for smaller LLMs, can quickly become a major issue in conversational BPMN-generating settings. In the current setting, the input given to the LLMs is already close to the limits of some LLMs like Phi4 with 14k context length. If the context length is exceeded, important information for the LLM is lost, causing the LLMs to generate unrelated content.
%. Further, we noticed that some LLMs generate unrelated content when the context length is exceeded. 
Therefore, more compact BPMN model representations are required \cite{brissard2025}.

Different LLMs show irregular variation in the generated BPMN model. While we intentionally aimed for some variety in the output of the LLMs to obtain different BPMN models, the variation was quite stark. This holds for the generated BPMN models within the same run, but also the generated BPMN models from the same LLM across the five runs. Some LLMs generated very different BPMN models with highly variable quality scores, even for the same input across the five runs. \textit{llama3.3:70b-instruct} was the LLMs performing most deterministically.

Overall, \textit{llama3.3:70b-instruct} performed best in our benchmark and significantly better than any other LLM if $Q_\text{total}$ is considered. Furthermore, it achieves high scores across all four dimensions, with only small gaps to the respective best values, and it clearly dominates in $Q_\text{val}$. However, smaller LLMs like \textit{llama3.1:8b-instruct} offer comparable BPMN quality at significantly lower computational cost, potentially compensating for the worse validity scores. In summary, the descriptive analysis highlights promising capabilities in syntactic and pragmatic dimensions while underscoring significant limitations in semantic accuracy, output consistency, and validity. These findings provide clear insights into areas requiring future LLM development and refinement.

\subsection{Comparison of BPMN-Modeling between LLMs and Human Experts}
Interestingly, the scores of the LLMs and human experts are largely comparable and fall within a similar range. They even show a similar tendency, i.e., the scores of both LLMs and experts tend to be low in the same areas. However, a significant difference can be found by comparing the scores for the individual BPMN models. The quality scores of LLM-generated BPMN models often exhibit a high variance, while the quality scores of human-created BPMN models show a relatively low variance. This suggests that human experts, in contrast to an LLM, tend to create process models with a relatively consistent level of quality. However, it is worth noting that even among human experts, certain metrics, such as partitionability, exhibit high variability in scores, suggesting that achieving consistency across some quality aspects can be challenging even among skilled process modelers.

Overall, the results indicate that LLMs can compete with the abilities of humans regarding the process model qualities measured within the BEF4LLM framework, with similar strengths and weaknesses. This result can also be interpreted as certain LLMs reaching the human "baseline" of modeling competencies. This can be explained by the fact that LLMs are also trained on human-generated BPMN models.

\subsection{Limitations}
Several limitations apply to the design of the framework and the conducted experiments. 
For now, the framework includes a considerable number of curated metrics. However, more metrics exist than can enhance the significance of the framework and extend it towards other aspects not covered for now, e.g., cyclicity metrics. 
The layout is not considered yet, but has an impact on the pragmatic quality since the positioning of elements has an influence on the readability of the BPMN model \cite{Ullrich_kompetenzorientiertes_2024}. However, we did not find established metrics with defined thresholds that could effectively assess the impact of the layout on the comprehensibility of BPMN models. This type of metric is therefore not consistent with the other metrics used in the pragmatic quality dimension, hindering its integration in the BEF4LLM framework. Further, not generating the layout with an LLM reduces the amount of tokens required significantly, since the layout makes up almost 70\% of a BPMN XML file \cite{brissard2025}. It can, by contrast, easily be added algorithmically to a BPMN XML file. Therefore, we argue that the layout is not a relevant aspect for LLM-driven BPMN modeling. 

Although the pragmatic scores are normalized by empirically validated thresholds, they are still affected by the overall size of the resulting BPMN model. If a complex process is described, resulting in a large BPMN model, this leads to a worse pragmatic score compared to a simple BPMN model. This tendency is also validated by the statistical tests with smaller LLMs, generating smaller BPMN models, reaching better pragmatic scores compared to larger LLMs generating larger BPMN models. Nevertheless, none of the 105 samples used in the experiments demands creating a disproportionately large BPMN model.

Further, the quality scores can only be calculated if the BPMN XML file is valid, which affects the interpretation of the results. Although the statistical tests using the Skillings-Mack test can deal with missing data and the results are therefore sound, the computed quality scores are still subject to being computed on potentially different subsets of BPMN models. We opted for having the validity as a separate dimension instead of rating invalid BPMN models with 0 in all quality dimensions to give a detailed picture of the ability of LLMs to generate valid BPMN models. This ability is indispensable for the practical applicability of LLMs. However, alternatives exist in existing research that propose process model formats that LLMs might understand more easily, e.g., in the form of a JSON \cite{kourani_promoai_2024}.

We aggregate the dimension scores into  $Q_{\mathrm{qual}}$ and $Q_{\mathrm{total}}$, mainly as transparent composite indices to summarize benchmarking results. They should, however, not be interpreted as a universally valid “general quality” score because we did not empirically validate a weighting scheme or the aggregation model against expert judgments or user studies.
Also, the BEF4LLM framework does not yet include LLM-specific metrics like the run-time of LLMs or memory requirements, which should be added in the future, since the response time is also an important factor for the applicability of such a solution. 

The experiments are also affected by several limitations. In general, the comparatively low number of available text-BPMN model pairs of only 105 samples limits the generalizability of the results. Similarly, the comparison with human experts is based on only nine different textual descriptions and 45 samples in total, limiting its significance. Therefore, one should interpret the results with these limitations in mind. The temperature, potentially affecting the quality of the generated BPMN model, was also fixed to $0.1$. Although initial experiments with other values did not result in better scores (neither lower temperature nor higher), the parameter has an effect on the quality of the generated BPMN models. Further research is required to determine the effect of this parameter. Due to the high token count of BPMN XML files, the input given to the LLMs might be exceeded for some LLMs, resulting in unpredictable results. This is especially relevant for cases when the refinement loop is triggered, where the input to the LLM also contains the currently faulty BPMN XML file. Also, a comparison to commercial LLMs like those from OpenAI or Google is missing.
Finally, a human interpretation of the generated BPMN models by the LLMs is missing. For now, the BPMN models are automatically inspected. However, since they are intended to be used by humans, a human judgment of the result is essential and might change the overall interpretation of the results.

\subsection{Practical Implications and Future Work}

The most prominent use of BEF4LLM is as a controlled benchmark to quantify the quality of generated BPMN models. In particular, the framework can be employed to compare alternative LLM prompting mechanisms (e.g., zero-shot vs. few-shot prompts, or prompts with explicit structural constraints) under otherwise identical conditions. Further, it allows to investigate the role of intermediate process representations. In that regard, BEF4LLM can be used to evaluate whether generating an intermediate structured format (such as JSON \cite{Larcardo_2025_BPMNAssistant_arXiv2509_24592}) improves the quality and/or the rate of valid BPMN outputs. 

Another usage scenario could be to extend it from passive evaluation to active improvement by integrating it in a refinement loop in which the BPMN generation is iteratively guided by validation feedback. This could help to improve the quality of the process model or guide the improvement of specific characteristics. 

Next to BPMN, several other process modeling languages, such as EPC or Petri nets, exist. A natural next step is to adapt the BEF4LLM framework to also support these languages. This requires revising the metric set to match the target notation: existing metrics must be transferred where applicable, BPMN-specific metrics must be excluded, and additional language-specific metrics (in particular, syntactic metrics derived from the respective modeling rules) must be introduced. 
Beyond language adaptation, BEF4LLM can be extended by incorporating further metrics that capture additional quality aspects not covered so far. To increase robustness, the semantic-quality assessment could be extended with additional metrics, for example, label similarity measures that support one-to-many alignments instead of enforcing strictly one-to-one label matching. Apart from that, the extension can also be made by covering more BPMN elements, like artifacts, to allow for a more exhaustive evaluation of supported elements. 
Finally, the framework would benefit from metric weighting (e.g., via expert elicitation or multi-criteria aggregation) to reflect differing evaluation priorities, e.g., to prefer semantic over pragmatic quality. The modular structure of our framework and implementation allows to switch the weighting between the metrics easily.

The framework only includes metrics that can be computed automatically.
Consequently, metrics that require human judgment, such as subjective readability or domain appropriateness, are excluded because they would hinder automated, scalable evaluation. To enable a more exhaustive assessment, integrating expert-in-the-loop evaluations constitutes a valuable extension. Such assessments could uncover human-centric issues specific to particular domains or user groups that may not be captured by purely automated metrics.

For practical use, one plausible deployment scenario is to integrate BEF4LLM into a quality assurance workflow that combines automated checks with expert oversight.
Concretely, organizations can operationalize this as a staged pipeline comprising generation, automated validation, and subsequent human review. This could filter out syntactically invalid artifacts early and ultimately enforce modeling conventions.

Moreover, benchmark-driven model selection enables organizations to select an LLM based on deployment priorities. This could include prioritizing XML validity in high-throughput settings versus prioritizing semantic fidelity in safety or compliance-critical processes. Our findings indicate a  trade-off between cost and quality in which larger LLMs tend to achieve stronger results, while also increasing purchase and operational costs. Thus, BEF4LLM can support evidence-based decisions on whether to deploy smaller LLMs broadly or larger models selectively for critical cases.

\section{Conclusion} \label{chap:conclusion}
This paper presented the BEF4LLM framework, accompanied by an extensive benchmark of recent and prominent open-source LLMs, to assess and compare their abilities in generating BPMN models from textual descriptions. Building upon the well-established SIQ framework and extending it to capture validity aspects unique to LLMs, our approach provides the first systematic and in-depth evaluation of LLMs’ capabilities for BPMN modeling.

Our findings reveal insightful distinctions regarding the strengths and weaknesses of open-source LLMs, particularly in comparison to a selected group of human modeling experts. Notably, LLM performance on key aspects, as measured by our framework, is often comparable to that of human modelers. Larger LLMs do not always yield better results; in our experiments, increased parameter counts sometimes coincided with lower pragmatic quality in the generated BPMN models, potentially because larger LLMs tend to produce more extensive and complex process models. Substantial differences exist between LLM families, with the larger LLMs from the Llama3 family achieving the highest overall scores. Nonetheless, generating valid BPMN XML files remains a significant hurdle for most LLMs, further amplified by context window limitations.
Our results also demonstrate that advances in benchmark scores for LLMs do not automatically translate to improved performance in the text-to-BPMN task. 
 
From an application standpoint, several LLMs show strong potential for practical use in text-to-BPMN tasks and conversational process modeling. Even without perfect results, these models can address the “blank-page” problem in process modeling, enabling more users to generate high-quality BPMN models with greater ease. In this way, AI-driven process design can substantially reduce barriers to entry for process modeling.

The results also aid future research. There is a great necessity for more compact and comprehensible BPMN representations to reduce token usage during generation—thereby making generation more efficient and increasing validity rates. Further, addressing LLMs’ deficiencies in semantic quality through targeted fine-tuning for BPMN tasks holds promise. The BEF4LLM framework can facilitate the development of such approaches by providing reliable feedback through its metrics. Finally, training LLMs with multi-turn conversational data could further enable robust deployment in chat-based process modeling scenarios.

\paragraph*{\textbf{Declaration of Generative AI and AI-assisted technologies in the writing process}}
During the preparation of this work, the authors used Llama3, Perplexity, ChatGPT o3 and Gemini to enhance language clarity, coherence, and rephrasing. After using this tool/service, the author reviewed and edited the content as needed and take full responsibility for the content of the publication.

\paragraph*{\textbf{Acknowledgment}}
Parts of this work were conducted within the project KICoPro (Grant 01IS24053C), funded by the Federal Ministry of Research, Technology and Space (BMFTR).
 
\newpage

\bibliographystyle{elsarticle-num} 
%\bibliography{bibliography}

\bibliography{bibliography}

@article{Krogstie_2006_quality_framework,
doi= {10.1057/palgrave.ejis.3000598},
author = {John Krogstie, Guttorm Sindre and Håvard Jørgensen},
title = {Process models representing knowledge for action: a revised quality framework},
journal = {European Journal of Information Systems},
volume = {15},
number = {1},
pages = {91--102},
year = {2006},
publisher = {Taylor \& Francis},
URL = {https://doi.org/10.1057/palgrave.ejis.3000598},
eprint = {https://doi.org/10.1057/palgrave.ejis.3000598}
}

@inproceedings{kourani_promoai_2024,
author = {Kourani, Humam and Berti, Alessandro and Schuster, Daniel and Van der Aalst, Wil M. P.},
title = {ProMoAI: Process Modeling with Generative AI},
year = {2024},
isbn = {978-1-956792-04-1},
url = {https://doi.org/10.24963/ijcai.2024/1014},
doi = {10.24963/ijcai.2024/1014},
booktitle = {Proceedings of the Thirty-Third International Joint Conference on Artificial Intelligence},
articleno = {1014},
numpages = {5},
location = {Jeju, Korea},
series = {IJCAI '24}
}

@inproceedings{klievtsova_conversational_2023,
	location = {Cham},
	title = {Conversational Process Modelling: State of the Art, Applications, and Implications in Practice},
	isbn = {978-3-031-41623-1},
	doi = {10.1007/978-3-031-41623-1_19},
	shorttitle = {Conversational Process Modelling},
	pages = {319--336},
	booktitle = {Business Process Management Forum},
	publisher = {Springer Nature Switzerland},
	author = {Klievtsova, Nataliia and Benzin, Janik-Vasily and Kampik, Timotheus and Mangler, Juergen and Rinderle-Ma, Stefanie},
	editor = {Di Francescomarino, Chiara and Burattin, Andrea and Janiesch, Christian and Sadiq, Shazia},
	date = {2023},
	langid = {english},
    year = {2023}
}

@InProceedings{grohs_large_2023,
	title = {Large Language Models can accomplish Business Process Management Tasks},
    booktitle="Business Process Management Workshops",
    editor="De Weerdt, Jochen and Pufahl, Luise",
    year="2024",
    publisher="Springer Nature Switzerland",
	doi = {10.48550/arXiv.2307.09923},
	author = {Grohs, Michael and Abb, Luka and Elsayed, Nourhan and Rehse, Jana-Rebecca},
}

@misc{kourani_process_2024,
	title = {Process Modeling With Large Language Models},
	url = {http://arxiv.org/abs/2403.07541},
	publisher="Springer Nature Switzerland",
    editor="van der Aa, Han
    and Bork, Dominik
    and Schmidt, Rainer
    and Sturm, Arnon",
	author = {Kourani, Humam and Berti, Alessandro and Schuster, Daniel and van der Aalst, Wil M. P.},
    year = {2024}
}

@Inbook{Reijers_Bsuiness_2010,
author="Reijers, Hajo A.
and Mendling, Jan
and Recker, Jan",
editor="Brocke, Jan vom
and Rosemann, Michael",
title="Business Process Quality Management",
publisher="Handbook on Business Process Management 1: Introduction, Methods, and Information Systems",
year="2010",
pages="167--185",
isbn="978-3-642-00416-2",
doi="10.1007/978-3-642-00416-2_8",
}

@Inbook{Mendling_Metrics_2008,
author="Mendling, Jan",
title="Metrics for Business Process Models",
bookTitle="Metrics for Process Models: Empirical Foundations of Verification, Error Prediction, and Guidelines for Correctness",
year="2008",
publisher="Springer Berlin Heidelberg",
address="Berlin, Heidelberg",
pages="103--133",
isbn="978-3-540-89224-3",
doi="10.1007/978-3-540-89224-3_4",
url="https://doi.org/10.1007/978-3-540-89224-3_4"
}

@article{Dijkman_Similarity_2011,
title = {Similarity of business process models: Metrics and evaluation},
journal = {Information Systems},
volume = {36},
number = {2},
pages = {498-516},
year = {2011},
note = {Special Issue: Semantic Integration of Data, Multimedia, and Services},
issn = {0306-4379},
doi = {https://doi.org/10.1016/j.is.2010.09.006},
url = {https://www.sciencedirect.com/science/article/pii/S0306437910001006},
author = {Remco Dijkman and Marlon Dumas and Boudewijn {van Dongen} and Reina Käärik and Jan Mendling}
}

@article{Schoknecht_SimilaritySOTA_2017,
  title={Similarity of Business Process Models—A State-of-the-Art Analysis},
  author={Andreas Schoknecht and Tom Thaler and Peter Fettke and Andreas Oberweis and Ralf Laue},
  journal={ACM Computing Surveys (CSUR)},
  year={2017},
  volume={50},
  pages={1 - 33},
  url={https://api.semanticscholar.org/CorpusID:9172088}
}

@Inbook{vanDongen_Measuring_2013,
author="van Dongen, Boudewijn
and Dijkman, Remco
and Mendling, Jan",
editor="Bubenko, Janis
and Krogstie, John
and Pastor, Oscar
and Pernici, Barbara
and Rolland, Colette
and S{\o}lvberg, Arne",
title="Measuring Similarity between Business Process Models",
publisher="Seminal Contributions to Information Systems Engineering: 25 Years of CAiSE",
year="2013",
pages="405--419",
isbn="978-3-642-36926-1",
doi="10.1007/978-3-642-36926-1_33",
url="https://doi.org/10.1007/978-3-642-36926-1_33"
}

@article{Becker_comparative_2012,
title = {A comparative survey of business process similarity measures},
journal = {Computers in Industry},
volume = {63},
number = {2},
pages = {148-167},
year = {2012},
issn = {0166-3615},
doi = {https://doi.org/10.1016/j.compind.2011.11.003},
url = {https://www.sciencedirect.com/science/article/pii/S0166361511001333},
author = {Michael Becker and Ralf Laue},
}

@techreort{Dijkman_2007Formal_SA,
  title={Formal semantics and automated analysis of BPMN process models},
  author={Remco M. Dijkman and Marlon Dumas and Chun Ouyang},
  year={2007},
institution  = {Queensland University of Technology},
  type         = {Technical Report Preprint},
  number       = {Vol. 5969},
  url={https://api.semanticscholar.org/CorpusID:7918800}
}

@InProceedings{Wong_2008_Process,
author="Wong, Peter Y. H.
and Gibbons, Jeremy",
editor="Liu, Shaoying
and Maibaum, Tom
and Araki, Keijiro",
title="A Process Semantics for BPMN",
booktitle="Formal Methods and Software Engineering",
year="2008",
publisher="Springer Berlin Heidelberg",
address="Berlin, Heidelberg",
pages="355--374",
}

@inproceedings{Snachez_2011_BPMImprovementMeasures,
  title={Business process model improvement based on measurement activities},
  author={S{\'a}nchez-Gonz{\'a}lez, Laura and Ruiz, Francisco and Garc{\'\i}a, F{\'e}lix and Piattini, Mario},
  booktitle={International Conference on Evaluation of Novel Software Approaches to Software Engineering},
  volume={2},
  pages={104--113},
  year={2011},
  organization={SCITEPRESS}
}

@mastersthesis{boomsma_2009_evaluation,
  title={An evaluation of thresholds for business process model metrics},
  author={Boomsma, R.D.},
  school={University of Twente},
  year={2009}
}

@article{Sanchez_2015_CaseStudyThresholds,
  title={A case study about the improvement of business process models driven by indicators},
  author={S{\'a}nchez-Gonz{\'a}lez, Laura and Garc{\'\i}a, F{\'e}lix and Ruiz, Francisco and Piattini, Mario},
  journal={Software \& Systems Modeling},
  volume={16},
  number={3},
  pages={759--788},
  year={2017},
  publisher={Springer}
}

@article{Berti_Benchmark_strategies_2024,
   author = {Alessandro Berti and Humam Kourani and Hannes Häfke and Chiao Yun Li and Daniel Schuster},
   doi = {10.1007/978-3-031-61007-3_2/TABLES/1},
   isbn = {9783031610066},
   issn = {18651356},
   journal = {Lecture Notes in Business Information Processing},
   pages = {13-21},
   publisher = {Springer Science and Business Media Deutschland GmbH},
   title = {Evaluating Large Language Models in Process Mining: Capabilities, Benchmarks, and Evaluation Strategies},
   volume = {511 LNBIP},
   url = {https://link.springer.com/chapter/10.1007/978-3-031-61007-3_2},
   year = {2024}
}

@article{Kourani_Benchmark_2024,
   author       = {Humam Kourani and
                  Alessandro Berti and
                  Daniel Schuster and
                  Wil M. P. van der Aalst},
  title        = {Evaluating Large Language Models on Business Process Modeling: Framework,
                  Benchmark, and Self-Improvement Analysis},
  journal      = {CoRR},
  volume       = {abs/2412.00023},
  year         = {2024},
  doi          = {10.48550/ARXIV.2412.00023},
  eprinttype    = {arXiv},
  eprint       = {2412.00023},
  bibsource    = {dblp computer science bibliography, https://dblp.org}
}

@article{Ziche2024,
   author = {Clara Ziche and Giovanni Apruzzese},
   doi = {10.1007/978-3-031-70445-1_35/FIGURES/6},
   isbn = {9783031704444},
   issn = {18651356},
   journal = {Lecture Notes in Business Information Processing},
   pages = {472-483},
   publisher = {Springer Science and Business Media Deutschland GmbH},
   title = {LLM4PM: A Case Study on Using Large Language Models for Process Modeling in Enterprise Organizations},
   volume = {527 LNBIP},
   url = {https://link.springer.com/chapter/10.1007/978-3-031-70445-1_35},
   year = {2024}
}

@InProceedings{Berti_Benchmark_evaluation_2024,
author="Berti, Alessandro
and Kourani, Humam
and van der Aalst, Wil M. P.",
editor="Delgado, Andrea
and Slaats, Tijs",
title="PM-LLM-Benchmark: Evaluating Large Language Models on Process Mining Tasks",
booktitle="Process Mining Workshops",
year="2025",
publisher="Springer Nature Switzerland",
address="Cham",
pages="610--623",
isbn="978-3-031-82225-4"
}

@article{Fournier_Benchmark_Causal_reasoning_2025,
   author = {Fabiana Fournier and Lior Limonad and Inna Skarbovsky},
   doi = {10.1007/978-3-031-78666-2_18/TABLES/7},
   isbn = {9783031786655},
   issn = {18651356},
   journal = {Lecture Notes in Business Information Processing},
   pages = {233-246},
   publisher = {Springer Science and Business Media Deutschland GmbH},
   title = {Towards a Benchmark for Causal Business Process Reasoning with LLMs},
   volume = {534 LNBIP},
   url = {https://link.springer.com/chapter/10.1007/978-3-031-78666-2_18},
   year = {2025}
}

@article{Busch_Benchmark_BPM_2024,
  title={Towards a benchmark for large language models for business process management tasks},
  author={Busch, Kiran and Leopold, Henrik},
  journal={arXiv preprint arXiv:2410.03255},
  year={2024}
}

@misc{BPMN_rule_set,
   author = {Object Management Group},
   title = {About the Business Process Model and Notation Specification Version 2.0.2},
   url = {https://www.omg.org/spec/BPMN},
   year = {2013}
}

@article{Becker_GOM_2000,
   author = {Jörg Becker and Michael Rosemann and Christoph von Uthmann},
   doi = {10.1007/3-540-45594-9_3},
   isbn = {978-3-540-45594-3},
   issn = {1611-3349},
   journal = {LNCS},
   pages = {30-49},
   publisher = {Springer, Berlin, Heidelberg},
   title = {Guidelines of Business Process Modeling},
   volume = {1806},
   url = {https://link.springer.com/chapter/10.1007/3-540-45594-9_3},
   year = {2000}
}

@article{Mendling_7PMG_2010,
   author = {J. Mendling and H. A. Reijers and W. M.P. van der Aalst},
   doi = {10.1016/J.INFSOF.2009.08.004},
   issn = {09505849},
   issue = {2},
   journal = {Information and Software Technology},
   month = {2},
   pages = {127-136},
   title = {Seven process modeling guidelines (7PMG)},
   volume = {52},
   url = {https://www.researchgate.net/publication/222694111_Seven_Process_Modeling_Guidelines_7PMG},
   year = {2010}
}

@article{Krogstie_sequal_2006_old,
   author = {John Krogstie and Guttorm Sindre and Håvard Jørgensen},
   doi = {10.1057/PALGRAVE.EJIS.3000598/FIGURES/3},
   issn = {14769344},
   issue = {1},
   journal = {European Journal of Information Systems},
   month = {2},
   pages = {91-102},
   publisher = {Palgrave Macmillan Ltd.},
   title = {Process models representing knowledge for action: A revised quality framework},
   volume = {15},
   url = {https://link.springer.com/article/10.1057/palgrave.ejis.3000598},
   year = {2006}
}

@inproceedings{Krogstie_sequal_1995,
  author    = {John Krogstie and Odd Ivar Lindland and Guttorm Sindre},
  title     = {Defining Quality Aspects for Conceptual Models},
  booktitle = {Proceedings of the IFIP 8.1 Working Conference on Information Systems Concepts (ISCO 3): Towards a Consolidation of Views},
  year      = {1995}
}

@incollection{Krogstie_sequal_BPM_2016,
  title={Quality of business process models},
  author={Krogstie, John},
  booktitle={Quality in business process modeling},
  pages={53--102},
  year={2016},
  publisher={Springer}
}

@article{Vaswani_attention_2023,
  title={Attention is all you need},
  author={Vaswani, Ashish and Shazeer, Noam and Parmar, Niki and Uszkoreit, Jakob and Jones, Llion and Gomez, Aidan N and Kaiser, {\L}ukasz and Polosukhin, Illia},
  journal={Advances in neural information processing systems},
  volume={30},
  year={2017}
}

@article{livebench,
  publtype={informal},
  author={Colin White and Samuel Dooley and Manley Roberts and Arka Pal and Benjamin Feuer and Siddhartha Jain and Ravid Shwartz-Ziv and Neel Jain and Khalid Saifullah and Siddartha Naidu and Chinmay Hegde and Yann LeCun and Tom Goldstein and Willie Neiswanger and Micah Goldblum},
  title={LiveBench: A Challenging, Contamination-Free LLM Benchmark},
  year={2024},
  cdate={1704067200000},
  journal={CoRR},
  volume={abs/2406.19314},
  url={https://doi.org/10.48550/arXiv.2406.19314}
}

@incollection{Dumas2018,
address = {Berlin, Heidelberg},
author = {Dumas, Marlon and {La Rosa}, Marcello and Mendling, Jan and Reijers, Hajo A},
booktitle = {Fundamentals of Business Process Management},
doi = {10.1007/978-3-662-56509-4_5},
isbn = {978-3-662-56509-4},
pages = {159--212},
publisher = {Springer Berlin Heidelberg},
title = {{Process Discovery}},
url = {http://link.springer.com/10.1007/978-3-662-56509-4_5},
year = {2018}
}

@article{Compagnucci2024,
author = {Compagnucci, Ivan and Corradini, Flavio and Fornari, Fabrizio and Re, Barbara},
doi = {10.1007/s12599-023-00818-7},
issn = {2363-7005},
journal = {Business \& Information Systems Engineering},
month = {feb},
number = {1},
pages = {43--66},
title = {{A Study on the Usage of the BPMN Notation for Designing Process Collaboration, Choreography, and Conversation Models}},
url = {https://doi.org/10.1007/s12599-023-00818-7 https://link.springer.com/10.1007/s12599-023-00818-7},
volume = {66},
year = {2024}
}

@inproceedings{Rolon_2008_Evaluation,
author = {Rolón, Elvira and Garcia, Felix and Ruiz, Francisco and Piattini, Mario and Visaggio, Corrado Aaron and Canfora, Gerardo},
year = {2008},
month = {01},
pages = {56-63},
title = {Evaluation of BPMN Models Quality - A Family of Experiments.},
booktitle={Third International Conference on Evaluation of Novel Approaches to Software Engineering},
}

@article{Vanderfeesten_crossconnectivity_2008,
   author = {Irene Vanderfeesten and Hajo A. Reijers and Jan Mendling and Wil M.P. Van Der Aalst and Jorge Cardoso},
   doi = {10.1007/978-3-540-69534-9_36},
   isbn = {978-3-540-69534-9},
   issn = {1611-3349},
   journal = {Lecture Notes in Computer Science},
   pages = {480-494},
   title = {On a Quest for Good Process Models: The Cross-Connectivity Metric},
   volume = {5074 LNCS},
   url = {https://link.springer.com/chapter/10.1007/978-3-540-69534-9_36},
   year = {2008}
}

@dataset{mangler_textual_2023,
  author       = {Mangler, Juergen and
                  Klievtsova, Nataliia},
  title        = {Textual Process Descriptions and Corresponding
                   BPMN Models },
  month        = mar,
  year         = 2023,
  publisher    = {Zenodo},
  doi          = {10.5281/zenodo.7783492},
}

@InProceedings{Norouzifar_Discovering_2024,
author="Norouzifar, Ali
and Kourani, Humam
and Dees, Marcus
and van der Aalst, Wil M. P.",
editor="Gdowska, Katarzyna
and G{\'o}mez-L{\'o}pez, Mar{\'i}a Teresa
and Rehse, Jana-Rebecca",
title="Bridging Domain Knowledge and Process Discovery Using Large Language Models",
booktitle="Business Process Management Workshops",
year="2025",
publisher="Springer Nature Switzerland",
address="Cham",
pages="44--56",
}

@article{Berti_Leveraging_2023,
  title={Leveraging large language models (LLMs) for process mining (Technical Report)},
  author={Berti, Alessandro and Qafari, Mahnaz Sadat},
  journal={arXiv preprint arXiv:2307.12701},
  year={2023}
}

@book{Jurafsky_speech_2024,
   author = {Daniel Jurafsky and James H. Martin},
   title = {Speech and Language Processing - An Introduction to Natural Language Processing,
Computational Linguistics, and Speech Recognition
with Language Models - Third Edition Draft},
   url = {https://web.stanford.edu/~jurafsky/slp3/},
   year = {2024}
}

@article{Casciani_AIAugmented_2024,
   author = {Angelo Casciani and Mario L. Bernardi and Marta Cimitile and Andrea Marrella},
   doi = {10.1007/978-3-031-59465-6_12/TABLES/2},
   isbn = {9783031594649},
   issn = {18651356},
   journal = {Lecture Notes in Business Information Processing},
   pages = {183-200},
   publisher = {Springer Science and Business Media Deutschland GmbH},
   title = {Conversational Systems for AI-Augmented Business Process Management},
   volume = {513},
   url = {https://link.springer.com/chapter/10.1007/978-3-031-59465-6_12},
   year = {2024}
}

@inproceedings{Makni_Modeling_2010,
   author = {Lobna Makni and Wiem Khlif and Nahla Zaaboub Haddar and Hanène Ben-Abdallah},
   booktitle = {INFORMATIK 2010 – Business Process and Service Science – Proceedings of ISSS and BPSC"},
   pages = {230--242},
   publisher = {Gesellschaft für Informatik e.V.},
   title = {A Tool forEvaluatingthe Quality of Business Process Models},
   year = {2010}
}

@phdthesis{Ullrich_kompetenzorientiertes_2024,
    author       = {Ullrich, Meike},
    year         = {2024},
    title        = {Kompetenzorientiertes E-Assessment für die grafische Modellierung in der Hochschullehre},
    doi          = {10.5445/IR/1000171737},
    publisher    = {{Karlsruher Institut für Technologie (KIT)}},
    pagetotal    = {315},
    school       = {Karlsruher Institut für Technologie (KIT)},
    language     = {german}
}

@article{Sanchez-Gonzlez_BPMIMA_2017,
   author = {Laura Sánchez-González and Félix García and Francisco Ruiz and Mario Piattini},
   doi = {10.1007/S10270-015-0482-0/TABLES/10},
   issn = {16191374},
   issue = {3},
   journal = {Software and Systems Modeling},
   month = {7},
   pages = {759-788},
   publisher = {Springer Verlag},
   title = {A case study about the improvement of business process models driven by indicators},
   volume = {16},
   url = {https://link.springer.com/article/10.1007/s10270-015-0482-0},
   year = {2017}
}

@article{Kopp_prag_quality_assess_2021,
   author = {Andrii Kopp and Dmytro Orlovskyi},
   doi = {10.1007/978-3-030-77592-6_5/FIGURES/11},
   isbn = {9783030775919},
   issn = {18650937},
   journal = {Communications in Computer and Information Science},
   pages = {93-118},
   publisher = {Springer Science and Business Media Deutschland GmbH},
   title = {Towards the Method and Information Technology for Evaluation of Business Process Model Quality},
   volume = {1308},
   url = {https://link.springer.com/chapter/10.1007/978-3-030-77592-6_5},
   year = {2021}
}

@inproceedings{Sorg_complexity_2025,
  title={Process Model Complexity Metrics, Cognitive Load and Visual Behavior: A Multi-granular Eye-Tracking Analysis},
  author={Sorg, Thierry and Abbad-Andaloussi, Amine and Kindler, Ekkart and Weber, Barbara},
  booktitle={International Conference on Business Process Modeling, Development and Support},
  pages={87--103},
  year={2025},
  organization={Springer}
}

@inproceedings{Lauer_Generating_2025,
  title={Conversational Business Process Modeling using LLMs: Initial Results and Challenges},
  author={Lauer, Chantale and Pfeiffer, Peter and Rombach, Alexander and Mehdiyev, Nijat},
  booktitle={EMISA 2025},
  year={2025},
  organization={Gesellschaft f{\"u}r Informatik eV}
}

@misc{brissard2025,
archivePrefix = {arXiv},
arxivId = {cs.CL/2507.11356},
author = {Brissard, Alexis and Cuppens, Fr{\'{e}}d{\'{e}}ric and Zouaq, Amal},
eprint = {2507.11356},
primaryClass = {cs.CL},
title = {{What is the Best Process Model Representation? A Comparative Analysis for Process Modeling with Large Language Models}},
url = {https://arxiv.org/abs/2507.11356},
year = {2025}
}

@inproceedings{Wornow_Wonderbread_2024,
 author = {Wornow, Michael and Narayan, Avanika and Viggiano, Ben and Khare, Ishan S. and Verma, Tathagat and Thompson, Tibor and Hernandez, Miguel Angel Fuentes and Sundar, Sudharsan and Trujillo, Chloe and Chawla, Krrish and Lu, Rongfei and Shen, Justin and Nagaraj, Divya and Martinez, Joshua and Agrawal, Vardhan and Hudson, Althea and Shah, Nigam H. and R\'{e}, Christopher},
 booktitle = {Advances in Neural Information Processing Systems},
 editor = {A. Globerson and L. Mackey and D. Belgrave and A. Fan and U. Paquet and J. Tomczak and C. Zhang},
 pages = {115963--116021},
 publisher = {Curran Associates, Inc.},
 title = {WONDERBREAD: A Benchmark for Evaluating Multimodal Foundation Models on Business Process Management Tasks},
 url = {https://proceedings.neurips.cc/paper_files/paper/2024/file/d1fa821312040303b089ae529dbf81a6-Paper-Datasets_and_Benchmarks_Track.pdf},
 volume = {37},
 year = {2024}
}

@inproceedings{Saber_formalization_2014,
  title={BPMN formalization and verification using Maude},
  author={El-Saber, Nissreen and Boronat, Artur},
  booktitle={Proceedings of the 2014 Workshop on Behaviour Modelling-Foundations and Applications},
  pages={1--12},
  year={2014}
}

@article{mendeling_thresholds_2012,
title = {Thresholds for error probability measures of business process models},
journal = {Journal of Systems and Software},
volume = {85},
number = {5},
pages = {1188-1197},
year = {2012},
issn = {0164-1212},
doi = {https://doi.org/10.1016/j.jss.2012.01.017},
url = {https://www.sciencedirect.com/science/article/pii/S0164121212000040},
author = {Jan Mendling and Laura Sánchez-González and Félix García and Marcello {La Rosa}},
}

@article{Peeperkorn_Temperature_2024,
  title={Is temperature the creativity parameter of large language models?},
  author={Peeperkorn, Max and Kouwenhoven, Tom and Brown, Dan and Jordanous, Anna},
  journal={arXiv preprint arXiv:2405.00492},
  year={2024}
}

@article{Dettmers_GPT3_2022,
   author = {Tim Dettmers and Mike Lewis and Younes and Belkada and Luke Zettlemoyer},
   journal = {Advances in Neural Information Processing Systems},
   month = {12},
   pages = {30318-30332},
   title = {GPT3.int8(): 8-bit Matrix Multiplication for Transformers at Scale},
   volume = {35},
   url = {https://github.com/TimDettmers/bitsandbytes},
   year = {2022}
}

@article{Chatfleld_Skillings_Mack_2009,
   author = {Mark Chatfleld and Adrian Mander},
   doi = {10.1177/1536867x0900900208},
   issn = {1536867X},
   issue = {2},
   journal = {The Stata journal},
   pages = {299},
   pmid = {19829764},
   publisher = {DPC Nederland},
   title = {The Skillings–Mack test (Friedman test when there are missing data)},
   volume = {9},
   url = {https://pmc.ncbi.nlm.nih.gov/articles/PMC2761045/},
   year = {2009}
}

@book{gibbons1993nonparametric,
  title={Nonparametric statistics: An introduction},
  author={Gibbons, Jean Dickinson},
  volume={9},
  year={1993},
  publisher={Sage}
}

@mastersthesis{ekstedt_2015_quality,
  title={Quality of business process models expressed in BPMN},
  author={Ekstedt, Joakim},
  school={KTH, School of Information and Communication Technology (ICT)},
  year={2015},
  url={https://www.diva-portal.org/smash/get/diva2:830545/FULLTEXT01.pdf}
}

@article{kourani_2024_promoai,
  author    = {Kourani, Humam and Berti, Alessandro and van der Aalst, Wil M. P.},
  title     = {Process Modeling With Large Language Models},
  journal   = {arXiv preprint arXiv:2403.07541},
  year      = {2024}
}

@misc{KopkeSafan2024EfficientConversationalProcessModeling,
  author = {{Julius K{\"o}pke and Aya Safan}},
  title  = {Efficient LLM-Based Conversational Process Modeling},
  year   = {2024},
  url    = {https://isys.uni-klu.ac.at/PDF/BPM_2024_paper_1442.pdf}
}

@misc{Larcardo_2025_BPMNAssistant_arXiv2509_24592,
  author        = {Larcardo, Josip Tomo and Tankovic, Nikola and Etinger, Darko},
  title         = {BPMN Assistant: An LLM-Based Approach to Business Process Modeling},
  year          = {2025},
  eprint        = {2509.24592},
  archivePrefix = {arXiv},
  url           = {https://arxiv.org/abs/2509.24592},
  doi           = {10.48550/arXiv.2509.24592}
}

@misc{Camunda_2026_BPMNCopilotDocs,
  author  = {{Camunda}},
  title   = {{BPMN Copilot (Camunda 8 Docs)}},
  year    = {2026},
  url     = {https://docs.camunda.io/docs/components/early-access/alpha/bpmn-copilot/},
  urldate = {2026-02-17}
}

@article{Guiterrez_2024_DirectManipulationInterfaceLLMProcessModeling,
  author = {Guitierrez, Sergio Pardo amd Alrabie, Lina and Di Fede, Guilia and Vitali, Monica and Andolina, Salvatore},
  title  = {A Direct Manipulation Interface for LLM-based Process Modeling},
  year   = {2025},
  doi    = {10.1145/3750069.3755960},
  url    = {https://dl.acm.org/doi/10.1145/3750069.3755960}
  }

@misc{Kourani_2024_EnhancedProcessModelComprehension,
  author        = {Kourani, Humama and Berti, Alessandro and Hennrich, Jasmin and Kratsch, Wolfgang and Weidlich, Robin and Li, Chiao-Yun and Arslan, Ahmad and Schuster, Daniel and van der Aalst, Wil M.P.},
  title         = {Leveraging Large Language Models for Enhanced Process Model Comprehension},
  year          = {2024},
  eprint        = {2408.08892},
  archivePrefix = {arXiv},
  url           = {https://arxiv.org/abs/2408.08892}
}

@article{Hoerner2026AutomaticallyGeneratingBPMN,
  author  = {H{\"o}rner, L. F. and M{\"o}ller, M. and Reichert, M.},
  title   = {Automatically Generating BPMN 2.0 Process Models from Natural Language Process Descriptions: Challenges, Framework, Quality Assessment},
  journal = {Business \& Information Systems Engineering},
  year    = {2026},
  doi     = {10.1007/s12599-025-00983-x},
  url     = {https://doi.org/10.1007/s12599-025-00983-x}
}
\appendix

\section{Additional information to the Framework}
\autoref{Tab:Criteria_syntactic_quality_long}--\autoref{Tab:Criteria_semantic_quality_long} list the metrics for each quality dimension, including the corresponding formulas and references.

We use the following notation: $\mathcal{E}^{S},\mathcal{E}^{E},\mathcal{E}^{I}$ start/end/intermediate events; $\mathcal{A}$ activities; $\mathcal{G}$ gateways with $\mathcal{G}^{S}$ (split) and $\mathcal{G}^{J}$ (join); $\mathcal{FO}$ flow objects; $\mathcal{F}^{S},\mathcal{F}^{M}$ sequence/message flows; $\mathcal{P}$ processes; $\mathcal{PO}$ pools; $\mathit{in}(x),\mathit{out}(x)$ incoming/outgoing sequence flows of $x$; $\operatorname{Matchgate}:\mathcal{G}^{S}\to\mathcal{G}^{J}$; $\textit{Excp}:\mathcal{E}^{I}\rightharpoonup\mathcal{A}$ with $\operatorname{dom}(\textit{Excp})$ interrupting events; $\operatorname{label}(t)$ task label ($\bot$ if missing); $\operatorname{norm}_{\mathrm{asc/desc}}$ ascending/descending normalization; $\textit{Paths}$ start-to-end node sequences; $p(l)$ gateway-type share for $l\in{\mathrm{AND},\mathrm{XOR},\mathrm{OR}}$; $\operatorname{depth}(fo)$ nesting depth; Cut-Vertex = articulation node; $M_c,M_g$ candidate/ground-truth models; $\mathcal{FO}_i,\mathcal{F}_i$ node/edge sets; $\tau_i:N_i\to\mathcal{Ty}$ type map with $\mathcal{G}\subset\mathcal{Ty}$ gateway types excluded from label similarity; $l$ node label; $\operatorname{ed}$ Levenshtein distance; $w_i,w_s$ label-similarity weights; $sn,se$ inserted/deleted nodes/edges; $CF_i$ causal-footprint relations; $DG_i$ dependency-graph edges.

\begin{table}[htbp]
\footnotesize
\centering
\begin{tabular}{C{0.01\linewidth}L{0.29\linewidth}L{0.54\linewidth}C{0.04\linewidth}}
\toprule
 & \textbf{Metric description} & \textbf{Calculation} & \textbf{Ref.}\\
\midrule
1 & Existence of a start event &
$\exists e\in\mathcal{E}^{\mathrm{S}}$ &
\cite{Dijkman_2007Formal_SA}\\
2 & Existence of an end event &
$\exists e\in\mathcal{E}^{\mathrm{E}}$ &
\cite{Dijkman_2007Formal_SA}\\
3 & One start event per process &
%$\dfrac{\lvert\mathcal{E}^{\mathrm{S}}\rvert}{\lvert\mathcal{P}\rvert}=1$ &
$\dfrac{\left| \left\{ p \in \mathcal{P} \mid \mathcal{E}^{\mathrm{S}} \neq \varnothing \right\} \right|}{|\mathcal{P}|}$ &
\cite{Dijkman_2007Formal_SA}\\
4 & One end event per process &
%$\dfrac{\lvert\mathcal{E}^{\mathrm{E}}\rvert}{\lvert\mathcal{P}\rvert}=1$ &
$\dfrac{\left| \left\{ p \in \mathcal{P} \mid \mathcal{E}^{\mathrm{E}} \neq \varnothing \right\} \right|}{|\mathcal{P}|}$ &
\cite{Dijkman_2007Formal_SA}\\
5 & Sequence‑flow connection rules &
\scalebox{0.9}{$\displaystyle
\frac{\bigl\lvert\{f=(x,y)\in\mathcal{F}^{\mathrm{S}}\mid
\substack{x\in\mathcal{E}^{\mathrm{S}}\cup\mathcal{E}^{\mathrm{I}}\cup
              \mathcal{A}\cup\mathcal{G},\\
          y\in\mathcal{E}^{\mathrm{E}}\cup\mathcal{E}^{\mathrm{I}}\cup
              \mathcal{A}\cup\mathcal{G}}\}\bigr\rvert}
     {\lvert\mathcal{F}^{\mathrm{S}}\rvert}$} &
\cite{BPMN_rule_set}\\
6 & Message‑flow connection rules &
\scalebox{0.9}{$\displaystyle
\frac{\bigl\lvert\{f=(x,y)\in\mathcal{F}^{\mathrm{M}}\mid
\substack{x\in\mathcal{PO}\cup\mathcal{A}\cup
          \mathcal{E}^{\mathrm{E}}_{M}\cup\mathcal{E}^{\mathrm{I}}_{M},\\
          y\in\mathcal{PO}\cup\mathcal{A}\cup
          \mathcal{E}^{\mathrm{S}}_{M}\cup\mathcal{E}^{\mathrm{I}}_{M}}\}\bigr\rvert}
     {\lvert\mathcal{F}^{\mathrm{M}}\rvert}$} &
\cite{BPMN_rule_set}\\
7 & Start event: $in=0$, $out=1$ &
$\dfrac{\lvert\{e\in\mathcal{E}^{\mathrm{S}}\mid
               |in(e)|=0\land|out(e)|=1\}\rvert}
      {\lvert\mathcal{E}^{\mathrm{S}}\rvert}$ &
\cite{Dijkman_2007Formal_SA}\\
8 & End event: $in=1$, $out=0$ &
$\dfrac{\lvert\{e\in\mathcal{E}^{\mathrm{E}}\mid
               |in(e)|=1\land|out(e)|=0\}\rvert}
      {\lvert\mathcal{E}^{\mathrm{E}}\rvert}$ &
\cite{Dijkman_2007Formal_SA}\\
9 & Split gateway has matching join gateway &
$\dfrac{\lvert\{g_{s}\in\mathcal{G}^{\mathrm{S}}\mid
\exists g_{j}\in\mathcal{G}^{\mathrm{J}}\!:
\operatorname{Matchgate}(g_{s})=g_{j}\}\rvert}
      {\lvert\mathcal{G}^{\mathrm{S}}\rvert}$ &
\cite{Dijkman_2007Formal_SA}\\
10 & Exactly one process per pool &
%$\forall po\in\mathcal{PO}:
%  \lvert\{P\in\mathcal{P}\mid P\subseteq po\}\rvert=1$ &
$\dfrac{\left| \left\{ po \in \mathcal{PO} \mid \lvert\{ P \in \mathcal{P} \mid P \subseteq po \} \rvert = 1 \right\} \right|}{|\mathcal{PO}|}$&
\cite{Wong_2008_Process}\\
11 & Each observable task has a label &
$\dfrac{\lvert\{t\in\mathcal{T}\mid \operatorname{label}(t)\neq\varnothing\}\rvert}
      {\lvert\mathcal{T}\rvert}$ &
\cite{BPMN_rule_set}\\
12 & Task: $in=1$, $out=1$ &
$\dfrac{\lvert\{t\in\mathcal{T}\mid |in(t)|=1\land|out(t)|=1\}\rvert}
      {\lvert\mathcal{T}\rvert}$ &
\cite{Dijkman_2007Formal_SA}\\
13 & Non‑exception intermediate event: $in=1$, $out=1$ &
$\dfrac{\lvert\{e\in\mathcal{E}^{\mathrm{I}}\setminus
           \operatorname{dom}(Excp)\mid
|in(e)|=1\land|out(e)|=1\}\rvert}
      {\lvert\mathcal{E}^{\mathrm{I}}\rvert}$ &
\cite{Dijkman_2007Formal_SA}\\
14 & Exception event: $in=0$, $out=1$ &
$\dfrac{\lvert\{e\in\operatorname{dom}(Excp)\mid
           |in(e)|=0\land|out(e)|=1\}\rvert}
      {\lvert\operatorname{dom}(Excp)\rvert}$ &
\cite{Dijkman_2007Formal_SA}\\
15 & Split gateway: $in=1$, $out>1$ &
$\dfrac{\lvert\{g\in\mathcal{G}^{\mathrm{S}}\mid
               |in(g)|=1\land|out(g)|>1\}\rvert}
      {\lvert\mathcal{G}^{\mathrm{S}}\rvert}$ &
\cite{Dijkman_2007Formal_SA}\\
16 & Join gateway: $in>1$, $out=1$ &
$\dfrac{\lvert\{g\in\mathcal{G}^{\mathrm{J}}\mid
               |in(g)|>1\land|out(g)|=1\}\rvert}
      {\lvert\mathcal{G}^{\mathrm{J}}\rvert}$ &
\cite{Dijkman_2007Formal_SA}\\
\bottomrule
\end{tabular}
%%\footnotetext{Exception events are excluded from the matching‑gateway rule.}
\caption{Metrics for syntactic quality in the \textsc{BEF4LLM} framework. 
%All ratios are normalised to $[0,1]$, so no further scaling is required when aggregating the syntactic quality score.
} 
\label{Tab:Criteria_syntactic_quality_long}
\end{table}

\begin{table}[htbp!]
\footnotesize
\centering
%\rowcolors{2}{}{gray!20}
\begin{tabular}{C{0.01\linewidth}L{0.29\linewidth}L{0.54\linewidth}C{0.04\linewidth}}
\toprule
 & \textbf{Metric} & \textbf{Calculation} & \textbf{Ref.} \\ 
\midrule
\multicolumn{4}{c}{\textbf{Size}} \\ \midrule
1 & TNN (total number of nodes) &
$\operatorname{norm}_{desc}\!\bigl(\lvert\mathcal{FO}\rvert\bigr)$ &
\cite{Mendling_Metrics_2008}\\
2 & TNG (total number of gateways) &
$\operatorname{norm}_{desc}\!\bigl(\lvert\mathcal{G}\rvert\bigr)$ &
\cite{Rolon_2008_Evaluation}\\
3 & TNSF (total number of sequence flows) &
$\operatorname{norm}_{desc}\!\bigl(\lvert\mathcal{F}^{\mathrm{S}}\rvert\bigr)$ &
\cite{Rolon_2008_Evaluation}\\
4 & TNMF (total number of message flows) &
$\operatorname{norm}_{desc}\!\bigl(\lvert\mathcal{F}^{\mathrm{M}}\rvert\bigr)$ &
\cite{Rolon_2008_Evaluation}\\
5 & Diameter &
\scalebox{0.9}{$\displaystyle
\operatorname{norm}_{desc}\!\bigl(\max\{\,\lvert p\rvert \mid p\in\text{Paths}\}\bigr)$} &
\cite{Mendling_Metrics_2008}\\
\midrule
\multicolumn{4}{c}{\textbf{Density}} \\ \midrule
6 & Density &
\scalebox{0.9}{$\displaystyle
\operatorname{norm}_{desc}\!\Bigl(
\dfrac{\lvert\mathcal{F}^{\mathrm{S}}\rvert}
      {\lvert\mathcal{FO}\rvert(\lvert\mathcal{FO}\rvert-1)}
\Bigr)$} &
\cite{Mendling_Metrics_2008}\\
7 & AGD (average gateway degree) &
\scalebox{0.9}{$\displaystyle
\operatorname{norm}_{desc}\!\Bigl(
\dfrac{\sum_{g\in\mathcal{G}}(\lvert\mathit{in}(g)\rvert+
      \lvert\mathit{out}(g)\rvert)}
      {\lvert\mathcal{G}\rvert}
\Bigr)$} &
\cite{Mendling_Metrics_2008}\\
8 & CNC (connectivity coefficient) &
$\operatorname{norm}_{desc}\!\bigl(
\lvert\mathcal{F}^{\mathrm{S}}\rvert/\lvert\mathcal{FO}\rvert\bigr)$ &
\cite{Mendling_Metrics_2008}\\
\midrule
\multicolumn{4}{c}{\textbf{Connector interplay}} \\ \midrule
9 & GH (gateway heterogeneity) &
$\operatorname{norm}_{desc}\!\Bigl(
-\!\!\sum_{l\in\{\mathrm{AND},\mathrm{XOR},\mathrm{OR}\}}
p(l)\log_{3}p(l)\Bigr)$ &
\cite{Mendling_Metrics_2008}\\
10 & CFC (control‑flow complexity) &
\scalebox{0.9}{$\displaystyle
\operatorname{norm}_{desc}\!\Bigl(
\sum_{g\in\mathcal{G}^{\mathrm{S}}_{\mathrm{AND}}}\!1
+\!\!\sum_{g\in\mathcal{G}^{\mathrm{S}}_{\mathrm{XOR}}}\!\lvert\mathit{out}(g)\rvert
+\!\!\sum_{g\in\mathcal{G}^{\mathrm{S}}_{\mathrm{OR}}}\!(2^{\lvert\mathit{out}(g)\rvert}-1)
\Bigr)$} &
\cite{Mendling_Metrics_2008}\\
11 & CC (cross‑connectivity) &
$\operatorname{norm}_{asc}\!\bigl(
\mathit{len}(\text{longest\_loop})\bigr)$ &
\cite{Vanderfeesten_crossconnectivity_2008}\\
\midrule
\multicolumn{4}{c}{\textbf{Partitionability}} \\ \midrule
12 & Sequentiality &
\scalebox{0.9}{$\displaystyle
\operatorname{norm}_{desc}\!\Bigl(
\dfrac{\sum_{g\in\mathcal{G}}(\lvert\mathit{in}(g)\rvert+
      \lvert\mathit{out}(g)\rvert)}
      {\lvert\mathcal{G}\rvert}
\Bigr)$} &
\cite{Mendling_Metrics_2008}\\
13 & Separability &
\scalebox{0.9}{$\displaystyle
\operatorname{norm}_{asc}\!\Bigl(
\dfrac{\lvert\{\,fo\in\mathcal{FO}\mid fo\text{ ist Cut‑Vertex}\,\}\rvert}
      {\lvert\mathcal{FO}\rvert-2}
\Bigr)$} &
\cite{Mendling_Metrics_2008}\\
14 & Depth &
$\operatorname{norm}_{desc}\!\bigl(
\max\{\operatorname{depth}(fo)\mid fo\in\mathcal{FO}\}\bigr)$ &
\cite{Mendling_Metrics_2008}\\
\midrule
\multicolumn{4}{c}{\textbf{Concurrency}} \\ \midrule
15 & TS (token split) &
\scalebox{0.9}{$\displaystyle
\operatorname{norm}_{desc}\!\Bigl(
\sum_{g\in
      (\mathcal{G}^{\mathrm{S}}_{\mathrm{OR}}\cup
       \mathcal{G}^{\mathrm{S}}_{\mathrm{AND}})}
(\lvert\mathit{out}(g)\rvert-1)
\Bigr)$} &
\cite{Mendling_Metrics_2008}\\
\bottomrule
\end{tabular}
\caption{Metric set for pragmatic quality in the \textsc{BEF4LLM} framework.  
}
\label{Tab:Criteria_pragmatic_quality_long}
\end{table}

All metrics in \autoref{Tab:Criteria_semantic_quality_long} rely on a common notation. $\mathcal{FO}$ and $\mathcal{F}$ denote the flow objects and connection objects sets of a model, while $\tau:\mathcal{FO}\!\rightarrow\!Type$ assigns BPMN element types; gateway types collected in $\mathcal G\subset Type$ are excluded from label‑based comparisons. Label strings are turned into word sets $w$ after tokenisation and stemming; $w_i$ and $w_s$ weight exact word overlap and synonym overlap, respectively, with synonyms detected by a function $s(\cdot,\cdot)$.  Optimal bipartite matchings $M_{\mathrm{SimX}}^{\mathrm{opt}}\subseteq \mathcal{FO}_{c}\times \mathcal{FO}_{g}$ maximize similarity under criterion X ($\mathrm{X}\in\{\mathrm{Syn},\mathrm{Sem},\mathrm{Ctx}\}$). For structural metrics $sn$ and $se$ denote inserted or deleted flow objects and connection objects, whereas $CF_i$ and $DG_i$ contain the causal‑footprint relations and dependency‑graph edges extracted from model $i$. 
Each metric is already scaled to the interval $[0,1]$, so their aggregation yields the overall semantic score $Q_{\text{sem}}$ without additional normalization.

\newcolumntype{F}[1]{>{\scriptsize\centering\arraybackslash}p{#1}}

\begin{sidewaystable}[htbp]
\footnotesize
\centering
%\rowcolors{2}{}{gray!20}
\begin{tabular}{L{0.01\textheight}L{0.2\textheight}F{0.28\textheight}F{0.37\textheight}C{0.02\textheight}}
\toprule
&\textbf{Metric} & \textbf{Calculation per node} &
\textbf{Calculation for the whole graph} & \textbf{Ref.} \\
\midrule
\multicolumn{5}{c}{\textbf{Natural‑language similarity}}\\\midrule
1 & Syntactic label similarity &
$1-\dfrac{\operatorname{ed}(label(x_1),label(x_2))}
        {\max(\lvert label(x_1)\rvert,\lvert label(x_2)\rvert)}$ &
$\displaystyle
\frac{2\sum_{(n,m)\in M_{\mathrm{SimSyn}}^{\mathrm{opt}}}
          \operatorname{SimSyn}(n,m)}
     {\lvert\{n\in \mathcal{FO}_{c}\mid\tau_{c}(n)\notin\mathcal G\}\rvert+
      \lvert\{n\in \mathcal{FO}_{g}\mid\tau_{g}(n)\notin\mathcal G\}\rvert}$ &
\hspace{2mm}\cite{Dijkman_Similarity_2011}\\[3pt]

2 & Semantic label similarity &
$\displaystyle
\frac{2w_{i}\lvert w_{1}\cap w_{2}\rvert+
       w_{s}\bigl(s(w_{1},w_{2})+s(w_{2},w_{1})\bigr)}
      {\lvert w_{1}\rvert+\lvert w_{2}\rvert}$ &
$\displaystyle
\frac{2\sum_{(n,m)\in M_{\mathrm{SimSem}}^{\mathrm{opt}}}
          \operatorname{SimSem}(n,m)}
     {\lvert\{n\in \mathcal{FO}_{c}\mid\tau_{c}(n)\notin\mathcal G\}\rvert+
      \lvert\{n\in \mathcal{FO}_{g}\mid\tau_{g}(n)\notin\mathcal G\}\rvert}$ &
\hspace{2mm}\cite{Dijkman_Similarity_2011}\\[4pt]

3 & Context similarity &
$\displaystyle
\frac{\lvert M_{\mathrm{Sim}}^{\mathrm{opt,in}}\rvert}
     {2\sqrt{\lvert n_{1}^{\mathrm{in}}\rvert\,\lvert n_{2}^{\mathrm{in}}\rvert}}
+
\frac{\lvert M_{\mathrm{Sim}}^{\mathrm{opt,out}}\rvert}
     {2\sqrt{\lvert n_{1}^{\mathrm{out}}\rvert\,\lvert n_{2}^{\mathrm{out}}\rvert}}$ &
$\displaystyle
\frac{2\sum_{(n,m)\in M_{\mathrm{SimCtx}}^{\mathrm{opt}}}
          \operatorname{SimCtx}(n,m)}
     {\lvert\{n\in \mathcal{FO}_{c}\mid\tau_{c}(n)\notin\mathcal G\}\rvert+
      \lvert\{n\in \mathcal{FO}_{g}\mid\tau_{g}(n)\notin\mathcal G\}\rvert}$ &
\hspace{2mm}\cite{Dijkman_Similarity_2011}\\
\midrule
\multicolumn{5}{c}{\textbf{Graph‑structure similarity}}\\\midrule
4 & Graph‑edit distance &
— &
$\displaystyle
%1-\operatorname{avg}(s_{nv},s_{ev},s_{bv}),\;
\begin{aligned}
1-\operatorname{avg}(s_{nv},s_{ev},s_{bv})\\
s_{nv}&=\frac{\lvert sn\rvert}{\lvert \mathcal{FO}_{c}\rvert+\lvert \mathcal{FO}_{g}\rvert},\\
s_{ev}&=\frac{\lvert se\rvert}{\lvert \mathcal{F}_{c}\rvert+\lvert \mathcal{F}_{g}\rvert},\\
s_{bv}&=\frac{2\sum_{(n,m)\in M}(1-\operatorname{Sim}(n,m))}
               {\lvert \mathcal{FO}_{c}\rvert+\lvert \mathcal{FO}_{g}\rvert-\lvert sn\rvert}
\end{aligned}$ &
\hspace{2mm}\cite{Dijkman_Similarity_2011}\\[5pt]

5 & Common nodes and edges &
— &
$\displaystyle
1-\frac{\lvert \mathcal{FO}_{c}\setminus \mathcal{FO}_{g}\rvert+\lvert \mathcal{FO}_{g}\setminus \mathcal{FO}_{c}\rvert+
        \lvert \mathcal{F}_{c}\setminus \mathcal{F}_{g}\rvert+\lvert \mathcal{F}_{g}\setminus \mathcal{F}_{c}\rvert}
       {\lvert \mathcal{F=}_{c}\rvert+\lvert \mathcal{FO}_{g}\rvert+\lvert \mathcal{F}_{c}\rvert+\lvert \mathcal{F}_{g}\rvert}$ &
\hspace{2mm}\cite{Becker_comparative_2012}\\
\midrule
\multicolumn{5}{c}{\textbf{Behavioural similarity}}\\\midrule
6 & Causal‑footprint overlap &
— &
$\displaystyle
\operatorname{Sim}_{\mathrm{CF}}=
\frac{\lvert CF_{c}\cap CF_{g}\rvert}{\lvert CF_{c}\cup CF_{g}\rvert}$ &
\hspace{2mm}\cite{Dijkman_Similarity_2011,vanDongen_Measuring_2013}\\[4pt]

7 & Dependency‑graph overlap &
— &
$\displaystyle
\operatorname{Sim}_{\mathrm{DG}}=
\frac{\lvert DG_{c}\cap DG_{g}\rvert}{\lvert DG_{c}\cup DG_{g}\rvert}$ &
\hspace{2mm}\cite{Dijkman_Similarity_2011,vanDongen_Measuring_2013}\\
\bottomrule
\end{tabular}
\caption{Metric set for semantic quality. }
\label{Tab:Criteria_semantic_quality_long}
\end{sidewaystable}

\autoref{Tab:thresholds} lists all thresholds for the pragmatic quality dimension and the corresponding paper that defines this threshold.

\begin{table}[htbp!]
\centering
\begin{tabular}{lccccc}
    \toprule
        \textbf{Metric} & \textbf{$t_1$} & \textbf{$t_2$} & \textbf{$t_3$} & \textbf{$t_4$} & \textbf{Ref.} \\
        \midrule
        TNN 
        & 29.9
        & 43.7
        & 58.1
        & 81.1
        & \cite{Snachez_2011_BPMImprovementMeasures}
        \\

        TNG 
        & 1.42
        & 3.36
        & 5.3
        & 6.49
        & \cite{Snachez_2011_BPMImprovementMeasures}
        \\

        TNSF 
        & 19.4
        & 34.8
        & 50.2
        & 74.8
        & \cite{Snachez_2011_BPMImprovementMeasures}
        \\

        TNMF 
        & 1.09
        & 7.15
        & 13.2
        & 22.8
        & \cite{Snachez_2011_BPMImprovementMeasures}
        \\

        Diameter 
        & 7.92
        & 12.2
        & 16.5
        & 23.4
        & \cite{Snachez_2011_BPMImprovementMeasures}
        \\

        Density 
        & 0.1361169 
        & 0.357143
        & 0.741667
        & 2.33333
        & \cite{ekstedt_2015_quality}
        \\

        AGD 
        & 3.67
        & 3.88
        & 4.06
        & 4.18
        &  \cite{Sanchez_2015_CaseStudyThresholds}
        \\

        CNC 
        & 0.37
        & 0.9
        & 1.43
        & 4.18
        &  \cite{boomsma_2009_evaluation}
        \\
        
        GH 
        & 0.62 
        & 0.79
        & 0.92
        & 0.94
        &  \cite{Sanchez_2015_CaseStudyThresholds}
        \\

        CFC 
        & 13
        & 22
        & 37
        & 51
        &  \cite{Sanchez_2015_CaseStudyThresholds}
        \\

        CC 
        & 0.007996
        & 0.030407
        & 0.061814
        & 0.112903
        &  \cite{ekstedt_2015_quality}
        \\

        Sequentiality 
        & 0.25
        & 0.48
        & 0.7
        & 1.07
        & \cite{Snachez_2011_BPMImprovementMeasures}
        \\

        Separability 
        & 0.03
        & 0.37
        & 0.71
        & 1.24
        &  \cite{boomsma_2009_evaluation}
        \\

        Depth
        & 0.42
        & 1.72
        & 3.02
        & 5.09
        &  \cite{Sanchez_2015_CaseStudyThresholds}
        \\

        TS
        & 0.12
        & 0.21
        & 0.6
        & 1.36
        & \cite{boomsma_2009_evaluation}
        \\
        \bottomrule
\end{tabular}
\caption{Thresholds for pragmatic quality metrics.}
\label{Tab:thresholds}
\end{table}

\newpage

\section{Complete Results}
\autoref{tab:results_full} lists the full results for all LLMs, including those that generated less than 30 valid BPMNs per run. Further, we extended the table with a column "timed-out", indicating the number of textual descriptions for which a time-out occurred, and therefore, no BPMN was generated.

\begin{table}[htbp!]
    \centering
    \tabcolsep=5pt
    
    \resizebox{\textwidth}{!}{
    \begin{tabular}{lcccccccc}
    \toprule
         & \multicolumn{2}{c}{\textbf{Validity}} & \multicolumn{3}{c}{\textbf{Process Model Quality}} & \multicolumn{2}{c}{\textbf{Total Scores}} & \\
        \cmidrule(lr){2-3} \cmidrule(lr){4-6}  \cmidrule(lr){7-8} \cmidrule{9-9}
        \textbf{LLM} & $\overline{Q_{\text{val}}}$ & AVBM & $\overline{Q_{\text{syn}}}$ & $\overline{Q_{\text{prag}}}$ & $\overline{Q_{\text{sem}}}$ & $\overline{Q_{\text{qual}}}$ & $\overline{Q_{\text{total}}}$ & time-out  \\ \midrule
        deepseek-r1:1.5b-qwen-distill & 0 & 0 & - & - & - & 0 & 2\\
        deepseek-r1:14b-qwen-distill & 0.6362 & 66.8 & 0.8548 & 0.8841 & 0.5270 & 0.7553 &  0.7225 & 0 \\
        deepseek-r1:8b-llama-distill & 0.2248 & 23.6 & 0.7497 & \textbf{0.9165} & 0.4404 & 0.7022 &  0.5828 & 0 \\
        deepseek-r1:70b-llama-distill & 0.7905 & 83 & 0.8708 & 0.8636 & 0.5609 & 0.7651 &0.7714 & 0 \\
        \midrule
        falcon3:3b-instruct & 0.0800 & 8.4 & 0.7387 & 0.9129 & 0.4412 & 0.6976 & 0.5432 & 30 \\
        falcon3:10b-instruct & 0.3067 & 32.2 & \textbf{0.9082} & 0.8837  & 0.5544 & 0.7821 & 0.6632 & 5 \\ \midrule
        llama3.2:1b-instruct & 0.0133 & 1.4 & 0.8243 & 0.9174 & 0.4738 & 0.7385 &0.2249 & 56 \\
        llama3.1:8b-instruct & 0.7067 & 74.2 & 0.8597 & 0.8954 & 0.5389 & 0.7647 & 0.7502 & 6 \\
        llama3.3:70b-instruct &\textbf{0.9733} & 102.2 & 0.8955 & 0.8721 & 0.5747 & 0.7808 & \textbf{0.8289} & 2 \\ \midrule
        phi4:14b & 0.5848 & 61.4 & 0.8580 & 0.8710 & 0.5666 & 0.7652 & 0.7201 & 0 \\ \midrule
        qwen2.5:1.5b-instruct & 0.1562 & 16.4 & 0.7838 & 0.9300 & 0.4841 & 0.7326 & 0.5885 & 124 \\
        qwen2.5:14b-instruct & 0.5467 & 57.4 & 0.8076 & 0.8907 & 0.5521 & 0.7501 & 0.6993 & 0 \\
        qwen2.5:32b-instruct & 0.5105 & 53.6 & 0.8782 & 0.8834 & \textbf{0.5768} & 0.7795 & 0.7122 & 0 \\ \midrule
        qwen3:1.7b & 0.2267 & 23.8 & 0.8770 & 0.8895 & 0.4933 & 0.7655 & 0.6241 & 32 \\
        qwen3:14b & 0.6533 & 68.6 & 0.8643 & 0.8670 & 0.5720 & 0.7678 & 0.7392 &  22 \\
        qwen3:30b-a3b & 0.4895 & 51.4 & 0.8792 & 0.8877 & 0.5485 & 0.7718 & 0.7012 & 2 \\
        qwen3:235b-a22b & 0.5771 & 60.6 & 0.8790 & 0.8537 & 0.5620 & 0.7649 & 0.7180 &  97\\
     \bottomrule
     \end{tabular}
     }
     \caption{Results of the first experiment, including all tested LLMs. The column timed-out indicated the total number of BPMNs that are not generated by the LLM due to a time-out across all 5 runs.}
     \label{tab:results_full}
\end{table}

\newpage

\section{Detailed Results with all Metric results}
The following tables show more detailed results for the three quality dimensions. \autoref{tab:results_temp_01_detail} shows the results for the subgroups of the quality dimensions per LLM, while \autoref{tab:results_temp_01_metric_syn_1} to \autoref{tab:results_temp_01_metric_sem} summarize the results for each metric per LLM. 

\begin{sidewaystable}[htbp]
    \tiny
    \centering

    \begin{tabular}{L{0.17\textheight}C{0.06\textheight}C{0.03\textheight}C{0.05\textheight}C{0.06\textheight}C{0.09\textheight}C{0.08\textheight}C{0.07\textheight}C{0.08\textheight}C{0.06\textheight}}
    \toprule
       \textbf{LLM} & \textbf{Syntactic quality} & \textbf{Size} & \textbf{Density} & \textbf{Connector interplay} & \textbf{Partionability} & \textbf{Concurrency} & \textbf{Natural language} & \textbf{Graph structure} & \textbf{Behaviour}  \\ \midrule
       %\cmidrule(lr){1-1} \cmidrule(lr){2-2} \cmidrule(lr){3-3} \cmidrule(lr){4-4} \cmidrule(lr){5-5} \cmidrule(lr){6-6}n\cmidrule(lr){7-7} \cmidrule(lr){8-8} \cmidrule(lr){9-9} \cmidrule(lr){10-10}
        deepseek-r1:1.5b-qwen-distill & - & - & - & - & -& - & - & - &   \\ 
        deepseek-r1:14b-qwen-distill & 0.8548 & 0.9517 & 0.8476 & 0.9815 & 0.7116 & 0.8806 & 0.3690 & 0.7720 & 0.5189  \\ 
        deepseek-r1:8b-llama-distill& 0.7497 & 0.9596 & 0.9318 & 0.9615 & 0.7691 & 0.9505 & 0.2776 & 0.7581 & 0.3670  \\  
        deepseek-r1:70b-llama-distill& 0.8708 & 0.9454 & 0.8349 & 0.9835 & 0.6993 & 0.6743 & 0.4015 & 0.7855 & 0.5754  \\ \midrule
        Falcon3:3b-instruct & 0.7387 & 0.9022 & 0.9302 & 0.9614 & 0.8581 & 0.9333 & 0.2816 & 0.7571 & 0.3645  \\ 
        falcon3:10b-instruct & 0.9082 & 0.9564 & 0.8309 & 0.9917 & 0.7023 & 0.8987 & 0.3919 & 0.7907 & 0.5619  \\ \midrule
        llama3.2:1b-instruct& 0.8243 & 0.9938 & 0.8507 & 1.0000 & 0.7465 & 1.0000 & 0.2972 & 0.7546 & 0.4579  \\ 
        llama3.1:8b-instruct& 0.8597 & 0.9543 & 0.8481 & 0.9941 & 0.7460 & 0.8942 & 0.3827 & 0.7827 & 0.5294  \\ 
        llama3.3:70b-instruct& 0.8955 & 0.9433 & 0.8294 & 0.9933 & 0.7014 & 0.7924 & 0.4212 & 0.7902 & 0.5894  \\ \midrule
        phi4:14b& 0.8580 & 0.9410 & 0.8402 & 0.9872 & 0.6933 & 0.8371 & 0.4086 & 0.7827 & 0.5873  \\ \midrule
        qwen2.5:1.5b-instruct& 0.7838 & 0.9746 & 0.8900 & 0.9862 & 0.7653 & 0.9538 & 0.3151 & 0.7556 & 0.4661  \\ 
        qwen2.5:14b-instruct& 0.8076 & 0.9472 & 0.8728 & 0.9809 & 0.7391 & 0.9488 & 0.4028 & 0.7846 & 0.5436  \\ 
        qwen2.5:32b-instruct& 0.8782 & 0.9351 & 0.8580 & 0.9942 & 0.7165 & 0.8999 & 0.4234 & 0.7909 & 0.5929  \\ \midrule
        qwen3:1.7b& 0.8770 & 0.9529 & 0.8520 & 0.9810 & 0.7386 & 0.9099 & 0.3075 & 0.7651 & 0.5004  \\ 
        qwen3:14b& 0.8643 & 0.9485 & 0.8379 & 0.9896 & 0.7063 & 0.8136 & 0.4326 & 0.7886 & 0.5645  \\ 
        qwen3:235b-a22b & 0.8790 & 0.9114 & 0.8467 & 0.9861 & 0.6823 & 0.7276 & 0.4126 & 0.7859 & 0.5622  \\ 
        qwen3:30b-a3b& 0.8792 & 0.9282 & 0.8408 & 0.9980 & 0.7055 & 0.9537 & 0.3886 & 0.7818 & 0.5553 \\ 
     \bottomrule
     \end{tabular}
     \caption{Scores for the categories in the different quality dimensions (each LLM is used with q\_8 quantization).}
     \label{tab:results_temp_01_detail}
\end{sidewaystable}

\newpage

\begin{sidewaystable}[htbp]
    \scriptsize
    \centering

    \begin{tabular}{L{0.2\textheight}C{0.08\textheight}C{0.08\textheight}C{0.07\textheight}C{0.07\textheight}C{0.07\textheight}C{0.07\textheight}C{0.07\textheight}C{0.09\textheight}}
    \toprule
        \textbf{LLM} & \textbf{Existence Start Event} & \textbf{Existence End Event} & \textbf{Start Event Degrees} & \textbf{End Event Degrees} & \textbf{Labeled Tasks} & \textbf{Connected Nodes} & \textbf{Tasks Degrees} & \textbf{Intermediate Event Degrees} \\
        \midrule
        deepseek-r1:1.5b-qwen-distill & - & - & - & - & - & - & - & -   \\ 
        deepseek-r1:14b-qwen-distill &  1.0000 &  0.9879 &  0.9339 &  0.8316 &  0.9939 &  0.7014 &  0.7391 &  1.0000 \\ 
        deepseek-r1:8b-llama-distill &  1.0000 &  0.9886 &  0.5423 &  0.4754 &  1.0000 &  0.2074 &  0.2497 &  1.0000 \\
        deepseek-r1:70b-llama-distill &  0.9978 &  0.9476 &  0.9603 &  0.8701 &  1.0000 &  0.8126 &  0.8262 &  0.9802 \\ 
         \midrule
         falcon3:3b-instruct &  1.0000 &  0.8690 &  0.3048 &  0.3810 &  1.0000 &  0.1826 &  0.2055 &  1.0000   \\
        falcon3:10b-instruct &  1.0000 &  0.9518 &  0.9701 &  0.9233 &  1.0000 &  0.8407 &  0.8462 &  1.0000 \\ 
         \midrule
         llama3.2:1b-instruct &  1.0000 &  0.6667 &  1.0000 &  1.0000 &  1.0000 &  0.5083 &  0.4762 &  1.0000   \\ 
        llama3.1:8b-instruct &  1.0000 &  0.9728 &  0.9346 &  0.8299 &  0.9889 &  0.7360 &  0.8050 &  0.9971  \\ 
        llama3.3:70b-instruct &  1.0000 &  0.8471 &  0.9765 &  0.9922 &  0.9705 &  0.8290 &  0.7428 &  0.9725  \\ 
        phi4:14b &  1.0000 &  0.9874 &  0.8376 &  0.9223 &  1.0000 &  0.8150 &  0.8047 &  0.9751   \\ \midrule
        qwen2.5: 1.5b-instruct &  1.0000 &  0.8630 &  0.7391 &  0.3622 &  0.9800 &  0.2931 &  0.5521 &  1.0000  \\ 
        qwen2.5:14b-instruct &  1.0000 &  0.9830 &  0.9015 &  0.6983 &  0.9845 &  0.6055 &  0.6383 &  0.9666  \\ 
        qwen2.5:32b-instruct &  1.0000 &  0.9602 &  0.9099 &  0.8200 &  1.0000 &  0.7713 &  0.7715 &  0.9891 \\ \midrule
        qwen3: 1.7b &  0.9592 &  0.9864 &  0.9643 &  0.8456 &  1.0000 &  0.6996 &  0.8051 &  1.0000 \\ 
        qwen3:14b &  0.9833 &  0.9367 &  0.9009 &  0.9053 &  1.0000 &  0.7199 &  0.7357 &  0.9608 \\ 
        qwen3:30b-a3b &  0.9971 &  0.9976 &  0.9012 &  0.9382 &  1.0000 &  0.8324 &  0.7848 &  0.9520 \\
        qwen3:235b-a22b &  0.9914 &  0.9853 &  0.9509 &  0.8924 &  0.9971 &  0.8283 &  0.7839 &  0.8601  \\ 
     \bottomrule
     \end{tabular}
     \caption{Scores for the syntactic quality metrics for the first experiment, part 1 (each LLM is used with q\_8 quantization).}
     \label{tab:results_temp_01_metric_syn_1}
\end{sidewaystable}

\normalsize

\begin{sidewaystable}[htbp]
    \scriptsize
    \centering
    \begin{tabular}{L{0.21\textheight}C{0.07\textheight}C{0.07\textheight}C{0.1\textheight}C{0.06\textheight}C{0.06\textheight}C{0.07\textheight}C{0.07\textheight}C{0.09\textheight}}
    \toprule
     \textbf{LLM} & \textbf{Gateway Degrees} & \textbf{Gateway Pairs} & \textbf{Event Gateway Pre- \& Successor} & \textbf{One Start Event} & \textbf{One End Event} & \textbf{Wrong Sequence Flow} & \textbf{Wrong Message Flow} & \textbf{One Process In Pool} \\
    \midrule
    deepseek-r1:1.5b-qwen-distill  & - & - & - & - & - & - & - &   \\ 
    deepseek-r1:14b-qwen-distill  &0.7047 &0.1852 &1.0000 &0.9940 &0.8611 &0.8207 &0.9870 &1.0000  \\ 
    deepseek-r1:8b-llama-distill  &0.1507 &0.2932 &1.0000 &0.9579 &0.9779 &0.9813 &1.0000 &1.0000  \\ 
    deepseek-r1:70b-llama-distill  &0.8129 &0.2080 &1.0000 &0.9971 &0.8701 &0.6654 &0.9754 &1.0000  \\ \midrule
    falcon3:3b-instruct  &0.1140 &0.5416 &1.0000 &1.0000 &1.0000 &0.9873 &0.8500 &1.0000  \\ 
    falcon3:10b-instruct  &0.8442 &0.2128 &1.0000 &0.9894 &0.9402 &0.9866 &1.0000 &1.0000  \\ \midrule
    llama3.2:1b-instruct  &1.0000 &1.0000 &1.0000 &1.0000 &1.0000 &1.0000 &1.0000 &1.0000  \\ 
    llama3.1:8b-instruct  &0.5638 &0.2059 &1.0000 &0.9582 &0.8903 &0.8831 &1.0000 &1.0000  \\ 
    llama3.3:70b-instruct  &0.9153 &0.3168 &1.0000 &0.9963 &0.8771 &0.9306 &0.9884 &1.0000  \\ \midrule
    phi4:14b  &0.8805 &0.1749 &1.0000 &0.9586 &0.8065 &0.4962 &0.9880 &1.0000  \\ \midrule
    qwen2.5:1.5b-instruct  &0.8230 &0.8804 &1.0000 &0.9042 &0.9966 &0.4594 &1.0000 &1.0000  \\ 
    qwen2.5:14b-instruct  &0.7599 &0.2430 &1.0000 &0.9750 &0.8115 &0.4572 &1.0000 &1.0000  \\ 
    qwen2.5:32b-instruct  &0.8159 &0.2624 &1.0000 &0.9906 &0.8134 &0.9521 &0.9264 &1.0000  \\ \midrule
    qwen3:1.7b  &0.5383 &0.4238 &1.0000 &0.9921 &0.9631 &0.9752 &1.0000 &1.0000  \\ 
    qwen3:14b  &0.8352 &0.2355 &0.9938 &0.9969 &0.8898 &0.9770 &0.8186 &1.0000  \\ 
    qwen3:30b-a3b &0.8517 &0.1309 &1.0000 &0.9982 &0.8464 &0.9539 &0.8607 &1.0000 \\ 
    qwen3:235b-a22b  &0.8729 &0.2716 &1.0000 &0.9900 &0.7865 &0.9511 &0.8372 &1.0000  \\ 
     \bottomrule
     \end{tabular}
     \caption{Scores for the syntactic quality metrics for the first experiment, part 2 (each LLM is used with q\_8 quantization).}
     \label{tab:results_temp_01_metric_syn_2}
\end{sidewaystable}

\begin{sidewaystable}[htbp]
    \scriptsize
    \centering
    \begin{tabular}{L{0.2\textheight}C{0.07\textheight}C{0.07\textheight}C{0.07\textheight}C{0.07\textheight}C{0.07\textheight}C{0.07\textheight}C{0.07\textheight}C{0.07\textheight}}
    \toprule
     \textbf{LLM} & \textbf{TNG} & \textbf{TNN} & \textbf{TNSF} & \textbf{TNMF} & \textbf{diameter} & \textbf{Density} & \textbf{AGD} & \textbf{CNC} \\
    \midrule
        deepseek-r1:1.5b-qwen-distill  & - & - & - & - & - & - & - & -   \\ 
        deepseek-r1:14b-qwen-distill  &0.7993 &1.0000 &0.9968 &0.9903 &0.9760 &0.9369 &0.9890 &0.6127  \\ 
        deepseek-r1:8b-llama-distill  &0.7954 &1.0000 &1.0000 &1.0000 &1.0000 &0.9869 &1.0000 &0.8295\\ 
        deepseek-r1:70b-llama-distill  &0.8166 &1.0000 &0.9962 &0.9877 &0.9300 &0.9447 &0.9906 &0.5682  \\ 
        \midrule
        falcon3:3b-instruct-  &0.4929 &1.0000 &1.0000 &1.0000 &1.0000 &0.9887 &1.0000 &0.8655   \\
        falcon3:10b-instruct-  &0.8649 &1.0000 &0.9975 &1.0000 &0.9169 &0.9270 &0.9813 &0.5715  \\ 
        \midrule 
        llama3.2:1b-instruct-  &1.0000 &1.0000 &1.0000 &1.0000 &1.0000 &0.8333 &1.0000 &0.5833  \\ 
        llama3.1:8b-instruct-  &0.7890 &0.9942 &0.9958 &1.0000 &0.9871 &0.9133 &1.0000 &0.6227 \\ 
        llama3.3:70b-instruct-  &0.8485 &0.9975 &0.9873 &0.9598 &0.9221 &0.9412 &0.9745 &0.5740  \\ \midrule
        phi4:14b-  &0.7991 &0.9993 &0.9837 &0.9960 &0.9034 &0.9593 &0.9891 &0.5718   \\ \midrule
        qwen2.5:1.5b-instruct-  &0.9203 &1.0000 &1.0000 &1.0000 &0.9492 &0.9483 &1.0000 &0.7155 \\ 
        qwen2.5:14b-instruct-  &0.7624 &0.9990 &0.9963 &0.9982 &0.9568 &0.9550 &0.9873 &0.6448  \\ 
        qwen2.5:32b-instruct-  &0.7656 &1.0000 &0.9941 &0.9676 &0.9299 &0.9753 &0.9937 &0.6127 \\ \midrule
        qwen3:1.7b-  &0.8469 &1.0000 &0.9982 &1.0000 &0.9462 &0.9378 &0.9935 &0.6414   \\ 
        qwen3:14b-  &0.8207 &1.0000 &0.9936 &0.9410 &0.9366 &0.9485 &0.9813 &0.5814   \\ 
        qwen3:30b-a3b- &0.8218 &1.0000 &0.9844 &0.9437 &0.8893 &0.9511 &1.0000 &0.5686  \\  
        qwen3:235b-a22b  &0.7584 &0.9980 &0.9782 &0.9207 &0.9035 &0.9700 &0.9861 &0.5813   \\ 
     \bottomrule
     \end{tabular}
     \caption{Scores for the pragmatic quality metrics for the first experiment, part 1 (each LLM is used with q\_8 quantization).}
     \label{tab:results_temp_01_metric_prag_1}
\end{sidewaystable}

\begin{sidewaystable}[htbp]
    \footnotesize
    \centering
    \begin{tabular}{L{0.23\textheight}C{0.06\textheight}C{0.06\textheight}C{0.12\textheight}C{0.06\textheight}C{0.12\textheight}C{0.07\textheight}C{0.07\textheight}}
    \toprule
     \textbf{LLM} & \textbf{GH} & \textbf{CFC} & \textbf{Sequentiality} & \textbf{CC} & \textbf{Seperatibility} & \textbf{Depth} & \textbf{TS} \\
    \midrule
        deepseek-r1:1.5b-qwen-distill  & - & - & - & - & - & - & - \\
        deepseek-r1:14b-qwen-distill  &0.9597 &1.0000 &0.5909 &0.9920 &0.5477 &0.9779 &0.8882  \\
        deepseek-r1:8b-llama-distill  &0.9614 &1.0000 &0.7022 &0.9340 &0.6265 &0.9759 &0.9505  \\
        deepseek-r1:70b-llama-distill  &0.9550 &0.9994 &0.5579 &0.9983 &0.5716 &0.9594 &0.6978  \\ \midrule
        falcon3:3b-instruct  &0.9577 &0.9917 &0.9536 &0.9863 &0.7940 &0.9774 &1.0000  \\
        falcon3:10b-instruct  &0.9768 &1.0000 &0.5823 &0.9963 &0.5180 &0.9885 &0.9013  \\ \midrule
        llama3.2:1b-instruct  &1.0000 &1.0000 &0.7500 &1.0000 &0.4167 &1.0000 &1.0000  \\
        llama3.1:8b-instruct  &0.9863 &1.0000 &0.6630 &0.9993 &0.5905 &0.9897 &0.8983  \\
        llama3.3:70b-instruct  &0.9789 &1.0000 &0.5765 &0.9995 &0.5534 &0.9745 &0.7880  \\ \midrule
        phi4:14b  &0.9732 &0.9978 &0.5419 &0.9969 &0.5330 &0.9631 &0.7847  \\ \midrule
        qwen2.5:1.5b-instruct  &1.0000 &1.0000 &0.7394 &0.9963 &0.6763 &1.0000 &1.0000  \\
        qwen2.5:14b-instruct  &0.9708 &0.9954 &0.6714 &0.9675 &0.5406 &0.9816 &0.9215  \\
        qwen2.5:32b-instruct  &0.9913 &0.9992 &0.5968 &0.9962 &0.5553 &0.9799 &0.8803  \\ \midrule
        qwen3:1.7b  &0.9143 &1.0000 &0.7056 &0.9960 &0.6365 &0.9446 &0.8857  \\
        qwen3:14b  &0.9759 &1.0000 &0.5530 &0.9859 &0.5351 &0.9632 &0.7588  \\
        qwen3:30b-a3b &0.9994 &1.0000 &0.5642 &0.9970 &0.5145 &1.0000 &0.9976  \\   
        qwen3:235b-a22b  &0.9756 &0.9963 &0.5550 &0.9859 &0.5422 &0.9427 &0.7701  \\
     \bottomrule
    \end{tabular}
    \caption{Scores for the pragmatic quality metrics for the first experiment, part 2 (each LLM is used with q\_8 quantization).}
    \label{tab:results_temp_01_metric_prag_2}
\end{sidewaystable}

\begin{sidewaystable}[htbp]
    \footnotesize
    \centering
    \tabcolsep=5pt

    \begin{tabular}{L{0.23\textheight}C{0.11\textheight}C{0.11\textheight}C{0.11\textheight}C{0.07\textheight}C{0.07\textheight}C{0.07\textheight}C{0.07\textheight}}
    \toprule
     \textbf{LLM} & \textbf{Node Matching (Syntactic Sim)} & \textbf{Node Matching (Semantic Sim)} & \textbf{Node Matching (Context Sim)} & \textbf{GED} & \textbf{PCN} & \textbf{CF} & \textbf{DGS} \\
    \midrule
        deepseek-r1:1.5b-qwen-distill  & - & - & - & - & - & - & - \\
        deepseek-r1:14b-qwen-distill  &0.4011 &0.3798 &0.3341 &0.5551 &0.9898 &0.8461 &0.1986 \\
        deepseek-r1:8b-llama-distill  &0.3574 &0.3107 &0.1418 &0.5098 &0.9979 &0.6932 &0.0316 \\
        deepseek-r1:70b-llama-distill  &0.4196 &0.4074 &0.3850 &0.5778 &0.9908 &0.8912 &0.2612 \\ \midrule
        falcon3:3b-instruct  &0.3945 &0.3495 &0.0974 &0.5356 &0.9746 &0.6725 &0.0470 \\
        falcon3:10b-instruct  &0.4171 &0.4018 &0.3611 &0.5899 &0.9912 &0.8876 &0.2330 \\ \midrule
        llama3.2:1b-instruct  &0.3090 &0.3500 &0.2879 &0.5602 &1.0000 &0.8267 &0.1616 \\
        llama3.1:8b-instruct  &0.4130 &0.4043 &0.3301 &0.5682 &0.9954 &0.8567 &0.2049 \\
        llama3.3:70b-instruct  &0.4274 &0.4240 &0.4117 &0.5858 &0.9938 &0.8686 &0.3073 \\ \midrule
        phi4:14b  &0.3996 &0.3951 &0.4189 &0.5687 &0.9935 &0.9017 &0.2790 \\ \midrule
        qwen2.5:1.5b-instruct  &0.3567 &0.3394 &0.2869 &0.5275 &0.9905 &0.8055 &0.1465 \\
        qwen2.5:14b-instruct  &0.4386 &0.4312 &0.3592 &0.5795 &0.9976 &0.8540 &0.2411 \\
        qwen2.5:32b-instruct  &0.4404 &0.4362 &0.3947 &0.5887 &0.9939 &0.8947 &0.2687 \\ \midrule
        qwen3:1.7b  &0.3055 &0.2883 &0.3078 &0.5419 &0.9936 &0.8286 &0.1953 \\
        qwen3:14b  &0.4630 &0.4449 &0.3960 &0.5886 &0.9949 &0.8793 &0.2669 \\
        qwen3:30b-a3b  &0.3962 &0.3862 &0.3732 &0.5769 &0.9900 &0.8731 &0.2311 \\
        qwen3:235b-a22b  &0.4295 &0.4159 &0.4253 &0.5821 &0.9947 &0.8893 &0.2731 \\
     \bottomrule
    \end{tabular}
    \caption{Scores for the semantic quality metrics for the first experiment (each LLM is used with q\_8 quantization).}
    \label{tab:results_temp_01_metric_sem}
\end{sidewaystable}

%--------------------------alt---------------
%\input{bib}
%\printbibliography

\end{document}